\documentclass[twocolumn,prl,amsmath,amssymb,superscriptaddress]{revtex4-2}

\usepackage{amsthm}
\usepackage{graphicx}
\usepackage{color}
\usepackage{dcolumn}
\usepackage{bm}
\usepackage{bbm}
\usepackage{mathtools}      
\usepackage{mathrsfs}       
\usepackage{braket}
\usepackage{float}
\usepackage[table]{xcolor}

\newcommand{\cE}{\ensuremath{\mathcal{E}}}
\newcommand{\cF}{\ensuremath{\mathcal{F}}}

\newcommand{\cV}{\ensuremath{\mathcal{V}}}

\newcommand{\cL}{\ensuremath{\mathcal{L}}}
\newcommand{\cR}{\ensuremath{\mathcal{R}}}

\newcommand{\cS}{\ensuremath{\mathcal{S}}}

\newcommand{\bx}{\ensuremath{\boldsymbol{x}}}
\newcommand{\by}{\ensuremath{\boldsymbol{y}}}

\newcommand{\bw}{\ensuremath{\boldsymbol{w}}}

\newcommand{\hsig}{\ensuremath{\hat{s}}}

\DeclareMathOperator{\tr}{\textup{Tr}}

\usepackage{hyperref}
\hypersetup{
    colorlinks = true,
    linkcolor = {blue},
    citecolor = {blue},
    urlcolor = {blue},
}

\begin{document}
\title{Learning Temporal Quantum Tomography}

\author{Quoc Hoan Tran}
\email{tran\_qh@ai.u-tokyo.ac.jp}
\affiliation{
	Graduate School of Information Science and Technology, The University of Tokyo, Tokyo 113-8656, Japan
}

\author{Kohei Nakajima}
\email{k\_nakajima@mech.t.u-tokyo.ac.jp}
\affiliation{
	Graduate School of Information Science and Technology, The University of Tokyo, Tokyo 113-8656, Japan
}
\affiliation{Next Generation Artificial Intelligence Research Center, The University of Tokyo, Tokyo 113-8656, Japan}
\date{\today}

\begin{abstract}
Quantifying and verifying the control level in preparing a quantum state are central challenges in building quantum devices.
The quantum state is characterized from experimental measurements, using a procedure known as tomography, which requires a vast number of resources. 
However, tomography for a quantum device with temporal processing, which is fundamentally different from standard tomography, has not been formulated.
We develop a practical and approximate tomography method using a recurrent machine learning framework for this intriguing situation.
The method is based on repeated quantum interactions between a system called quantum reservoir with a stream of quantum states.
Measurement data from the reservoir are connected to a linear readout to train a recurrent relation between quantum channels applied to the input stream.
We demonstrate our algorithms for representative quantum learning tasks, followed by the proposal of a quantum memory capacity to evaluate the temporal processing ability of near-term quantum devices.
\end{abstract}

\pacs{Valid PACS appear here}

\maketitle
\textit{Introduction.---}
The impressive progress in realizing quantum-enhanced technologies places a demand on the characterization and validation of quantum hardware.
One of the most quintessential parts of building quantum devices is quantum process tomography (QPT),
which is used in verifying quantum devices via the reconstruction of an unknown quantum channel from measurement data~\cite{nielsen:2011:QCQI,mohseni:2008:qpt}.
Standard QPT approaches, which have been focused recently on small system size~\cite{brien:2004:prl:qpt,riebe:2006:prl:qpt,Bialczak:2010:natphys:qpt,shabani:2011:qpt:sensing,Govia:2020:natcom:qpt}, assume the quantum device processes input states separately in a time-independent manner.
In the envisioned picture of quantum time-series processing, the quantum device may output the states in a sequence where the current output depends on the past inputs and outputs.
For example, the quantum device may generate temporal and input-dependent noise or fluctuations,
which may have effects on the output states~\cite{gehrig:2002:temporal,martinez:2021:temporal}.
Moreover, optical quantum states defined in temporal modes can be manipulated in a time-dependent and input-dependent manner using nonlinear optical processes~\cite{brecht:2015:photon,raymer:2020:temporal}.
Performing tomography for such devices differs from standard QPT, because the memory effects need to be taken into account.

\begin{figure*}
		\includegraphics[width=17cm]{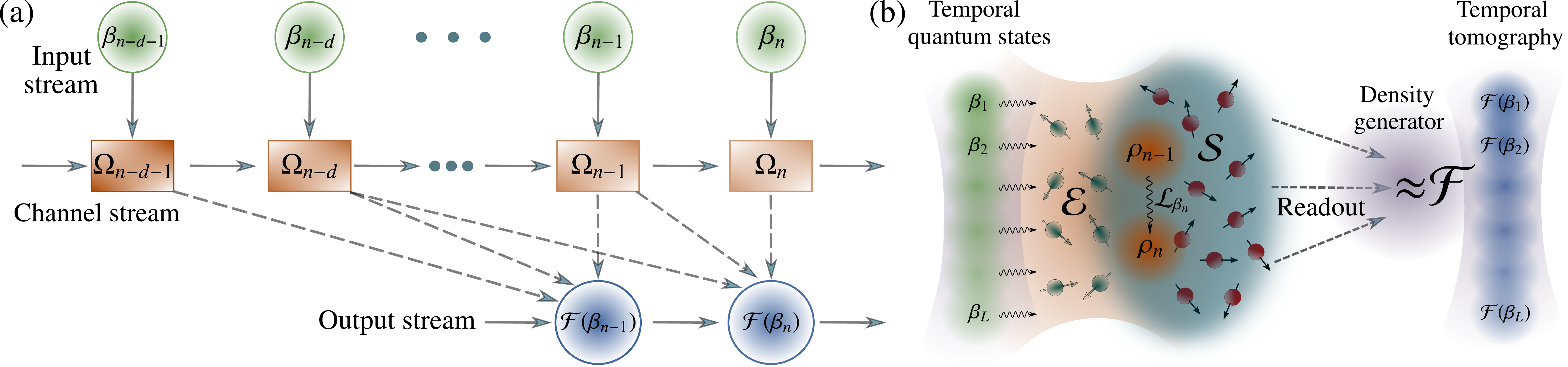}
		\protect\caption{Our framework can learn the tomography of a device that is supposed to implement an unknown temporal quantum map $\cF$; or emulate a predefined $\cF$.
		(a) A stream of quantum channels $\{\Omega_n\}$ applies to the input stream $\{\beta_n\}$ where each channel's output is determined by finite input history and past channels. The device's output is a function of input history such as a coherent superposition or a convex mixture of channels.
		(b) Our framework consists of a quantum reservoir $\cS$ interacting with $\{\beta_n\}$ with memory effects.
		The reservoir's internal state is evolved via CPTP maps $\cL_{\beta_n}$, which transfer the information from the input states to the reservoir.
		Measurement results in $\cS$ are used in a readout layer to reconstruct $\cF$.
		\label{fig:qpt:overview}}
\end{figure*}

Given a sequence of quantum states $\beta_1, \beta_2, \ldots$ in a $D_A$-dimensional Hilbert space, 
a quantum device processes this sequence via a temporal map $\cF$ to output quantum states $\cF(\beta_1), \cF(\beta_2), \ldots$ in a $D_B$-dimensional Hilbert space with a temporal dependency behavior: $\cF(\beta_n)$ only depends on a finite input history.
An intriguing example is the temporal depolarizing channel $\cF(\beta_n)=p_n\frac{I}{D} + (1-p_n)\beta_n$, which replaces $\beta_n$ with a completely mixed state $I/D$ with probability $p_n$ and leaves the state untouched otherwise ($D_A=D_B=D$ for notational simplicity).
The temporal dependency can be established if $p_n$ depends on the recent inputs.
As similar to the model of quantum channels with memory~\cite{kretschmann:2005:qmemory,caruso:2014:qmemory}, we model $\cF$ via a stream of quantum channels $\{\Omega_n\}$ that applies to $\{\beta_n\}$ and introduces the correlations between outputs.
$\cF$ can be considered as a coherent superposition or a convex mixture of channels at different times [Fig.~\ref{fig:qpt:overview}(a)].
The output of $\Omega_{n}$ is independent of future inputs; it is determined by finite input history and past channels to make the temporal dependency in $\cF$.
For example, $\Omega_n$ corresponds to the experimental studies on quantum processors where the dynamics of decoherence effects act as time-varying quantum channels~\cite{martinez:2021:temporal,burnett:2019:npj,klimov:2018:prl:fluct,steffen:2019:prl:corr,wang:2019:cavity,stelli:2020:aip:coherent}.
The effect of previous quantum inputs to the output has also been demonstrated experimentally for optical fiber channels~\cite{ball:2004:pra:corr,banaszek:2004:prl:corr}.
Here, $\cF$ envisions a typical case in the future realization of quantum communication and quantum internet~\cite{kimble:2008:quinternet,simon:2017:nat:quinternet,wehner:2018:science:quinternet} where quantum data can be transmitted via time-dependent and delayed channels.

In this Letter, we propose a supervised learning framework to perform the approximate tomography of $\cF$.
A naive approach is to perform state tomography of $\cF(\beta_n)$ for every $n$. This requires many repetition experiments on copies of $\cF(\beta_n)$ and the inversion of a huge linear system for every $n$~\cite{supp}.
Our idea is to simplify the experimental protocol and reduce the implementation cost
under the assumption of the temporal dependency in $\cF$.
We assume that it is possible to perform state tomography at some time steps as $\cF(\beta_1),\ldots,\cF(\beta_L)$.
We consider a quantum system $\cS$, called a \textit{quantum reservoir} (QR), interacting with the input stream  belonging to an auxiliary system $\cE$.
Each $\beta_n$ interacts for a certain time with $\cS$ before being replaced by another one.
Between two consecutive interactions by $\beta_n$ and $\beta_{n+1}$, the QR's internal evolution is described by a completely positive and trace-preserving (CPTP) map $\cL_{\beta_n}$ whose role is to effectively transfer the information of $\beta_n$ from $\cE$ into $\cS$.
With an effective setting of these CPTP maps, the QR's state after applying $\cL_{\beta_n}$ will depend more on the recent past inputs than distant past inputs.
This scheme ensures the fading memory property~\cite{boyd:1985:fading}, which is the ability to retain information about recent inputs for proper learning of functions of the past inputs.
Therefore, the results of measurements in $\cS$ can be used as high-dimensional quantum features to train a regression model to output density matrices that approximate $\cF(\beta_1),\ldots,\cF(\beta_L)$ [Fig.~\ref{fig:qpt:overview}(b)].
After this procedure, we are able to reconstruct $\cF(\beta_n)$ for $n > L$ from the trained parameters by performing a single measurement protocol in $\cS$.

Our framework is a quantum extension of classical reservoir computing (RC) to perform quantum tasks (Section I.A-B in~\cite{supp}). 
The crucial principle of RC is to represent the input sequence by feeding the input into a dynamical system, called the reservoir, to encode all relevant nonlinear dynamics in  high-dimensional trajectories~\cite{jaeger:2001:echo,maass:2002:reservoir,lukoeviius:2009:reservoir,nakajima:2021:RCbook}. Our proposal exploits quantum dynamics as a reservoir in the time-series processing of quantum data. This idea develops the initial proposal of harnessing disordered quantum dynamics for machine learning with classical time-series data~\cite{fujii:2017:qrc,nakajima:2019:qrc,fujii:2021:rcbook}.
While the RC approaches in tomography tasks focused on a static quantum state~\cite{ghosh:2019:quantum,ghosh:2020:reconstruct,ghosh:2021:quantum:adv}, our approach can process time-dependent quantum states.
We further propose the concept of quantum memory capacity to uncover the temporal processing ability of near-term quantum devices.

\textit{Model.---}
Assume that the initial state of the coupled system $(\cS, \cE)$ is a product state $\varrho = \rho \otimes \beta$, where $\rho$ and $\beta$ are the state of $\cS$ and $\cE$, respectively.
The coupled system is evolved under a unitary evolution $U$ and the state $\rho$ of $\cS$ is transformed via the CPTP \textit{reduced dynamics map} $\cL_{\beta}$, where
$
    \cL_{\beta}(\rho) = \tr_{\cE}[U(\rho\otimes\beta)U^{\dagger}]
$.
The successive interactions are described as
\begin{align}
    \rho_n = \cL_{\beta_n}(\rho_{n-1}) = \tr_{\cE}[U(\rho_{n-1}\otimes \beta_n)U^{\dagger}],\label{eqn:recurr}
\end{align}
where $\rho_n$ is the state of $\cS$ for the $n$th interaction.
We measure local observables $O_1,\ldots,O_K$ on $\rho_n$
to obtain a high-dimensional feature vector called \textit{reservoir state} $\bx_n$.
The $k$th element in $\bx_n$ can be calculated as 
$
    x_{nk} = \tr[O_k\rho_n] = \langle O_k\rangle_{\rho_n},
$
which is the expectation of the measurement result via $O_k$.
Between two inputs, $M$ cycles of the unitary evolution are processed and each of them is followed by  measurements.
$M$ is called the measurement multiplexity, thus we obtain $MK$ elements in $\bx_n$.

\begin{figure*}
	\includegraphics[width=17.0cm]{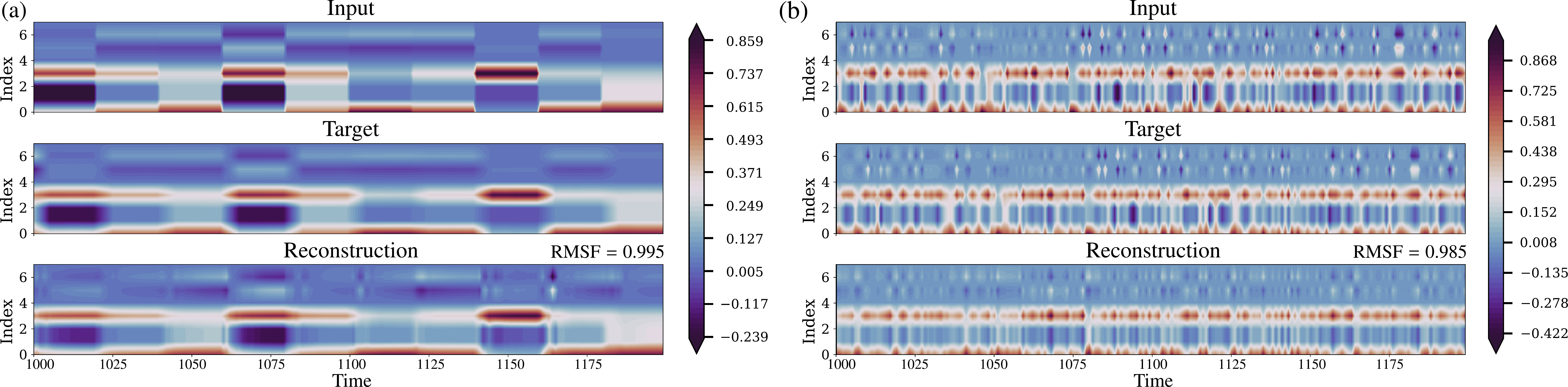}
	\protect\caption{Temporal tomography for (a) the quantum simple moving average filter and (b) the delayed depolarizing map with $d=5$, $N_m=5$, $N_e=1$,
		$\alpha=1.0$, $J/B=1.0$, and $\tau B = 2.6$ (for (a)) and $\tau B = 2.0$ (for (b)).
		At each time point, the density matrix is vectorized by stacking the real and imaginary parts, where the range of values is indicated in the color bars.
		\label{fig:qpt:multiplex-filter}}
\end{figure*}

In the training stage, we are given an input sequence $\{\beta_1,\ldots,\beta_L\}$ and the target sequence $\{\hat{\by}_1,\ldots,\hat{\by}_L\}$ where $\hat{\by}_k$ is the real vector form to stack the real and imaginary elements of $\cF(\beta_k)$.
Our framework includes a readout map $h$, which is simply taken as a linear combination of the reservoir states as $\by_n = h(\bx_n) = \bw^\top\bx_n$.
Here, $\bw$ is the parameter to be optimized by minimizing the mean-square error between $\by_n$ and $\hat{\by_n}$ over $n=1,\ldots,L$. 
In the evaluation stage, we are given an input sequence $\{\beta_{L+1}, \ldots, \beta_{L+T}\}$ with the target $\{\hat{\sigma}_{L+1},\ldots,\hat{\sigma}_{L+T}\}$ where $\hat{\sigma}_i = \cF(\beta_i)$.
The reconstructed output sequence is $\{{\by}_{L+1}, \ldots, {\by}_{L+T}\}$, which is rearranged in the matrix form $\{{\sigma}_{L+1}, \ldots, {\sigma}_{L+T}\}$~\footnote{For sufficient training samples, our framework can reconstruct the density matrices, which are positive semidefinite. However, due to statistical fluctuations, there are some cases in which the reconstructed matrix $A$ is not positive semidefinite. We project $A$ onto the spectrahedron to obtain a positive semidefinite matrix $\hat{A}$ such that the trace of $\hat{A}$ is equal to 1 and the Frobenius norm between $A$ and $\hat{A}$ is minimized~\cite{chen:2011:projection}}.
Since targets are density matrices, we use the fidelity $F(\rho,\sigma )=\tr[\sqrt{\sqrt{\sigma}\rho\sqrt{\sigma}}]$ to estimate the reconstruction error.
In error-free tomography, $F=1$, and $F<1$ otherwise.
We calculate the root mean square of fidelities in the evaluation stage as 
$
    \textup{RMSF} = \sqrt{\dfrac{1}{T}\sum_{i=L+1}^{i=L+T} F^2(\hat{\sigma}_{i}, {\sigma}_{i})}.
$

The reservoir's response to the same input sequence may differ with different initial states of the reservoir and may result in the loss of reproducibility in the temporal processing. To prevent this effect, the time evolution of the reservoir must mostly depend on the input sequence after enough transient time. 
This property is known as the echo state property~\cite{jaeger:2001:echo} in classical RC or the quantum echo state property (QESP) in quantum RC~\cite{chen:2019:dissipative,tran:2020:higherorder} to ensure the fading memory.
Based on the spectrum of reduced dynamics maps, we can evaluate the QESP time scale, which indicates the transient time to forget the QR's initial state before learning (Section II in~\cite{supp}).

\textit{Results.---}We present concrete applications of learning temporal tomography. We consider 
 $(\cS, \cE)$ as a closed system of the transverse field Ising model with the unitary $U=\exp(i\tau H)$, where $H = \sum_{i> j=1}^NJ_{i,j}\hsig^x_i\hsig^x_j + B\sum_j^N\hsig^z_j$ is unchanged during interaction time $\tau$.
Here, $B$ is the natural frequency and $\hsig_j^\gamma$ $(\gamma \in \{x, y, z\})$ are the Pauli operators measuring the qubit $j$ along the $\gamma$ direction.
We consider the \textit{power-law decaying} for $J_{ij}=J|i-j|^{-\alpha}/N(\alpha)$ with an interaction strength $J$, power coefficient $\alpha$ ($0 < \alpha < 3$), and 
$
N(\alpha) = \sum_{i>j}|i-j|^{-\alpha} / (N-1)
$
~\cite{porras:2004:trapped,kim:2009:trapped,jurcevic:2014:trapped}.
$\cE$ includes the first $N_e$ qubits
where the remaining $N_m=N-N_e$ qubits form the reservoir $\cS$. 

In our demonstrations, the number of observables is set to $K=N_m$ if we select observables as spin projections $\hsig^z_j$ over the $z$-axis for all $j$, and to $K=N_m(N_m+1)/2$ if we further select observables as two-spin correlations $\hsig^z_i\hsig^z_j$ for all $i < j$.
We consider the time-dependent depolarizing quantum channel $\Omega_n(\beta) =p_n\frac{I}{D} + (1-p_n)\beta$.
We introduce a temporal dependency in $\Omega_n$ by formulating $p_n$ as the $r$th-order nonlinear sequence: 
$
p_n = \kappa p_{n-1} + \eta p_{n-1}\left( \sum_{j=0}^{r-1} p_{n-j-1}\right) + \gamma u_{n-r+1}u_n + \delta,
$
where $r=10$, $\kappa=0.3$, $\eta=0.04$, $\gamma=1.5$, and $\delta=0.1$.
Here, $\{u_n\}$ is considered depending on $\{\beta_n\}$. It is randomly generated with the same random seed used to generate $\{\beta_n\}$~\footnote{The channels $\Omega_n$ can be considered quantum noises applying to the input states where there is a temporal correlation in these noises. The sequence $\{p_n\}$ resembles the NARMA benchmark~\cite{atiya:2000:narma}, which is commonly used for evaluating the computational capability of temporal processing with long time dependence. Here, $\{u_n\}$ is a random sequence of scalar values in $[0, 1]$ but is rescaled into $[0, 0.2]$ before creating $p_n$ to set $p_n$ into the stable range in $[0, 1]$.}. The QR can encode $\{\beta_n\}$ to reconstruct the output states, including the reconstruction of the nonlinear sequence $\{p_n\}$. This reconstruction is possible since observables in the QR become nonlinear functions of the input history due to the mixing of higher-order correlations in the quantum-chaotic dynamics.

We first consider $\cF$ as a quantum simple moving average filter $\cF(\beta_n)=\frac{1}{d+1}\sum_{i=0}^d\Omega_{n-i}(\beta_{n-i})$ [Fig.~\ref{fig:qpt:multiplex-filter}(a)] or a delayed depolarizing map $\cF(\beta_n)=\Omega_{n-d}(\beta_{n-d})$ [Fig.~\ref{fig:qpt:multiplex-filter}(b)] for a delay $d\geq 0$.
We refer to Section V.~B-C in~\cite{supp} for tomography of coherent superposition of channels in times and temporal entanglers. The QR must memorize the previous inputs and learn the properties of previous channels.
We set $d=5$, $K=N_m=5$, $N_e=1$, and $M=5$. 
Other model parameters are $\alpha=1.0$, $J/B=1.0$, and the normalized time $\tau B=2.6$ [Fig.~\ref{fig:qpt:multiplex-filter}(a)] and $\tau B = 2.0$ [Fig.~\ref{fig:qpt:multiplex-filter}(b)].
The number of time steps used in the initial transients, training, and evaluation stages are 500, 500, and 200, respectively.
At each time point, the density matrix is represented as a $2D^2$-dimensional vector ($D=2^{N_e}$) by stacking the real and imaginary parts.
In Fig.~\ref{fig:qpt:multiplex-filter}(a), the inputs jump to a new random quantum state every 20 time steps, thus introducing temporal dependencies between inputs.
Alternately, we consider a sequence of i.i.d. random inputs in Fig.~\ref{fig:qpt:multiplex-filter}(b).
The target sequences (middle panels in Fig.~\ref{fig:qpt:multiplex-filter}) in the evaluation stage can be almost perfectly reconstructed  since the RMSF values are above 98\% (bottom panels).

\begin{figure}
	\includegraphics[width=8.7cm]{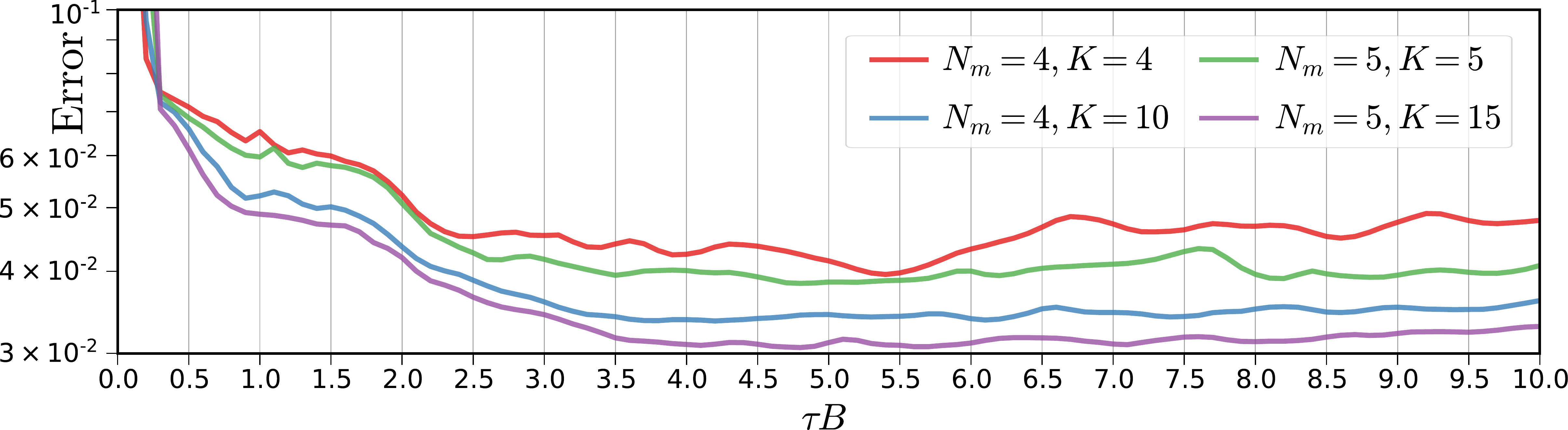}
	\protect\caption{Tomography errors according to $\tau B$ for $\cF(\beta_n)=\Omega_{n-1}(\beta_{n-1})$ with $\alpha=1.0$ and $J/B=1.0$.
		\label{fig:qpt:nspins}}
\end{figure}

Next, we investigate the dependency of the task's performance on the QR's parameters.
Figure~\ref{fig:qpt:nspins} illustrates the tomography errors (1.0 - RMSF) according to $\tau B$ in the reconstruction of $\cF(\beta_n)=\Omega_{n-1}(\beta_{n-1})$ at $N_e=2$ and $N_m=4,5$ qubits.
The transients, training, and evaluation time steps are 1000, 3000, and 1000, respectively.
The errors are averaged over 10 different runs with random trials of the input sequence and initial state.
The errors reduce quickly at low values of $\tau B$ and then settle to the stable lower values.
Table~\ref{tab:baseline:acc} presents the average errors along with their standard deviations at $\tau B = 10.0$ and $M=5$.
Particularly, with $N_m=5$ and $K=15$, the errors are lower than 1\%, 4\%, 6\%, and 8\% for $N_e=1,2,3,$ and 4 qubits, respectively.
We further compare our method with a classical baseline method, in which we assume that a full tomography of input states can be obtained. Instead of using measurements, the reservoir state $\bx_n$ is constructed directly from $\beta_n$ by stacking the real and imaginary parts in the corresponding density matrix to construct the vector form.
Table~\ref{tab:baseline:acc} shows that our method outperforms the classical baseline, which does not have memory effects.

We further investigate the \textit{short-term memory} (STM) of the QR via the delay-reconstruction task $\cF(\beta_n) = \beta_{n-d}$.
The STM in the classical context is defined via the coefficient of determination to measure how much variance of the delay inputs can be recovered from outputs~\cite{jaeger:2001:short}.
Since the input and output of our framework are density matrices, we define the $d$-delay STM of the QR by the squared distance correlation~\cite{szkely:2007:corr} between the output $\{\sigma_n\}$ and the target $\{\hat{\sigma}_n\}=\{\beta_{n-d}\}$:
\begin{align}
    \cR^2(d) = \dfrac{\cV^2(\{\sigma_n\}, \{\hat{\sigma}_n\})}{\sqrt{\cV^2(\{\sigma_n\}, \{\sigma_n\})\cV^2(\{\hat{\sigma}_n\}, \{\hat{\sigma}_n\})}}.
\end{align}
Here, $\cV^2$ represents the squared distance covariance of two random sequences of density matrices~\cite{supp}.
$\cR^2(d)$ is between 0 and 1, and it represents the QR's ability at the input $\beta_n$ to reconstruct the previous input $\beta_{n-d}$.
The behavior of $\cR^2(d)$ is indicated by the forgetting curve, which approaches to a small value for large values of $d$, thus realizing the STM.
We then define the quantum memory capacity,
$
    \textup{QMC} = \sum_{d=0}^{\infty}\cR^2(d),
$
to measure how much information of the delay input states can be recovered from output states, summed over all delays.
If the value of QMC increases, so does the duration of quantum inputs that can be memorized via the QR.
Quantifying QMC provides insights into the ability of the QR to reconstruct a temporal function of quantum inputs.

\begin{table}
	\caption{\label{tab:baseline:acc} Average and standard deviation (mean$\pm$sd) of the tomography errors (\%) for $\cF(\beta_n)=\Omega_{n-1}(\beta_{n-1})$ of the baseline and our method at $\tau B = 10.0$ and $M=5$.}
		\begin{ruledtabular}
	    \begin{tabular}{lcccc} 
		{$(N_m, K)$} & $N_e=1$ & $N_e=2$ & $N_e=3$ & $N_e=4$ \\ \hline
		{$(4, 4)$} & 0.9$\pm$0.0 & 4.8$\pm$0.1 & 8.0$\pm$0.0 & 8.9$\pm$0.1 \\
		{$(5, 5)$} & 0.9$\pm$0.0 & 4.0$\pm$0.1 & 7.6$\pm$0.0 & 8.6$\pm$0.1 \\
		{$(4, 10)$} & 0.9$\pm$0.0 & 3.6$\pm$0.1 & 6.4$\pm$0.0 & 8.3$\pm$0.1 \\
		{$(5, 15)$} & 0.9$\pm$0.0 & 3.3$\pm$0.0 & 5.6$\pm$0.0 & 7.7$\pm$0.0 \\
		{Baseline} & 3.2$\pm$0.1 & 9.2$\pm$0.1 & 10.0$\pm$0.1 & 10.2$\pm$0.1 \\
        \end{tabular}
        \end{ruledtabular}
\end{table}

\begin{figure}
		\includegraphics[width=8.7cm]{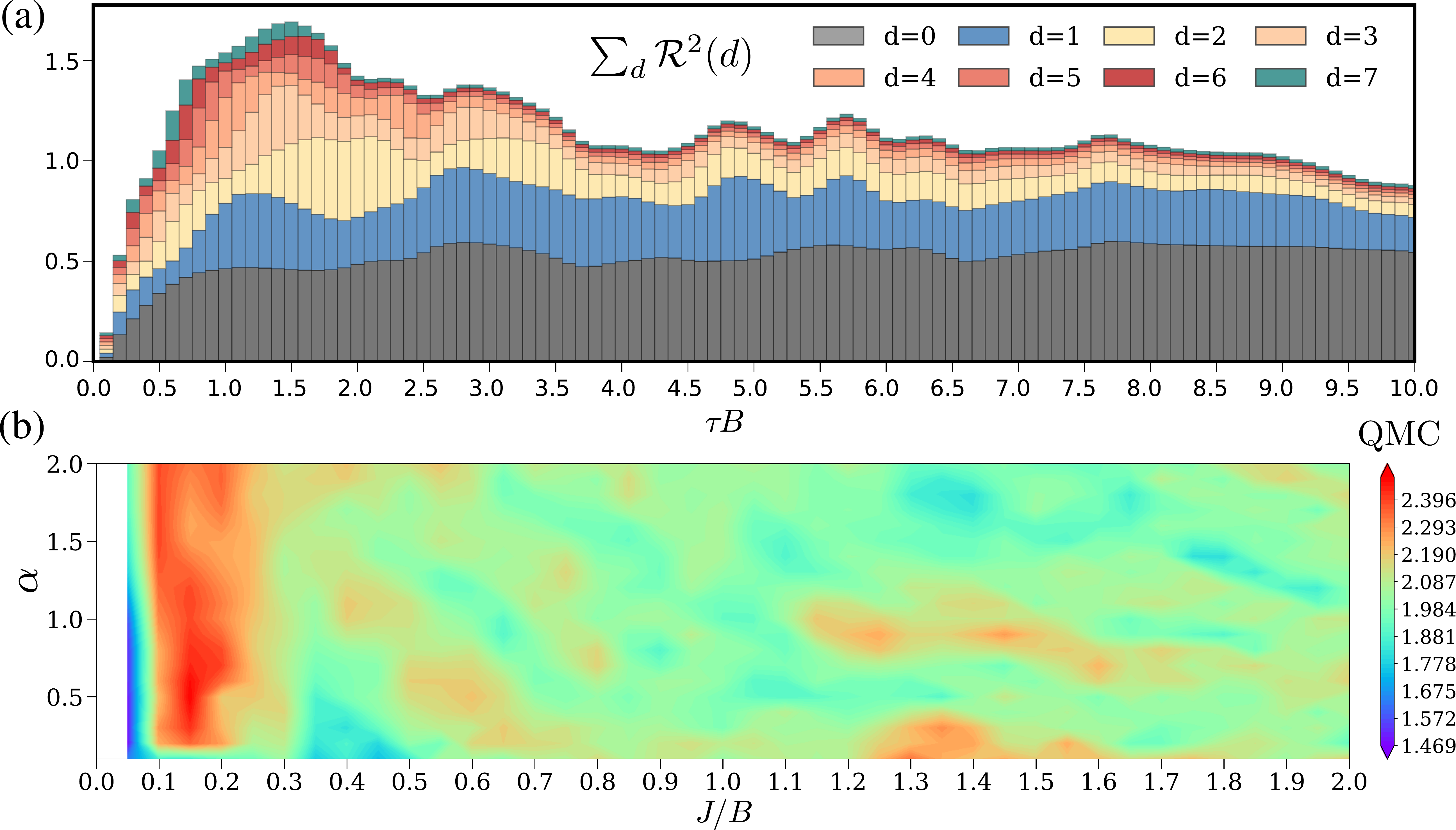}
		\protect\caption{(a) QMC broken down in delays $d$ according to $\tau B$ with $N_e=2$, $N_m=4$ with $\alpha=1.0$ and $J/B=1.0$. (b) The color map of QMC as the function of $\alpha$ and $J/B$ at $\tau B = 10.0$.
		\label{fig:qmc:abc}}
\end{figure}

Figure~\ref{fig:qmc:abc}(a) shows QMC as a function of $\tau B$ broken down in values of $d$ ($0\leq d \leq 7$) with the model parameters $\alpha=1, J/B=1.0, K=N_m=4$, $N_e=2$, and $M=1$ (other results in Section IV in~\cite{supp}).
$\cR^2(d)$ is averaged over different runs with 10 random trials of the initial state and input sequence. 
The total $\sum_{d=0}^{d=7}\cR^2(d)$ increases and obtains the peak value at the onset of the dynamical transition region ($1.5 < \tau B < 2.5$) of the reduced dynamics map $\cL_{\beta}$ (Fig.~S2(a) in~\cite{supp}).
We further examine the relation between QMC and other model parameters in Fig.~\ref{fig:qmc:abc}(b), which displays the average QMC (calculated until $d_{\textup{max}}=10$) as the function of $\alpha$ and $J/B$  at $\tau B=10.0$.
Interestingly, QMC achieves highest values in the region $0.1 < J/B < 0.2$, which is referred to the dynamical transition in Figs.~S3 and S6 in~\cite{supp}.
These observations can be explained by the difference in the eigenvalues' distribution of $\cL_{\beta}$ (Figs.~S2(a) and S3 in~\cite{supp}).
The regime with eigenvalues concentrated near the border of the unit disk leads toward a unitary behavior.
Therefore, only a little information of the input state $\beta$ is remained after applying $\cL_{\beta}$, which is unfavorable on temporal learning tasks.
In contrast, the regime with eigenvalues concentrated near the center of the unit disk guarantees enough information in the input states to be entangled with the QR.
Here, the dynamics becomes ergodic and the local observables become functions of a finite number of past inputs.
We anticipate that QMC builds up first as the dynamics moves from more unitary to more ergodic regime and obtains the peak at the transition between these regimes
\footnote{For temporal classical tasks in quantum spin networks, similar observations~\cite{pena:2021:qrc:dynamic} have also been investigated recently to address the impact of the transition between localization and thermalization manifest. Our theoretical investigation is not limited to quantum spin networks, but is more generally associated with the spectrum analysis of reduced dynamics maps. Furthermore, the intriguing results in QMC under varying model parameters  remind us of the well-known phenomenon in classical RC called the edge of chaos, where in some situations, a memory capacity achieves the maximum values at the edge of stability between different dynamics regimes~\cite{bertschinger:2004:edgeNN,toyoizumi:2011:edgepre,haruna:2019:shortterm}.}.

\textit{Conclusion and Discussion.---}
We formulate and propose the general framework for learning tomography of temporal quantum maps acting on quantum data.
We establish the concept of quantum memory capacity, which opens opportunities in developing the theoretical magnitude on the quantum time-series processing~\cite{mujal:2021:opportunities}.

The measurement protocol on the reservoir 
may lead to the effect of back-action, which is the problem of changes in quantum states due to measurement.
Each physical implementation can have measures that may successfully work around this problem~\cite{negoro:2021:rcbook:NMR,chen:2020:temporal}.
One can consider weak measurements on multiple copies of the same systems, such as a huge ensemble of identical molecules in a solid~\cite{fujii:2017:qrc,negoro:2021:rcbook:NMR}.
We can explore the implementation based on ion traps since it is possible to experimentally exploit nontrivial degrees of freedom with the measurements of the spin projections and correlations~\cite{richerme:2014:nature:corr,zhang:2017:nature:ion}
\footnote{In this platform, one can surpass the overhead on the large number of measurements repeated at each time step by utilizing the shadow tomography protocol to estimate many observables from a reasonable number of projective measurements~\cite{huang:2020:nat:measurements}.}.

To perform tomography of the temporal quantum map depending on long distant past inputs, we need to increase the quantum memory capacity of the reservoir, which can be a bottleneck with more scale of resources such as qubits, observables, and measurements. Therefore, it can be helpful in the resource design if we can quantify in advance the required information processing ability of a temporal quantum map, such as how far and what combinations of the past inputs are processed in this map. This directly relates to the  information processing framework in input-driven dynamical systems~\cite{dambre:2012:nonlinear,kubota:2021:prr:IPC} but presents further challenges in the quantum context. 

\begin{acknowledgments}
The authors acknowledge Shumpei Kobayashi for fruitful discussions.
This work is supported by MEXT Quantum Leap Flagship Program (MEXT Q-LEAP) Grant Nos. JPMXS0118067394 and JPMXS0120319794.
\end{acknowledgments}


\begin{thebibliography}{85}%
	\makeatletter
	\providecommand \@ifxundefined [1]{%
		\@ifx{#1\undefined}
	}%
	\providecommand \@ifnum [1]{%
		\ifnum #1\expandafter \@firstoftwo
		\else \expandafter \@secondoftwo
		\fi
	}%
	\providecommand \@ifx [1]{%
		\ifx #1\expandafter \@firstoftwo
		\else \expandafter \@secondoftwo
		\fi
	}%
	\providecommand \natexlab [1]{#1}%
	\providecommand \enquote  [1]{``#1''}%
	\providecommand \bibnamefont  [1]{#1}%
	\providecommand \bibfnamefont [1]{#1}%
	\providecommand \citenamefont [1]{#1}%
	\providecommand \href@noop [0]{\@secondoftwo}%
	\providecommand \href [0]{\begingroup \@sanitize@url \@href}%
	\providecommand \@href[1]{\@@startlink{#1}\@@href}%
	\providecommand \@@href[1]{\endgroup#1\@@endlink}%
	\providecommand \@sanitize@url [0]{\catcode `\\12\catcode `\$12\catcode
		`\&12\catcode `\#12\catcode `\^12\catcode `\_12\catcode `\%12\relax}%
	\providecommand \@@startlink[1]{}%
	\providecommand \@@endlink[0]{}%
	\providecommand \url  [0]{\begingroup\@sanitize@url \@url }%
	\providecommand \@url [1]{\endgroup\@href {#1}{\urlprefix }}%
	\providecommand \urlprefix  [0]{URL }%
	\providecommand \Eprint [0]{\href }%
	\providecommand \doibase [0]{https://doi.org/}%
	\providecommand \selectlanguage [0]{\@gobble}%
	\providecommand \bibinfo  [0]{\@secondoftwo}%
	\providecommand \bibfield  [0]{\@secondoftwo}%
	\providecommand \translation [1]{[#1]}%
	\providecommand \BibitemOpen [0]{}%
	\providecommand \bibitemStop [0]{}%
	\providecommand \bibitemNoStop [0]{.\EOS\space}%
	\providecommand \EOS [0]{\spacefactor3000\relax}%
	\providecommand \BibitemShut  [1]{\csname bibitem#1\endcsname}%
	\let\auto@bib@innerbib\@empty
	\bibitem [{\citenamefont {Nielsen}\ and\ \citenamefont
		{Chuang}(2011)}]{nielsen:2011:QCQI}%
	\BibitemOpen
	\bibfield  {author} {\bibinfo {author} {\bibfnamefont {M.~A.}\ \bibnamefont
			{Nielsen}}\ and\ \bibinfo {author} {\bibfnamefont {I.~L.}\ \bibnamefont
			{Chuang}},\ }\href@noop {} {\emph {\bibinfo {title} {Quantum Computation and
				Quantum Information: 10th Anniversary Edition}}},\ \bibinfo {edition} {10th}\
	ed.\ (\bibinfo  {publisher} {Cambridge University Press},\ \bibinfo {address}
	{USA},\ \bibinfo {year} {2011})\BibitemShut {NoStop}%
	\bibitem [{\citenamefont {Mohseni}\ \emph {et~al.}(2008)\citenamefont
		{Mohseni}, \citenamefont {Rezakhani},\ and\ \citenamefont
		{Lidar}}]{mohseni:2008:qpt}%
	\BibitemOpen
	\bibfield  {author} {\bibinfo {author} {\bibfnamefont {M.}~\bibnamefont
			{Mohseni}}, \bibinfo {author} {\bibfnamefont {A.~T.}\ \bibnamefont
			{Rezakhani}},\ and\ \bibinfo {author} {\bibfnamefont {D.~A.}\ \bibnamefont
			{Lidar}},\ }\bibfield  {title} {\bibinfo {title} {Quantum-process tomography:
			Resource analysis of different strategies},\ }\href
	{https://doi.org/10.1103/PhysRevA.77.032322} {\bibfield  {journal} {\bibinfo
			{journal} {Phys. Rev. A}\ }\textbf {\bibinfo {volume} {77}},\ \bibinfo
		{pages} {032322} (\bibinfo {year} {2008})}\BibitemShut {NoStop}%
	\bibitem [{\citenamefont {O'Brien}\ \emph {et~al.}(2004)\citenamefont
		{O'Brien}, \citenamefont {Pryde}, \citenamefont {Gilchrist}, \citenamefont
		{James}, \citenamefont {Langford}, \citenamefont {Ralph},\ and\ \citenamefont
		{White}}]{brien:2004:prl:qpt}%
	\BibitemOpen
	\bibfield  {author} {\bibinfo {author} {\bibfnamefont {J.~L.}\ \bibnamefont
			{O'Brien}}, \bibinfo {author} {\bibfnamefont {G.~J.}\ \bibnamefont {Pryde}},
		\bibinfo {author} {\bibfnamefont {A.}~\bibnamefont {Gilchrist}}, \bibinfo
		{author} {\bibfnamefont {D.~F.~V.}\ \bibnamefont {James}}, \bibinfo {author}
		{\bibfnamefont {N.~K.}\ \bibnamefont {Langford}}, \bibinfo {author}
		{\bibfnamefont {T.~C.}\ \bibnamefont {Ralph}},\ and\ \bibinfo {author}
		{\bibfnamefont {A.~G.}\ \bibnamefont {White}},\ }\bibfield  {title} {\bibinfo
		{title} {Quantum process tomography of a controlled-not gate},\ }\href
	{https://doi.org/10.1103/PhysRevLett.93.080502} {\bibfield  {journal}
		{\bibinfo  {journal} {Phys. Rev. Lett.}\ }\textbf {\bibinfo {volume} {93}},\
		\bibinfo {pages} {080502} (\bibinfo {year} {2004})}\BibitemShut {NoStop}%
	\bibitem [{\citenamefont {Riebe}\ \emph {et~al.}(2006)\citenamefont {Riebe},
		\citenamefont {Kim}, \citenamefont {Schindler}, \citenamefont {Monz},
		\citenamefont {Schmidt}, \citenamefont {K\"orber}, \citenamefont {H\"ansel},
		\citenamefont {H\"affner}, \citenamefont {Roos},\ and\ \citenamefont
		{Blatt}}]{riebe:2006:prl:qpt}%
	\BibitemOpen
	\bibfield  {author} {\bibinfo {author} {\bibfnamefont {M.}~\bibnamefont
			{Riebe}}, \bibinfo {author} {\bibfnamefont {K.}~\bibnamefont {Kim}}, \bibinfo
		{author} {\bibfnamefont {P.}~\bibnamefont {Schindler}}, \bibinfo {author}
		{\bibfnamefont {T.}~\bibnamefont {Monz}}, \bibinfo {author} {\bibfnamefont
			{P.~O.}\ \bibnamefont {Schmidt}}, \bibinfo {author} {\bibfnamefont {T.~K.}\
			\bibnamefont {K\"orber}}, \bibinfo {author} {\bibfnamefont {W.}~\bibnamefont
			{H\"ansel}}, \bibinfo {author} {\bibfnamefont {H.}~\bibnamefont {H\"affner}},
		\bibinfo {author} {\bibfnamefont {C.~F.}\ \bibnamefont {Roos}},\ and\
		\bibinfo {author} {\bibfnamefont {R.}~\bibnamefont {Blatt}},\ }\bibfield
	{title} {\bibinfo {title} {Process tomography of ion trap quantum gates},\
	}\href {https://doi.org/10.1103/PhysRevLett.97.220407} {\bibfield  {journal}
		{\bibinfo  {journal} {Phys. Rev. Lett.}\ }\textbf {\bibinfo {volume} {97}},\
		\bibinfo {pages} {220407} (\bibinfo {year} {2006})}\BibitemShut {NoStop}%
	\bibitem [{\citenamefont {Bialczak}\ \emph {et~al.}(2010)\citenamefont
		{Bialczak}, \citenamefont {Ansmann}, \citenamefont {Hofheinz}, \citenamefont
		{Lucero}, \citenamefont {Neeley}, \citenamefont {O'Connell}, \citenamefont
		{Sank}, \citenamefont {Wang}, \citenamefont {Wenner}, \citenamefont
		{Steffen}, \citenamefont {Cleland},\ and\ \citenamefont
		{Martinis}}]{Bialczak:2010:natphys:qpt}%
	\BibitemOpen
	\bibfield  {author} {\bibinfo {author} {\bibfnamefont {R.~C.}\ \bibnamefont
			{Bialczak}}, \bibinfo {author} {\bibfnamefont {M.}~\bibnamefont {Ansmann}},
		\bibinfo {author} {\bibfnamefont {M.}~\bibnamefont {Hofheinz}}, \bibinfo
		{author} {\bibfnamefont {E.}~\bibnamefont {Lucero}}, \bibinfo {author}
		{\bibfnamefont {M.}~\bibnamefont {Neeley}}, \bibinfo {author} {\bibfnamefont
			{A.~D.}\ \bibnamefont {O'Connell}}, \bibinfo {author} {\bibfnamefont
			{D.}~\bibnamefont {Sank}}, \bibinfo {author} {\bibfnamefont {H.}~\bibnamefont
			{Wang}}, \bibinfo {author} {\bibfnamefont {J.}~\bibnamefont {Wenner}},
		\bibinfo {author} {\bibfnamefont {M.}~\bibnamefont {Steffen}}, \bibinfo
		{author} {\bibfnamefont {A.~N.}\ \bibnamefont {Cleland}},\ and\ \bibinfo
		{author} {\bibfnamefont {J.~M.}\ \bibnamefont {Martinis}},\ }\bibfield
	{title} {\bibinfo {title} {Quantum process tomography of a universal
			entangling gate implemented with {Josephson} phase qubits},\ }\href
	{https://doi.org/10.1038/nphys1639} {\bibfield  {journal} {\bibinfo
			{journal} {Nat. Phys.}\ }\textbf {\bibinfo {volume} {6}},\ \bibinfo {pages}
		{409} (\bibinfo {year} {2010})}\BibitemShut {NoStop}%
	\bibitem [{\citenamefont {Shabani}\ \emph {et~al.}(2011)\citenamefont
		{Shabani}, \citenamefont {Kosut}, \citenamefont {Mohseni}, \citenamefont
		{Rabitz}, \citenamefont {Broome}, \citenamefont {Almeida}, \citenamefont
		{Fedrizzi},\ and\ \citenamefont {White}}]{shabani:2011:qpt:sensing}%
	\BibitemOpen
	\bibfield  {author} {\bibinfo {author} {\bibfnamefont {A.}~\bibnamefont
			{Shabani}}, \bibinfo {author} {\bibfnamefont {R.~L.}\ \bibnamefont {Kosut}},
		\bibinfo {author} {\bibfnamefont {M.}~\bibnamefont {Mohseni}}, \bibinfo
		{author} {\bibfnamefont {H.}~\bibnamefont {Rabitz}}, \bibinfo {author}
		{\bibfnamefont {M.~A.}\ \bibnamefont {Broome}}, \bibinfo {author}
		{\bibfnamefont {M.~P.}\ \bibnamefont {Almeida}}, \bibinfo {author}
		{\bibfnamefont {A.}~\bibnamefont {Fedrizzi}},\ and\ \bibinfo {author}
		{\bibfnamefont {A.~G.}\ \bibnamefont {White}},\ }\bibfield  {title} {\bibinfo
		{title} {Efficient measurement of quantum dynamics via compressive sensing},\
	}\href {https://doi.org/10.1103/PhysRevLett.106.100401} {\bibfield  {journal}
		{\bibinfo  {journal} {Phys. Rev. Lett.}\ }\textbf {\bibinfo {volume} {106}},\
		\bibinfo {pages} {100401} (\bibinfo {year} {2011})}\BibitemShut {NoStop}%
	\bibitem [{\citenamefont {Govia}\ \emph {et~al.}(2020)\citenamefont {Govia},
		\citenamefont {Ribeill}, \citenamefont {Rist{\`{e}}}, \citenamefont {Ware},\
		and\ \citenamefont {Krovi}}]{Govia:2020:natcom:qpt}%
	\BibitemOpen
	\bibfield  {author} {\bibinfo {author} {\bibfnamefont {L.~C.~G.}\
			\bibnamefont {Govia}}, \bibinfo {author} {\bibfnamefont {G.~J.}\ \bibnamefont
			{Ribeill}}, \bibinfo {author} {\bibfnamefont {D.}~\bibnamefont
			{Rist{\`{e}}}}, \bibinfo {author} {\bibfnamefont {M.}~\bibnamefont {Ware}},\
		and\ \bibinfo {author} {\bibfnamefont {H.}~\bibnamefont {Krovi}},\ }\bibfield
	{title} {\bibinfo {title} {Bootstrapping quantum process tomography via a
			perturbative ansatz},\ }\href {https://doi.org/10.1038/s41467-020-14873-1}
	{\bibfield  {journal} {\bibinfo  {journal} {Nat. Commun.}\ }\textbf {\bibinfo
			{volume} {11}},\ \bibinfo {pages} {1084} (\bibinfo {year}
		{2020})}\BibitemShut {NoStop}%
	\bibitem [{\citenamefont {Gehrig}\ and\ \citenamefont
		{Hess}(2002)}]{gehrig:2002:temporal}%
	\BibitemOpen
	\bibfield  {author} {\bibinfo {author} {\bibfnamefont {E.}~\bibnamefont
			{Gehrig}}\ and\ \bibinfo {author} {\bibfnamefont {O.~G.}\ \bibnamefont
			{Hess}},\ }\bibfield  {title} {\bibinfo {title} {{Spatio-temporal dynamics
				and fluctuations in quantum dot lasers: mesoscopic theory and modeling}},\
	}in\ \href {https://doi.org/10.1117/12.460803} {\emph {\bibinfo {booktitle}
			{Quantum Dot Devices and Computing}}},\ Vol.\ \bibinfo {volume} {4656},\
	\bibinfo {editor} {edited by\ \bibinfo {editor} {\bibfnamefont {J.~A.}\
			\bibnamefont {Lott}}, \bibinfo {editor} {\bibfnamefont {N.~N.}\ \bibnamefont
			{Ledentsov}}, \bibinfo {editor} {\bibfnamefont {K.~J.}\ \bibnamefont
			{Malloy}}, \bibinfo {editor} {\bibfnamefont {B.~E.}\ \bibnamefont {Kane}},\
		and\ \bibinfo {editor} {\bibfnamefont {T.~W.}\ \bibnamefont {Sigmon}}},\
	\bibinfo {organization} {International Society for Optics and Photonics}\
	(\bibinfo  {publisher} {SPIE},\ \bibinfo {year} {2002})\ pp.\ \bibinfo
	{pages} {69--78}\BibitemShut {NoStop}%
	\bibitem [{\citenamefont {Martinez}\ \emph {et~al.}(2021)\citenamefont
		{Martinez}, \citenamefont {Fuentes}, \citenamefont {Crespo},\ and\
		\citenamefont {Garcia-Frias}}]{martinez:2021:temporal}%
	\BibitemOpen
	\bibfield  {author} {\bibinfo {author} {\bibfnamefont {J.~E.}\ \bibnamefont
			{Martinez}}, \bibinfo {author} {\bibfnamefont {P.}~\bibnamefont {Fuentes}},
		\bibinfo {author} {\bibfnamefont {P.}~\bibnamefont {Crespo}},\ and\ \bibinfo
		{author} {\bibfnamefont {J.}~\bibnamefont {Garcia-Frias}},\ }\bibfield
	{title} {\bibinfo {title} {{Time-varying quantum channel models for
				superconducting qubits}},\ }\href
	{https://doi.org/10.1038/s41534-021-00448-5} {\bibfield  {journal} {\bibinfo
			{journal} {npj Quantum Inf.}\ }\textbf {\bibinfo {volume} {7}},\ \bibinfo
		{pages} {115} (\bibinfo {year} {2021})}\BibitemShut {NoStop}%
	\bibitem [{\citenamefont {Brecht}\ \emph {et~al.}(2015)\citenamefont {Brecht},
		\citenamefont {Reddy}, \citenamefont {Silberhorn},\ and\ \citenamefont
		{Raymer}}]{brecht:2015:photon}%
	\BibitemOpen
	\bibfield  {author} {\bibinfo {author} {\bibfnamefont {B.}~\bibnamefont
			{Brecht}}, \bibinfo {author} {\bibfnamefont {D.~V.}\ \bibnamefont {Reddy}},
		\bibinfo {author} {\bibfnamefont {C.}~\bibnamefont {Silberhorn}},\ and\
		\bibinfo {author} {\bibfnamefont {M.~G.}\ \bibnamefont {Raymer}},\ }\bibfield
	{title} {\bibinfo {title} {Photon temporal modes: A complete framework for
			quantum information science},\ }\href
	{https://doi.org/10.1103/PhysRevX.5.041017} {\bibfield  {journal} {\bibinfo
			{journal} {Phys. Rev. X}\ }\textbf {\bibinfo {volume} {5}},\ \bibinfo {pages}
		{041017} (\bibinfo {year} {2015})}\BibitemShut {NoStop}%
	\bibitem [{\citenamefont {Raymer}\ and\ \citenamefont
		{Walmsley}(2020)}]{raymer:2020:temporal}%
	\BibitemOpen
	\bibfield  {author} {\bibinfo {author} {\bibfnamefont {M.~G.}\ \bibnamefont
			{Raymer}}\ and\ \bibinfo {author} {\bibfnamefont {I.~A.}\ \bibnamefont
			{Walmsley}},\ }\bibfield  {title} {\bibinfo {title} {{Temporal modes in
				quantum optics: then and now}},\ }\href
	{https://doi.org/10.1088/1402-4896/ab6153} {\bibfield  {journal} {\bibinfo
			{journal} {Physica Scripta}\ }\textbf {\bibinfo {volume} {95}},\ \bibinfo
		{pages} {064002} (\bibinfo {year} {2020})}\BibitemShut {NoStop}%
	\bibitem [{\citenamefont {Kretschmann}\ and\ \citenamefont
		{Werner}(2005)}]{kretschmann:2005:qmemory}%
	\BibitemOpen
	\bibfield  {author} {\bibinfo {author} {\bibfnamefont {D.}~\bibnamefont
			{Kretschmann}}\ and\ \bibinfo {author} {\bibfnamefont {R.~F.}\ \bibnamefont
			{Werner}},\ }\bibfield  {title} {\bibinfo {title} {Quantum channels with
			memory},\ }\href {https://doi.org/10.1103/PhysRevA.72.062323} {\bibfield
		{journal} {\bibinfo  {journal} {Phys. Rev. A}\ }\textbf {\bibinfo {volume}
			{72}},\ \bibinfo {pages} {062323} (\bibinfo {year} {2005})}\BibitemShut
	{NoStop}%
	\bibitem [{\citenamefont {Caruso}\ \emph {et~al.}(2014)\citenamefont {Caruso},
		\citenamefont {Giovannetti}, \citenamefont {Lupo},\ and\ \citenamefont
		{Mancini}}]{caruso:2014:qmemory}%
	\BibitemOpen
	\bibfield  {author} {\bibinfo {author} {\bibfnamefont {F.}~\bibnamefont
			{Caruso}}, \bibinfo {author} {\bibfnamefont {V.}~\bibnamefont {Giovannetti}},
		\bibinfo {author} {\bibfnamefont {C.}~\bibnamefont {Lupo}},\ and\ \bibinfo
		{author} {\bibfnamefont {S.}~\bibnamefont {Mancini}},\ }\bibfield  {title}
	{\bibinfo {title} {Quantum channels and memory effects},\ }\href
	{https://doi.org/10.1103/RevModPhys.86.1203} {\bibfield  {journal} {\bibinfo
			{journal} {Rev. Mod. Phys.}\ }\textbf {\bibinfo {volume} {86}},\ \bibinfo
		{pages} {1203} (\bibinfo {year} {2014})}\BibitemShut {NoStop}%
	\bibitem [{\citenamefont {Burnett}\ \emph {et~al.}(2019)\citenamefont
		{Burnett}, \citenamefont {Bengtsson}, \citenamefont {Scigliuzzo},
		\citenamefont {Niepce}, \citenamefont {Kudra}, \citenamefont {Delsing},\ and\
		\citenamefont {Bylander}}]{burnett:2019:npj}%
	\BibitemOpen
	\bibfield  {author} {\bibinfo {author} {\bibfnamefont {J.~J.}\ \bibnamefont
			{Burnett}}, \bibinfo {author} {\bibfnamefont {A.}~\bibnamefont {Bengtsson}},
		\bibinfo {author} {\bibfnamefont {M.}~\bibnamefont {Scigliuzzo}}, \bibinfo
		{author} {\bibfnamefont {D.}~\bibnamefont {Niepce}}, \bibinfo {author}
		{\bibfnamefont {M.}~\bibnamefont {Kudra}}, \bibinfo {author} {\bibfnamefont
			{P.}~\bibnamefont {Delsing}},\ and\ \bibinfo {author} {\bibfnamefont
			{J.}~\bibnamefont {Bylander}},\ }\bibfield  {title} {\bibinfo {title}
		{Decoherence benchmarking of superconducting qubits},\ }\href
	{https://doi.org/10.1038/s41534-019-0168-5} {\bibfield  {journal} {\bibinfo
			{journal} {npj Quantum Inf.}\ }\textbf {\bibinfo {volume} {5}},\ \bibinfo
		{pages} {54} (\bibinfo {year} {2019})}\BibitemShut {NoStop}%
	\bibitem [{\citenamefont {Klimov}\ \emph {et~al.}(2018)\citenamefont {Klimov},
		\citenamefont {Kelly}, \citenamefont {Chen}, \citenamefont {Neeley},
		\citenamefont {Megrant}, \citenamefont {Burkett}, \citenamefont {Barends},
		\citenamefont {Arya}, \citenamefont {Chiaro}, \citenamefont {Chen},
		\citenamefont {Dunsworth}, \citenamefont {Fowler}, \citenamefont {Foxen},
		\citenamefont {Gidney}, \citenamefont {Giustina}, \citenamefont {Graff},
		\citenamefont {Huang}, \citenamefont {Jeffrey}, \citenamefont {Lucero},
		\citenamefont {Mutus}, \citenamefont {Naaman}, \citenamefont {Neill},
		\citenamefont {Quintana}, \citenamefont {Roushan}, \citenamefont {Sank},
		\citenamefont {Vainsencher}, \citenamefont {Wenner}, \citenamefont {White},
		\citenamefont {Boixo}, \citenamefont {Babbush}, \citenamefont {Smelyanskiy},
		\citenamefont {Neven},\ and\ \citenamefont
		{Martinis}}]{klimov:2018:prl:fluct}%
	\BibitemOpen
	\bibfield  {author} {\bibinfo {author} {\bibfnamefont {P.~V.}\ \bibnamefont
			{Klimov}}, \bibinfo {author} {\bibfnamefont {J.}~\bibnamefont {Kelly}},
		\bibinfo {author} {\bibfnamefont {Z.}~\bibnamefont {Chen}}, \bibinfo {author}
		{\bibfnamefont {M.}~\bibnamefont {Neeley}}, \bibinfo {author} {\bibfnamefont
			{A.}~\bibnamefont {Megrant}}, \bibinfo {author} {\bibfnamefont
			{B.}~\bibnamefont {Burkett}}, \bibinfo {author} {\bibfnamefont
			{R.}~\bibnamefont {Barends}}, \bibinfo {author} {\bibfnamefont
			{K.}~\bibnamefont {Arya}}, \bibinfo {author} {\bibfnamefont {B.}~\bibnamefont
			{Chiaro}}, \bibinfo {author} {\bibfnamefont {Y.}~\bibnamefont {Chen}},
		\bibinfo {author} {\bibfnamefont {A.}~\bibnamefont {Dunsworth}}, \bibinfo
		{author} {\bibfnamefont {A.}~\bibnamefont {Fowler}}, \bibinfo {author}
		{\bibfnamefont {B.}~\bibnamefont {Foxen}}, \bibinfo {author} {\bibfnamefont
			{C.}~\bibnamefont {Gidney}}, \bibinfo {author} {\bibfnamefont
			{M.}~\bibnamefont {Giustina}}, \bibinfo {author} {\bibfnamefont
			{R.}~\bibnamefont {Graff}}, \bibinfo {author} {\bibfnamefont
			{T.}~\bibnamefont {Huang}}, \bibinfo {author} {\bibfnamefont
			{E.}~\bibnamefont {Jeffrey}}, \bibinfo {author} {\bibfnamefont
			{E.}~\bibnamefont {Lucero}}, \bibinfo {author} {\bibfnamefont {J.~Y.}\
			\bibnamefont {Mutus}}, \bibinfo {author} {\bibfnamefont {O.}~\bibnamefont
			{Naaman}}, \bibinfo {author} {\bibfnamefont {C.}~\bibnamefont {Neill}},
		\bibinfo {author} {\bibfnamefont {C.}~\bibnamefont {Quintana}}, \bibinfo
		{author} {\bibfnamefont {P.}~\bibnamefont {Roushan}}, \bibinfo {author}
		{\bibfnamefont {D.}~\bibnamefont {Sank}}, \bibinfo {author} {\bibfnamefont
			{A.}~\bibnamefont {Vainsencher}}, \bibinfo {author} {\bibfnamefont
			{J.}~\bibnamefont {Wenner}}, \bibinfo {author} {\bibfnamefont {T.~C.}\
			\bibnamefont {White}}, \bibinfo {author} {\bibfnamefont {S.}~\bibnamefont
			{Boixo}}, \bibinfo {author} {\bibfnamefont {R.}~\bibnamefont {Babbush}},
		\bibinfo {author} {\bibfnamefont {V.~N.}\ \bibnamefont {Smelyanskiy}},
		\bibinfo {author} {\bibfnamefont {H.}~\bibnamefont {Neven}},\ and\ \bibinfo
		{author} {\bibfnamefont {J.~M.}\ \bibnamefont {Martinis}},\ }\bibfield
	{title} {\bibinfo {title} {Fluctuations of energy-relaxation times in
			superconducting qubits},\ }\href
	{https://doi.org/10.1103/PhysRevLett.121.090502} {\bibfield  {journal}
		{\bibinfo  {journal} {Phys. Rev. Lett.}\ }\textbf {\bibinfo {volume} {121}},\
		\bibinfo {pages} {090502} (\bibinfo {year} {2018})}\BibitemShut {NoStop}%
	\bibitem [{\citenamefont {Schl\"or}\ \emph {et~al.}(2019)\citenamefont
		{Schl\"or}, \citenamefont {Lisenfeld}, \citenamefont {M\"uller},
		\citenamefont {Bilmes}, \citenamefont {Schneider}, \citenamefont {Pappas},
		\citenamefont {Ustinov},\ and\ \citenamefont
		{Weides}}]{steffen:2019:prl:corr}%
	\BibitemOpen
	\bibfield  {author} {\bibinfo {author} {\bibfnamefont {S.}~\bibnamefont
			{Schl\"or}}, \bibinfo {author} {\bibfnamefont {J.}~\bibnamefont {Lisenfeld}},
		\bibinfo {author} {\bibfnamefont {C.}~\bibnamefont {M\"uller}}, \bibinfo
		{author} {\bibfnamefont {A.}~\bibnamefont {Bilmes}}, \bibinfo {author}
		{\bibfnamefont {A.}~\bibnamefont {Schneider}}, \bibinfo {author}
		{\bibfnamefont {D.~P.}\ \bibnamefont {Pappas}}, \bibinfo {author}
		{\bibfnamefont {A.~V.}\ \bibnamefont {Ustinov}},\ and\ \bibinfo {author}
		{\bibfnamefont {M.}~\bibnamefont {Weides}},\ }\bibfield  {title} {\bibinfo
		{title} {Correlating decoherence in transmon qubits: Low frequency noise by
			single fluctuators},\ }\href {https://doi.org/10.1103/PhysRevLett.123.190502}
	{\bibfield  {journal} {\bibinfo  {journal} {Phys. Rev. Lett.}\ }\textbf
		{\bibinfo {volume} {123}},\ \bibinfo {pages} {190502} (\bibinfo {year}
		{2019})}\BibitemShut {NoStop}%
	\bibitem [{\citenamefont {Wang}\ \emph {et~al.}(2019)\citenamefont {Wang},
		\citenamefont {Shankar}, \citenamefont {Minev}, \citenamefont
		{Campagne-Ibarcq}, \citenamefont {Narla},\ and\ \citenamefont
		{Devoret}}]{wang:2019:cavity}%
	\BibitemOpen
	\bibfield  {author} {\bibinfo {author} {\bibfnamefont {Z.}~\bibnamefont
			{Wang}}, \bibinfo {author} {\bibfnamefont {S.}~\bibnamefont {Shankar}},
		\bibinfo {author} {\bibfnamefont {Z. K.}~\bibnamefont {Minev}}, \bibinfo
		{author} {\bibfnamefont {P.}~\bibnamefont {Campagne-Ibarcq}}, \bibinfo
		{author} {\bibfnamefont {A.}~\bibnamefont {Narla}},\ and\ \bibinfo {author}
		{\bibfnamefont {M. H.}~\bibnamefont {Devoret}},\ }\bibfield  {title} {\bibinfo
		{title} {Cavity attenuators for superconducting qubits},\ }\href
	{https://doi.org/10.1103/PhysRevApplied.11.014031} {\bibfield  {journal}
		{\bibinfo  {journal} {Phys. Rev. Applied}\ }\textbf {\bibinfo {volume}
			{11}},\ \bibinfo {pages} {014031} (\bibinfo {year} {2019})}\BibitemShut
	{NoStop}%
	\bibitem [{\citenamefont {Stehli}\ \emph {et~al.}(2020)\citenamefont {Stehli},
		\citenamefont {Brehm}, \citenamefont {Wolz}, \citenamefont {Baity},
		\citenamefont {Danilin}, \citenamefont {Seferai}, \citenamefont {Rotzinger},
		\citenamefont {Ustinov},\ and\ \citenamefont
		{Weides}}]{stelli:2020:aip:coherent}%
	\BibitemOpen
	\bibfield  {author} {\bibinfo {author} {\bibfnamefont {A.}~\bibnamefont
			{Stehli}}, \bibinfo {author} {\bibfnamefont {J.~D.}\ \bibnamefont {Brehm}},
		\bibinfo {author} {\bibfnamefont {T.}~\bibnamefont {Wolz}}, \bibinfo {author}
		{\bibfnamefont {P.}~\bibnamefont {Baity}}, \bibinfo {author} {\bibfnamefont
			{S.}~\bibnamefont {Danilin}}, \bibinfo {author} {\bibfnamefont
			{V.}~\bibnamefont {Seferai}}, \bibinfo {author} {\bibfnamefont
			{H.}~\bibnamefont {Rotzinger}}, \bibinfo {author} {\bibfnamefont {A.~V.}\
			\bibnamefont {Ustinov}},\ and\ \bibinfo {author} {\bibfnamefont
			{M.}~\bibnamefont {Weides}},\ }\bibfield  {title} {\bibinfo {title} {Coherent
			superconducting qubits from a subtractive junction fabrication process},\
	}\href {https://doi.org/10.1063/5.0023533} {\bibfield  {journal} {\bibinfo
			{journal} {Appl. Phys. Lett.}\ }\textbf {\bibinfo {volume} {117}},\ \bibinfo
		{pages} {124005} (\bibinfo {year} {2020})}\BibitemShut {NoStop}%
	\bibitem [{\citenamefont {Ball}\ \emph {et~al.}(2004)\citenamefont {Ball},
		\citenamefont {Dragan},\ and\ \citenamefont {Banaszek}}]{ball:2004:pra:corr}%
	\BibitemOpen
	\bibfield  {author} {\bibinfo {author} {\bibfnamefont {J.}~\bibnamefont
			{Ball}}, \bibinfo {author} {\bibfnamefont {A.}~\bibnamefont {Dragan}},\ and\
		\bibinfo {author} {\bibfnamefont {K.}~\bibnamefont {Banaszek}},\ }\bibfield
	{title} {\bibinfo {title} {Exploiting entanglement in communication channels
			with correlated noise},\ }\href {https://doi.org/10.1103/PhysRevA.69.042324}
	{\bibfield  {journal} {\bibinfo  {journal} {Phys. Rev. A}\ }\textbf {\bibinfo
			{volume} {69}},\ \bibinfo {pages} {042324} (\bibinfo {year}
		{2004})}\BibitemShut {NoStop}%
	\bibitem [{\citenamefont {Banaszek}\ \emph {et~al.}(2004)\citenamefont
		{Banaszek}, \citenamefont {Dragan}, \citenamefont {Wasilewski},\ and\
		\citenamefont {Radzewicz}}]{banaszek:2004:prl:corr}%
	\BibitemOpen
	\bibfield  {author} {\bibinfo {author} {\bibfnamefont {K.}~\bibnamefont
			{Banaszek}}, \bibinfo {author} {\bibfnamefont {A.}~\bibnamefont {Dragan}},
		\bibinfo {author} {\bibfnamefont {W.}~\bibnamefont {Wasilewski}},\ and\
		\bibinfo {author} {\bibfnamefont {C.}~\bibnamefont {Radzewicz}},\ }\bibfield
	{title} {\bibinfo {title} {Experimental demonstration of
			entanglement-enhanced classical communication over a quantum channel with
			correlated noise},\ }\href {https://doi.org/10.1103/PhysRevLett.92.257901}
	{\bibfield  {journal} {\bibinfo  {journal} {Phys. Rev. Lett.}\ }\textbf
		{\bibinfo {volume} {92}},\ \bibinfo {pages} {257901} (\bibinfo {year}
		{2004})}\BibitemShut {NoStop}%
	\bibitem [{\citenamefont {Kimble}(2008)}]{kimble:2008:quinternet}%
	\BibitemOpen
	\bibfield  {author} {\bibinfo {author} {\bibfnamefont {H.~J.}\ \bibnamefont
			{Kimble}},\ }\bibfield  {title} {\bibinfo {title} {{The quantum internet}},\
	}\href {https://doi.org/10.1038/nature07127} {\bibfield  {journal} {\bibinfo
			{journal} {Nature}\ }\textbf {\bibinfo {volume} {453}},\ \bibinfo {pages}
		{1023} (\bibinfo {year} {2008})}\BibitemShut {NoStop}%
	\bibitem [{\citenamefont {Simon}(2017)}]{simon:2017:nat:quinternet}%
	\BibitemOpen
	\bibfield  {author} {\bibinfo {author} {\bibfnamefont {C.}~\bibnamefont
			{Simon}},\ }\bibfield  {title} {\bibinfo {title} {{Towards a global quantum
				network}},\ }\href {https://doi.org/10.1038/s41566-017-0032-0} {\bibfield
		{journal} {\bibinfo  {journal} {Nat. Photonics}\ }\textbf {\bibinfo {volume}
			{11}},\ \bibinfo {pages} {678} (\bibinfo {year} {2017})},\ \Eprint
	{https://arxiv.org/abs/1710.11585} {1710.11585} \BibitemShut {NoStop}%
	\bibitem [{\citenamefont {Wehner}\ \emph {et~al.}(2018)\citenamefont {Wehner},
		\citenamefont {Elkouss},\ and\ \citenamefont
		{Hanson}}]{wehner:2018:science:quinternet}%
	\BibitemOpen
	\bibfield  {author} {\bibinfo {author} {\bibfnamefont {S.}~\bibnamefont
			{Wehner}}, \bibinfo {author} {\bibfnamefont {D.}~\bibnamefont {Elkouss}},\
		and\ \bibinfo {author} {\bibfnamefont {R.}~\bibnamefont {Hanson}},\
	}\bibfield  {title} {\bibinfo {title} {{Quantum internet: A vision for the
				road ahead}},\ }\href {https://doi.org/10.1126/science.aam9288} {\bibfield
		{journal} {\bibinfo  {journal} {Science}\ }\textbf {\bibinfo {volume}
			{362}},\ \bibinfo {pages} {eaam9288} (\bibinfo {year} {2018})}\BibitemShut
	{NoStop}%
	\bibitem [{sup()}]{supp}%
	\BibitemOpen
	\href@noop {} {}\bibinfo {note} {See Supplemental Materials for detailed
		explanations of the reservoir computing, the quantum reservoir computing, the
		standard quantum process tomography, the temporal tomography, the convergence
		and metastability analysis of the quantum reservoir, the quantum memory
		capacity, and other results on the temporal tomography tasks, which include
		Refs.~\cite{nokkala:2020:gaussian,govia:2020:quantum,dasgupta:2020:designing,abram:1993:quantum,bantysh:2020:quantum,quek:2021:adaptive,bruneau:2010:infinite,nechita:2010:random,movassagh:2019:ergodic,movassagh:2020:theory,steele:1989:subadd,hennion:1997:product,macieszczak:2016:meta,lasserre:ineq:trace,misra:1977:zeno,chiribella:2012:qswitch,procopio:2015:qswitch:exp,rubino:2017:qswitch:exp,goswami:2018:qswitch:exp,wei:2019:qswtich:exp,guo:2020:qswitch:exp,chiribella:2013:qswitch,ebler:2018:qswitch,meschede:1985:prl:maser,varcoe:2000:nat:photon}}\BibitemShut
	{NoStop}%
	\bibitem [{\citenamefont {Boyd}\ and\ \citenamefont
		{Chua}(1985)}]{boyd:1985:fading}%
	\BibitemOpen
	\bibfield  {author} {\bibinfo {author} {\bibfnamefont {S.}~\bibnamefont
			{Boyd}}\ and\ \bibinfo {author} {\bibfnamefont {L.}~\bibnamefont {Chua}},\
	}\bibfield  {title} {\bibinfo {title} {Fading memory and the problem of
			approximating nonlinear operators with volterra series},\ }\href
	{https://doi.org/10.1109/tcs.1985.1085649} {\bibfield  {journal} {\bibinfo
			{journal} {IEEE Trans. Circuits Syst.}\ }\textbf {\bibinfo {volume} {32}},\
		\bibinfo {pages} {1150} (\bibinfo {year} {1985})}\BibitemShut {NoStop}%
	\bibitem [{\citenamefont {Jaeger}(2001{\natexlab{a}})}]{jaeger:2001:echo}%
	\BibitemOpen
	\bibfield  {author} {\bibinfo {author} {\bibfnamefont {H.}~\bibnamefont
			{Jaeger}},\ }\bibfield  {title} {\bibinfo {title} {The “echo state”
			approach to analysing and training recurrent neural networks-with an erratum
			note},\ }\href
	{http://www.faculty.jacobs-university.de/hjaeger/pubs/EchoStatesTechRep.pdf}
	{\bibfield  {journal} {\bibinfo  {journal} {Bonn, Germany: German National
				Research Center for Information Technology GMD Technical Report}\ }\textbf
		{\bibinfo {volume} {148}},\ \bibinfo {pages} {13} (\bibinfo {year}
		{2001}{\natexlab{a}})}\BibitemShut {NoStop}%
	\bibitem [{\citenamefont {Maass}\ \emph {et~al.}(2002)\citenamefont {Maass},
		\citenamefont {Natschl\"{a}ger},\ and\ \citenamefont
		{Markram}}]{maass:2002:reservoir}%
	\BibitemOpen
	\bibfield  {author} {\bibinfo {author} {\bibfnamefont {W.}~\bibnamefont
			{Maass}}, \bibinfo {author} {\bibfnamefont {T.}~\bibnamefont
			{Natschl\"{a}ger}},\ and\ \bibinfo {author} {\bibfnamefont {H.}~\bibnamefont
			{Markram}},\ }\bibfield  {title} {\bibinfo {title} {Real-time computing
			without stable states: A new framework for neural computation based on
			perturbations},\ }\href {https://doi.org/10.1162/089976602760407955}
	{\bibfield  {journal} {\bibinfo  {journal} {Neural Computation}\ }\textbf
		{\bibinfo {volume} {14}},\ \bibinfo {pages} {2531} (\bibinfo {year}
		{2002})}\BibitemShut {NoStop}%
	\bibitem [{\citenamefont {Luko{\v{s}}evi{\v{c}}ius}\ and\ \citenamefont
		{Jaeger}(2009)}]{lukoeviius:2009:reservoir}%
	\BibitemOpen
	\bibfield  {author} {\bibinfo {author} {\bibfnamefont {M.}~\bibnamefont
			{Luko{\v{s}}evi{\v{c}}ius}}\ and\ \bibinfo {author} {\bibfnamefont
			{H.}~\bibnamefont {Jaeger}},\ }\bibfield  {title} {\bibinfo {title}
		{Reservoir computing approaches to recurrent neural network training},\
	}\href {https://doi.org/10.1016/j.cosrev.2009.03.005} {\bibfield  {journal}
		{\bibinfo  {journal} {Comput. Sci. Rev.}\ }\textbf {\bibinfo {volume} {3}},\
		\bibinfo {pages} {127} (\bibinfo {year} {2009})}\BibitemShut {NoStop}%
	\bibitem [{\citenamefont {Nakajima}\ and\ \citenamefont
		{Fischer}(2021)}]{nakajima:2021:RCbook}%
	\BibitemOpen
	\bibinfo {editor} {\bibfnamefont {K.}~\bibnamefont {Nakajima}}\ and\ \bibinfo
	{editor} {\bibfnamefont {I.}~\bibnamefont {Fischer}},\ eds.,\ \href
	{https://doi.org/10.1007/978-981-13-1687-6} {\emph {\bibinfo {title}
			{Reservoir Computing: Theory, Physical Implementations, and Applications}}}\
	(\bibinfo  {publisher} {Springer Singapore},\ \bibinfo {address}
	{Singapore},\ \bibinfo {year} {2021})\BibitemShut {NoStop}%
	\bibitem [{\citenamefont {Fujii}\ and\ \citenamefont
		{Nakajima}(2017)}]{fujii:2017:qrc}%
	\BibitemOpen
	\bibfield  {author} {\bibinfo {author} {\bibfnamefont {K.}~\bibnamefont
			{Fujii}}\ and\ \bibinfo {author} {\bibfnamefont {K.}~\bibnamefont
			{Nakajima}},\ }\bibfield  {title} {\bibinfo {title} {Harnessing
			disordered-ensemble quantum dynamics for machine learning},\ }\href
	{https://doi.org/10.1103/PhysRevApplied.8.024030} {\bibfield  {journal}
		{\bibinfo  {journal} {Phys. Rev. Applied}\ }\textbf {\bibinfo {volume} {8}},\
		\bibinfo {pages} {024030} (\bibinfo {year} {2017})}\BibitemShut {NoStop}%
	\bibitem [{\citenamefont {Nakajima}\ \emph {et~al.}(2019)\citenamefont
		{Nakajima}, \citenamefont {Fujii}, \citenamefont {Negoro}, \citenamefont
		{Mitarai},\ and\ \citenamefont {Kitagawa}}]{nakajima:2019:qrc}%
	\BibitemOpen
	\bibfield  {author} {\bibinfo {author} {\bibfnamefont {K.}~\bibnamefont
			{Nakajima}}, \bibinfo {author} {\bibfnamefont {K.}~\bibnamefont {Fujii}},
		\bibinfo {author} {\bibfnamefont {M.}~\bibnamefont {Negoro}}, \bibinfo
		{author} {\bibfnamefont {K.}~\bibnamefont {Mitarai}},\ and\ \bibinfo {author}
		{\bibfnamefont {M.}~\bibnamefont {Kitagawa}},\ }\bibfield  {title} {\bibinfo
		{title} {Boosting computational power through spatial multiplexing in quantum
			reservoir computing},\ }\href
	{https://doi.org/10.1103/PhysRevApplied.11.034021} {\bibfield  {journal}
		{\bibinfo  {journal} {Phys. Rev. Applied}\ }\textbf {\bibinfo {volume}
			{11}},\ \bibinfo {pages} {034021} (\bibinfo {year} {2019})}\BibitemShut
	{NoStop}%
	\bibitem [{\citenamefont {Fujii}\ and\ \citenamefont
		{Nakajima}(2021)}]{fujii:2021:rcbook}%
	\BibitemOpen
	\bibfield  {author} {\bibinfo {author} {\bibfnamefont {K.}~\bibnamefont
			{Fujii}}\ and\ \bibinfo {author} {\bibfnamefont {K.}~\bibnamefont
			{Nakajima}},\ }\bibinfo {title} {Quantum reservoir computing: A reservoir
		approach toward quantum machine learning on near-term quantum devices},\ in\
	\href {https://doi.org/10.1007/978-981-13-1687-6_18} {\emph {\bibinfo
			{booktitle} {Reservoir Computing: Theory, Physical Implementations, and
				Applications}}},\ \bibinfo {editor} {edited by\ \bibinfo {editor}
		{\bibfnamefont {K.}~\bibnamefont {Nakajima}}\ and\ \bibinfo {editor}
		{\bibfnamefont {I.}~\bibnamefont {Fischer}}}\ (\bibinfo  {publisher}
	{Springer Singapore},\ \bibinfo {address} {Singapore},\ \bibinfo {year}
	{2021})\ pp.\ \bibinfo {pages} {423--450}\BibitemShut {NoStop}%
	\bibitem [{\citenamefont {Ghosh}\ \emph {et~al.}(2019)\citenamefont {Ghosh},
		\citenamefont {Opala}, \citenamefont {Matuszewski}, \citenamefont {Paterek},\
		and\ \citenamefont {Liew}}]{ghosh:2019:quantum}%
	\BibitemOpen
	\bibfield  {author} {\bibinfo {author} {\bibfnamefont {S.}~\bibnamefont
			{Ghosh}}, \bibinfo {author} {\bibfnamefont {A.}~\bibnamefont {Opala}},
		\bibinfo {author} {\bibfnamefont {M.}~\bibnamefont {Matuszewski}}, \bibinfo
		{author} {\bibfnamefont {T.}~\bibnamefont {Paterek}},\ and\ \bibinfo {author}
		{\bibfnamefont {T.~C.}\ \bibnamefont {Liew}},\ }\bibfield  {title} {\bibinfo
		{title} {Quantum reservoir processing},\ }\href
	{https://doi.org/10.1038/s41534-019-0149-8} {\bibfield  {journal} {\bibinfo
			{journal} {npj Quantum Inf.}\ }\textbf {\bibinfo {volume} {5}},\ \bibinfo
		{pages} {35} (\bibinfo {year} {2019})}\BibitemShut {NoStop}%
	\bibitem [{\citenamefont {Ghosh}\ \emph {et~al.}(2020)\citenamefont {Ghosh},
		\citenamefont {Opala}, \citenamefont {Matuszewski}, \citenamefont {Paterek},\
		and\ \citenamefont {Liew}}]{ghosh:2020:reconstruct}%
	\BibitemOpen
	\bibfield  {author} {\bibinfo {author} {\bibfnamefont {S.}~\bibnamefont
			{Ghosh}}, \bibinfo {author} {\bibfnamefont {A.}~\bibnamefont {Opala}},
		\bibinfo {author} {\bibfnamefont {M.}~\bibnamefont {Matuszewski}}, \bibinfo
		{author} {\bibfnamefont {T.}~\bibnamefont {Paterek}},\ and\ \bibinfo {author}
		{\bibfnamefont {T.~C.~H.}\ \bibnamefont {Liew}},\ }\bibfield  {title}
	{\bibinfo {title} {Reconstructing quantum states with quantum reservoir
			networks},\ }\href {https://doi.org/10.1109/tnnls.2020.3009716} {\bibfield
		{journal} {\bibinfo  {journal} {{IEEE} Trans. Neural Netw. Learn. Syst.}\
		}\textbf {\bibinfo {volume} {32}},\ \bibinfo {pages} {3148} (\bibinfo {year}
		{2020})}\BibitemShut {NoStop}%
	\bibitem [{\citenamefont {Ghosh}\ \emph {et~al.}(2021)\citenamefont {Ghosh},
		\citenamefont {Nakajima}, \citenamefont {Krisnanda}, \citenamefont {Fujii},\
		and\ \citenamefont {Liew}}]{ghosh:2021:quantum:adv}%
	\BibitemOpen
	\bibfield  {author} {\bibinfo {author} {\bibfnamefont {S.}~\bibnamefont
			{Ghosh}}, \bibinfo {author} {\bibfnamefont {K.}~\bibnamefont {Nakajima}},
		\bibinfo {author} {\bibfnamefont {T.}~\bibnamefont {Krisnanda}}, \bibinfo
		{author} {\bibfnamefont {K.}~\bibnamefont {Fujii}},\ and\ \bibinfo {author}
		{\bibfnamefont {T.~C.~H.}\ \bibnamefont {Liew}},\ }\bibfield  {title}
	{\bibinfo {title} {{Quantum Neuromorphic Computing with Reservoir Computing
				Networks}},\ }\href {https://doi.org/10.1002/qute.202100053} {\bibfield
		{journal} {\bibinfo  {journal} {Adv. Quantum Technol.}\ ,\ \bibinfo {pages}
			{2100053}} (\bibinfo {year} {2021})}\BibitemShut {NoStop}%
	\bibitem [{Note1()}]{Note1}%
	\BibitemOpen
	\bibinfo {note} {For sufficient training samples, our framework can
		reconstruct the density matrices, which are positive semidefinite. However,
		due to statistical fluctuations, there are some cases in which the
		reconstructed matrix $A$ is not positive semidefinite. We project $A$ onto
		the spectrahedron to obtain a positive semidefinite matrix $\protect \hat
		{A}$ such that the trace of $\protect \hat {A}$ is equal to 1 and the
		Frobenius norm between $A$ and $\protect \hat {A}$ is minimized~\cite
		{chen:2011:projection}}\BibitemShut {NoStop}%
	\bibitem [{\citenamefont {Chen}\ and\ \citenamefont
		{Nurdin}(2019)}]{chen:2019:dissipative}%
	\BibitemOpen
	\bibfield  {author} {\bibinfo {author} {\bibfnamefont {J.}~\bibnamefont
			{Chen}}\ and\ \bibinfo {author} {\bibfnamefont {H.~I.}\ \bibnamefont
			{Nurdin}},\ }\bibfield  {title} {\bibinfo {title} {Learning nonlinear
			input{\textendash}output maps with dissipative quantum systems},\ }\href
	{https://doi.org/10.1007/s11128-019-2311-9} {\bibfield  {journal} {\bibinfo
			{journal} {Quantum Inf. Process.}\ }\textbf {\bibinfo {volume} {18}},\
		\bibinfo {pages} {198} (\bibinfo {year} {2019})}\BibitemShut {NoStop}%
	\bibitem [{\citenamefont {Tran}\ and\ \citenamefont
		{Nakajima}(2020)}]{tran:2020:higherorder}%
	\BibitemOpen
	\bibfield  {author} {\bibinfo {author} {\bibfnamefont {Q.~H.}\ \bibnamefont
			{Tran}}\ and\ \bibinfo {author} {\bibfnamefont {K.}~\bibnamefont
			{Nakajima}},\ }\bibfield  {title} {\bibinfo {title} {Higher-order quantum
			reservoir computing},\ }\href {https://arxiv.org/abs/2006.08999} {\bibfield
		{journal} {\bibinfo  {journal} {Preprint at arXiv:2006.08999}\ } (\bibinfo
		{year} {2020})}\BibitemShut {NoStop}%
	\bibitem [{\citenamefont {Porras}\ and\ \citenamefont
		{Cirac}(2004)}]{porras:2004:trapped}%
	\BibitemOpen
	\bibfield  {author} {\bibinfo {author} {\bibfnamefont {D.}~\bibnamefont
			{Porras}}\ and\ \bibinfo {author} {\bibfnamefont {J.~I.}\ \bibnamefont
			{Cirac}},\ }\bibfield  {title} {\bibinfo {title} {Effective quantum spin
			systems with trapped ions},\ }\href
	{https://doi.org/10.1103/PhysRevLett.92.207901} {\bibfield  {journal}
		{\bibinfo  {journal} {Phys. Rev. Lett.}\ }\textbf {\bibinfo {volume} {92}},\
		\bibinfo {pages} {207901} (\bibinfo {year} {2004})}\BibitemShut {NoStop}%
	\bibitem [{\citenamefont {Kim}\ \emph {et~al.}(2009)\citenamefont {Kim},
		\citenamefont {Chang}, \citenamefont {Islam}, \citenamefont {Korenblit},
		\citenamefont {Duan},\ and\ \citenamefont {Monroe}}]{kim:2009:trapped}%
	\BibitemOpen
	\bibfield  {author} {\bibinfo {author} {\bibfnamefont {K.}~\bibnamefont
			{Kim}}, \bibinfo {author} {\bibfnamefont {M.-S.}\ \bibnamefont {Chang}},
		\bibinfo {author} {\bibfnamefont {R.}~\bibnamefont {Islam}}, \bibinfo
		{author} {\bibfnamefont {S.}~\bibnamefont {Korenblit}}, \bibinfo {author}
		{\bibfnamefont {L.-M.}\ \bibnamefont {Duan}},\ and\ \bibinfo {author}
		{\bibfnamefont {C.}~\bibnamefont {Monroe}},\ }\bibfield  {title} {\bibinfo
		{title} {Entanglement and tunable spin-spin couplings between trapped ions
			using multiple transverse modes},\ }\href
	{https://doi.org/10.1103/PhysRevLett.103.120502} {\bibfield  {journal}
		{\bibinfo  {journal} {Phys. Rev. Lett.}\ }\textbf {\bibinfo {volume} {103}},\
		\bibinfo {pages} {120502} (\bibinfo {year} {2009})}\BibitemShut {NoStop}%
	\bibitem [{\citenamefont {Jurcevic}\ \emph {et~al.}(2014)\citenamefont
		{Jurcevic}, \citenamefont {Lanyon}, \citenamefont {Hauke}, \citenamefont
		{Hempel}, \citenamefont {Zoller}, \citenamefont {Blatt},\ and\ \citenamefont
		{Roos}}]{jurcevic:2014:trapped}%
	\BibitemOpen
	\bibfield  {author} {\bibinfo {author} {\bibfnamefont {P.}~\bibnamefont
			{Jurcevic}}, \bibinfo {author} {\bibfnamefont {B.~P.}\ \bibnamefont
			{Lanyon}}, \bibinfo {author} {\bibfnamefont {P.}~\bibnamefont {Hauke}},
		\bibinfo {author} {\bibfnamefont {C.}~\bibnamefont {Hempel}}, \bibinfo
		{author} {\bibfnamefont {P.}~\bibnamefont {Zoller}}, \bibinfo {author}
		{\bibfnamefont {R.}~\bibnamefont {Blatt}},\ and\ \bibinfo {author}
		{\bibfnamefont {C.~F.}\ \bibnamefont {Roos}},\ }\bibfield  {title} {\bibinfo
		{title} {Quasiparticle engineering and entanglement propagation in a quantum
			many-body system},\ }\href {https://doi.org/10.1038/nature13461} {\bibfield
		{journal} {\bibinfo  {journal} {Nature}\ }\textbf {\bibinfo {volume} {511}},\
		\bibinfo {pages} {202} (\bibinfo {year} {2014})}\BibitemShut {NoStop}%
	\bibitem [{Note2()}]{Note2}%
	\BibitemOpen
	\bibinfo {note} {The channels $\Omega _n$ can be considered quantum noises
		applying to the input states where there is a temporal correlation in these
		noises. The sequence $\protect \{p_n\protect \}$ resembles the NARMA
		benchmark~\cite {atiya:2000:narma}, which is commonly used for evaluating the
		computational capability of temporal processing with long time dependence.
		Here, $\protect \{u_n\protect \}$ is a random sequence of scalar values in
		$[0, 1]$ but is rescaled into $[0, 0.2]$ before creating $p_n$ to set $p_n$
		into the stable range in $[0, 1]$.}\BibitemShut {Stop}%
	\bibitem [{\citenamefont {Jaeger}(2001{\natexlab{b}})}]{jaeger:2001:short}%
	\BibitemOpen
	\bibfield  {author} {\bibinfo {author} {\bibfnamefont {H.}~\bibnamefont
			{Jaeger}},\ }\bibinfo {title} {Short term memory in echo state networks}\
	(\bibinfo  {publisher} {GMD-Forschungszentrum Informationstechnik},\ \bibinfo
	{year} {2001})\ p.~\bibinfo {pages} {60}\BibitemShut {NoStop}%
	\bibitem [{\citenamefont {Sz{\'{e}}kely}\ \emph {et~al.}(2007)\citenamefont
		{Sz{\'{e}}kely}, \citenamefont {Rizzo},\ and\ \citenamefont
		{Bakirov}}]{szkely:2007:corr}%
	\BibitemOpen
	\bibfield  {author} {\bibinfo {author} {\bibfnamefont {G.~J.}\ \bibnamefont
			{Sz{\'{e}}kely}}, \bibinfo {author} {\bibfnamefont {M.~L.}\ \bibnamefont
			{Rizzo}},\ and\ \bibinfo {author} {\bibfnamefont {N.~K.}\ \bibnamefont
			{Bakirov}},\ }\bibfield  {title} {\bibinfo {title} {Measuring and testing
			dependence by correlation of distances},\ }\href
	{https://doi.org/10.1214/009053607000000505} {\bibfield  {journal} {\bibinfo
			{journal} {Ann. Stat.}\ }\textbf {\bibinfo {volume} {35}},\ \bibinfo {pages}
		{2769} (\bibinfo {year} {2007})}\BibitemShut {NoStop}%
	\bibitem [{Note3()}]{Note3}%
	\BibitemOpen
	\bibinfo {note} {For temporal classical tasks in quantum spin networks,
		similar observations~\cite {pena:2021:qrc:dynamic} have also been
		investigated recently to address the impact of the transition between
		localization and thermalization manifest. Our theoretical investigation is
		not limited to quantum spin networks, but is more generally associated with
		the spectrum analysis of reduced dynamics maps. Furthermore, the intriguing
		results in QMC under varying model parameters remind us of the well-known
		phenomenon in classical RC called the edge of chaos, where in some
		situations, a memory capacity achieves the maximum values at the edge of
		stability between different dynamics regimes~\cite
		{bertschinger:2004:edgeNN,toyoizumi:2011:edgepre,haruna:2019:shortterm}.}\BibitemShut
	{Stop}%
	\bibitem [{\citenamefont {Mujal}\ \emph {et~al.}(2021)\citenamefont {Mujal},
		\citenamefont {Mart{\'{\i}}nez-Pe{\~{n}}a}, \citenamefont {Nokkala},
		\citenamefont {Garc{\'{\i}}a-Beni}, \citenamefont {Giorgi}, \citenamefont
		{Soriano},\ and\ \citenamefont {Zambrini}}]{mujal:2021:opportunities}%
	\BibitemOpen
	\bibfield  {author} {\bibinfo {author} {\bibfnamefont {P.}~\bibnamefont
			{Mujal}}, \bibinfo {author} {\bibfnamefont {R.}~\bibnamefont
			{Mart{\'{\i}}nez-Pe{\~{n}}a}}, \bibinfo {author} {\bibfnamefont
			{J.}~\bibnamefont {Nokkala}}, \bibinfo {author} {\bibfnamefont
			{J.}~\bibnamefont {Garc{\'{\i}}a-Beni}}, \bibinfo {author} {\bibfnamefont
			{G.~L.}\ \bibnamefont {Giorgi}}, \bibinfo {author} {\bibfnamefont {M.~C.}\
			\bibnamefont {Soriano}},\ and\ \bibinfo {author} {\bibfnamefont
			{R.}~\bibnamefont {Zambrini}},\ }\bibfield  {title} {\bibinfo {title}
		{Opportunities in quantum reservoir computing and extreme learning
			machines},\ }\href {https://doi.org/10.1002/qute.202100027} {\bibfield
		{journal} {\bibinfo  {journal} {Adv. Quantum Technol.}\ ,\ \bibinfo {pages}
			{2100027}} (\bibinfo {year} {2021})}\BibitemShut {NoStop}%
	\bibitem [{\citenamefont {Negoro}\ \emph {et~al.}(2021)\citenamefont {Negoro},
		\citenamefont {Mitarai}, \citenamefont {Nakajima},\ and\ \citenamefont
		{Fujii}}]{negoro:2021:rcbook:NMR}%
	\BibitemOpen
	\bibfield  {author} {\bibinfo {author} {\bibfnamefont {M.}~\bibnamefont
			{Negoro}}, \bibinfo {author} {\bibfnamefont {K.}~\bibnamefont {Mitarai}},
		\bibinfo {author} {\bibfnamefont {K.}~\bibnamefont {Nakajima}},\ and\
		\bibinfo {author} {\bibfnamefont {K.}~\bibnamefont {Fujii}},\ }\bibinfo
	{title} {Toward \text{NMR} quantum reservoir computing},\ in\ \href
	{https://doi.org/10.1007/978-981-13-1687-6_19} {\emph {\bibinfo {booktitle}
			{Reservoir Computing: Theory, Physical Implementations, and Applications}}},\
	\bibinfo {editor} {edited by\ \bibinfo {editor} {\bibfnamefont
			{K.}~\bibnamefont {Nakajima}}\ and\ \bibinfo {editor} {\bibfnamefont
			{I.}~\bibnamefont {Fischer}}}\ (\bibinfo  {publisher} {Springer Singapore},\
	\bibinfo {address} {Singapore},\ \bibinfo {year} {2021})\ pp.\ \bibinfo
	{pages} {451--458}\BibitemShut {NoStop}%
	\bibitem [{\citenamefont {Chen}\ \emph {et~al.}(2020)\citenamefont {Chen},
		\citenamefont {Nurdin},\ and\ \citenamefont {Yamamoto}}]{chen:2020:temporal}%
	\BibitemOpen
	\bibfield  {author} {\bibinfo {author} {\bibfnamefont {J.}~\bibnamefont
			{Chen}}, \bibinfo {author} {\bibfnamefont {H.~I.}\ \bibnamefont {Nurdin}},\
		and\ \bibinfo {author} {\bibfnamefont {N.}~\bibnamefont {Yamamoto}},\
	}\bibfield  {title} {\bibinfo {title} {Temporal information processing on
			noisy quantum computers},\ }\href
	{https://doi.org/10.1103/PhysRevApplied.14.024065} {\bibfield  {journal}
		{\bibinfo  {journal} {Phys. Rev. Applied}\ }\textbf {\bibinfo {volume}
			{14}},\ \bibinfo {pages} {024065} (\bibinfo {year} {2020})}\BibitemShut
	{NoStop}%
	\bibitem [{\citenamefont {Richerme}\ \emph {et~al.}(2014)\citenamefont
		{Richerme}, \citenamefont {Gong}, \citenamefont {Lee}, \citenamefont {Senko},
		\citenamefont {Smith}, \citenamefont {Foss-Feig}, \citenamefont {Michalakis},
		\citenamefont {Gorshkov},\ and\ \citenamefont
		{Monroe}}]{richerme:2014:nature:corr}%
	\BibitemOpen
	\bibfield  {author} {\bibinfo {author} {\bibfnamefont {P.}~\bibnamefont
			{Richerme}}, \bibinfo {author} {\bibfnamefont {Z.-X.}\ \bibnamefont {Gong}},
		\bibinfo {author} {\bibfnamefont {A.}~\bibnamefont {Lee}}, \bibinfo {author}
		{\bibfnamefont {C.}~\bibnamefont {Senko}}, \bibinfo {author} {\bibfnamefont
			{J.}~\bibnamefont {Smith}}, \bibinfo {author} {\bibfnamefont
			{M.}~\bibnamefont {Foss-Feig}}, \bibinfo {author} {\bibfnamefont
			{S.}~\bibnamefont {Michalakis}}, \bibinfo {author} {\bibfnamefont {A.~V.}\
			\bibnamefont {Gorshkov}},\ and\ \bibinfo {author} {\bibfnamefont
			{C.}~\bibnamefont {Monroe}},\ }\bibfield  {title} {\bibinfo {title}
		{{Non-local propagation of correlations in quantum systems with long-range
				interactions}},\ }\href {https://doi.org/10.1038/nature13450} {\bibfield
		{journal} {\bibinfo  {journal} {Nature}\ }\textbf {\bibinfo {volume} {511}},\
		\bibinfo {pages} {198} (\bibinfo {year} {2014})},\ \Eprint
	{https://arxiv.org/abs/1401.5088} {1401.5088} \BibitemShut {NoStop}%
	\bibitem [{\citenamefont {Zhang}\ \emph {et~al.}(2017)\citenamefont {Zhang},
		\citenamefont {Pagano}, \citenamefont {Hess}, \citenamefont {Kyprianidis},
		\citenamefont {Becker}, \citenamefont {Kaplan}, \citenamefont {Gorshkov},
		\citenamefont {Gong},\ and\ \citenamefont {Monroe}}]{zhang:2017:nature:ion}%
	\BibitemOpen
	\bibfield  {author} {\bibinfo {author} {\bibfnamefont {J.}~\bibnamefont
			{Zhang}}, \bibinfo {author} {\bibfnamefont {G.}~\bibnamefont {Pagano}},
		\bibinfo {author} {\bibfnamefont {P.~W.}\ \bibnamefont {Hess}}, \bibinfo
		{author} {\bibfnamefont {A.}~\bibnamefont {Kyprianidis}}, \bibinfo {author}
		{\bibfnamefont {P.}~\bibnamefont {Becker}}, \bibinfo {author} {\bibfnamefont
			{H.}~\bibnamefont {Kaplan}}, \bibinfo {author} {\bibfnamefont {A.~V.}\
			\bibnamefont {Gorshkov}}, \bibinfo {author} {\bibfnamefont {Z.-X.}\
			\bibnamefont {Gong}},\ and\ \bibinfo {author} {\bibfnamefont
			{C.}~\bibnamefont {Monroe}},\ }\bibfield  {title} {\bibinfo {title}
		{{Observation of a many-body dynamical phase transition with a 53-qubit
				quantum simulator}},\ }\href {https://doi.org/10.1038/nature24654} {\bibfield
		{journal} {\bibinfo  {journal} {Nature}\ }\textbf {\bibinfo {volume}
			{551}},\ \bibinfo {pages} {601} (\bibinfo {year} {2017})},\ \Eprint
	{https://arxiv.org/abs/1708.01044} {1708.01044} \BibitemShut {NoStop}%
	\bibitem [{Note4()}]{Note4}%
	\BibitemOpen
	\bibinfo {note} {In this platform, one can surpass the overhead on the large
		number of measurements repeated at each time step by utilizing the shadow
		tomography protocol to estimate many observables from a reasonable number of
		projective measurements~\cite {huang:2020:nat:measurements}.}\BibitemShut
	{Stop}%
	\bibitem [{\citenamefont {Dambre}\ \emph {et~al.}(2012)\citenamefont {Dambre},
		\citenamefont {Verstraeten}, \citenamefont {Schrauwen},\ and\ \citenamefont
		{Massar}}]{dambre:2012:nonlinear}%
	\BibitemOpen
	\bibfield  {author} {\bibinfo {author} {\bibfnamefont {J.}~\bibnamefont
			{Dambre}}, \bibinfo {author} {\bibfnamefont {D.}~\bibnamefont {Verstraeten}},
		\bibinfo {author} {\bibfnamefont {B.}~\bibnamefont {Schrauwen}},\ and\
		\bibinfo {author} {\bibfnamefont {S.}~\bibnamefont {Massar}},\ }\bibfield
	{title} {\bibinfo {title} {Information processing capacity of dynamical
			systems},\ }\href {https://doi.org/10.1038/srep00514} {\bibfield  {journal}
		{\bibinfo  {journal} {Sci. Rep.}\ }\textbf {\bibinfo {volume} {2}},\ \bibinfo
		{pages} {514} (\bibinfo {year} {2012})}\BibitemShut {NoStop}%
	\bibitem [{\citenamefont {Kubota}\ \emph {et~al.}(2021)\citenamefont {Kubota},
		\citenamefont {Takahashi},\ and\ \citenamefont
		{Nakajima}}]{kubota:2021:prr:IPC}%
	\BibitemOpen
	\bibfield  {author} {\bibinfo {author} {\bibfnamefont {T.}~\bibnamefont
			{Kubota}}, \bibinfo {author} {\bibfnamefont {H.}~\bibnamefont {Takahashi}},\
		and\ \bibinfo {author} {\bibfnamefont {K.}~\bibnamefont {Nakajima}},\
	}\bibfield  {title} {\bibinfo {title} {Unifying framework for information
			processing in stochastically driven dynamical systems},\ }\href
	{https://doi.org/10.1103/PhysRevResearch.3.043135} {\bibfield  {journal}
		{\bibinfo  {journal} {Phys. Rev. Research}\ }\textbf {\bibinfo {volume}
			{3}},\ \bibinfo {pages} {043135} (\bibinfo {year} {2021})}\BibitemShut
	{NoStop}%
	\bibitem [{\citenamefont {Nokkala}\ \emph {et~al.}(2021)\citenamefont
		{Nokkala}, \citenamefont {Mart{\'{\i}}nez-Pe{\~{n}}a}, \citenamefont
		{Giorgi}, \citenamefont {Parigi}, \citenamefont {Soriano},\ and\
		\citenamefont {Zambrini}}]{nokkala:2020:gaussian}%
	\BibitemOpen
	\bibfield  {author} {\bibinfo {author} {\bibfnamefont {J.}~\bibnamefont
			{Nokkala}}, \bibinfo {author} {\bibfnamefont {R.}~\bibnamefont
			{Mart{\'{\i}}nez-Pe{\~{n}}a}}, \bibinfo {author} {\bibfnamefont {G.~L.}\
			\bibnamefont {Giorgi}}, \bibinfo {author} {\bibfnamefont {V.}~\bibnamefont
			{Parigi}}, \bibinfo {author} {\bibfnamefont {M.~C.}\ \bibnamefont
			{Soriano}},\ and\ \bibinfo {author} {\bibfnamefont {R.}~\bibnamefont
			{Zambrini}},\ }\bibfield  {title} {\bibinfo {title} {Gaussian states of
			continuous-variable quantum systems provide universal and versatile reservoir
			computing},\ }\href {https://doi.org/10.1038/s42005-021-00556-w} {\bibfield
		{journal} {\bibinfo  {journal} {Commun. Phys.}\ }\textbf {\bibinfo {volume}
			{4}},\ \bibinfo {pages} {53} (\bibinfo {year} {2021})}\BibitemShut {NoStop}%
	\bibitem [{\citenamefont {Govia}\ \emph {et~al.}(2021)\citenamefont {Govia},
		\citenamefont {Ribeill}, \citenamefont {Rowlands}, \citenamefont {Krovi},\
		and\ \citenamefont {Ohki}}]{govia:2020:quantum}%
	\BibitemOpen
	\bibfield  {author} {\bibinfo {author} {\bibfnamefont {L.~C.~G.}\
			\bibnamefont {Govia}}, \bibinfo {author} {\bibfnamefont {G.~J.}\ \bibnamefont
			{Ribeill}}, \bibinfo {author} {\bibfnamefont {G.~E.}\ \bibnamefont
			{Rowlands}}, \bibinfo {author} {\bibfnamefont {H.~K.}\ \bibnamefont
			{Krovi}},\ and\ \bibinfo {author} {\bibfnamefont {T.~A.}\ \bibnamefont
			{Ohki}},\ }\bibfield  {title} {\bibinfo {title} {Quantum reservoir computing
			with a single nonlinear oscillator},\ }\href
	{https://doi.org/10.1103/PhysRevResearch.3.013077} {\bibfield  {journal}
		{\bibinfo  {journal} {Phys. Rev. Research}\ }\textbf {\bibinfo {volume}
			{3}},\ \bibinfo {pages} {013077} (\bibinfo {year} {2021})}\BibitemShut
	{NoStop}%
	\bibitem [{\citenamefont {Dasgupta}\ \emph {et~al.}(2020)\citenamefont
		{Dasgupta}, \citenamefont {Hamilton},\ and\ \citenamefont
		{Banerjee}}]{dasgupta:2020:designing}%
	\BibitemOpen
	\bibfield  {author} {\bibinfo {author} {\bibfnamefont {S.}~\bibnamefont
			{Dasgupta}}, \bibinfo {author} {\bibfnamefont {K.~E.}\ \bibnamefont
			{Hamilton}},\ and\ \bibinfo {author} {\bibfnamefont {A.}~\bibnamefont
			{Banerjee}},\ }\bibfield  {title} {\bibinfo {title} {Designing a {NISQ}
			reservoir with maximal memory capacity for volatility forecasting},\ }\href
	{https://arxiv.org/abs/2004.08240} {\bibfield  {journal} {\bibinfo  {journal}
			{Preprint at arXiv:2004.08240}\ } (\bibinfo {year} {2020})}\BibitemShut
	{NoStop}%
	\bibitem [{\citenamefont {Abrams}\ and\ \citenamefont
		{Lloyd}(1999)}]{abram:1993:quantum}%
	\BibitemOpen
	\bibfield  {author} {\bibinfo {author} {\bibfnamefont {D.~S.}\ \bibnamefont
			{Abrams}}\ and\ \bibinfo {author} {\bibfnamefont {S.}~\bibnamefont {Lloyd}},\
	}\bibfield  {title} {\bibinfo {title} {Quantum algorithm providing
			exponential speed increase for finding eigenvalues and eigenvectors},\ }\href
	{https://doi.org/10.1103/PhysRevLett.83.5162} {\bibfield  {journal} {\bibinfo
			{journal} {Phys. Rev. Lett.}\ }\textbf {\bibinfo {volume} {83}},\ \bibinfo
		{pages} {5162} (\bibinfo {year} {1999})}\BibitemShut {NoStop}%
	\bibitem [{\citenamefont {Bantysh}\ \emph {et~al.}(2020)\citenamefont
		{Bantysh}, \citenamefont {Chernyavskiy},\ and\ \citenamefont
		{Bogdanov}}]{bantysh:2020:quantum}%
	\BibitemOpen
	\bibfield  {author} {\bibinfo {author} {\bibfnamefont {B.~I.}\ \bibnamefont
			{Bantysh}}, \bibinfo {author} {\bibfnamefont {A.~Y.}\ \bibnamefont
			{Chernyavskiy}},\ and\ \bibinfo {author} {\bibfnamefont {Y.~I.}\ \bibnamefont
			{Bogdanov}},\ }\bibfield  {title} {\bibinfo {title} {Quantum tomography
			benchmarking},\ }\href {https://arxiv.org/abs/2012.15656} {\bibfield
		{journal} {\bibinfo  {journal} {Preprint at arXiv:2012.15656}\ } (\bibinfo
		{year} {2020})}\BibitemShut {NoStop}%
	\bibitem [{\citenamefont {Quek}\ \emph {et~al.}(2021)\citenamefont {Quek},
		\citenamefont {Fort},\ and\ \citenamefont {Ng}}]{quek:2021:adaptive}%
	\BibitemOpen
	\bibfield  {author} {\bibinfo {author} {\bibfnamefont {Y.}~\bibnamefont
			{Quek}}, \bibinfo {author} {\bibfnamefont {S.}~\bibnamefont {Fort}},\ and\
		\bibinfo {author} {\bibfnamefont {H.~K.}\ \bibnamefont {Ng}},\ }\bibfield
	{title} {\bibinfo {title} {{Adaptive quantum state tomography with neural
				networks}},\ }\href {https://doi.org/10.1038/s41534-021-00436-9} {\bibfield
		{journal} {\bibinfo  {journal} {npj Quantum Inf.}\ }\textbf {\bibinfo
			{volume} {7}},\ \bibinfo {pages} {105} (\bibinfo {year} {2021})}\BibitemShut
	{NoStop}%
	\bibitem [{\citenamefont {Bruneau}\ \emph {et~al.}(2010)\citenamefont
		{Bruneau}, \citenamefont {Joye},\ and\ \citenamefont
		{Merkli}}]{bruneau:2010:infinite}%
	\BibitemOpen
	\bibfield  {author} {\bibinfo {author} {\bibfnamefont {L.}~\bibnamefont
			{Bruneau}}, \bibinfo {author} {\bibfnamefont {A.}~\bibnamefont {Joye}},\ and\
		\bibinfo {author} {\bibfnamefont {M.}~\bibnamefont {Merkli}},\ }\bibfield
	{title} {\bibinfo {title} {Infinite products of random matrices and repeated
			interaction dynamics},\ }\href {https://doi.org/10.1214/09-aihp211}
	{\bibfield  {journal} {\bibinfo  {journal} {Ann. Inst. H. Poincar{\'{e}}
				Probab. Statist.}\ }\textbf {\bibinfo {volume} {46}},\ \bibinfo {pages} {442}
		(\bibinfo {year} {2010})}\BibitemShut {NoStop}%
	\bibitem [{\citenamefont {Nechita}\ and\ \citenamefont
		{Pellegrini}(2010)}]{nechita:2010:random}%
	\BibitemOpen
	\bibfield  {author} {\bibinfo {author} {\bibfnamefont {I.}~\bibnamefont
			{Nechita}}\ and\ \bibinfo {author} {\bibfnamefont {C.}~\bibnamefont
			{Pellegrini}},\ }\bibfield  {title} {\bibinfo {title} {Random repeated
			quantum interactions and random invariant states},\ }\href
	{https://doi.org/10.1007/s00440-010-0323-6} {\bibfield  {journal} {\bibinfo
			{journal} {Probab. Theory Relat. Fields}\ }\textbf {\bibinfo {volume}
			{152}},\ \bibinfo {pages} {299} (\bibinfo {year} {2010})}\BibitemShut
	{NoStop}%
	\bibitem [{\citenamefont {Movassagh}\ and\ \citenamefont
		{Schenker}(2019)}]{movassagh:2019:ergodic}%
	\BibitemOpen
	\bibfield  {author} {\bibinfo {author} {\bibfnamefont {R.}~\bibnamefont
			{Movassagh}}\ and\ \bibinfo {author} {\bibfnamefont {J.}~\bibnamefont
			{Schenker}},\ }\bibfield  {title} {\bibinfo {title} {An ergodic theorem for
			homogeneously distributed quantum channels with applications to matrix
			product states},\ }\href {https://arxiv.org/abs/1909.11769} {\bibfield
		{journal} {\bibinfo  {journal} {Preprint at arXiv:1909.11769}\ } (\bibinfo
		{year} {2019})}\BibitemShut {NoStop}%
	\bibitem [{\citenamefont {Movassagh}\ and\ \citenamefont
		{Schenker}(2021)}]{movassagh:2020:theory}%
	\BibitemOpen
	\bibfield  {author} {\bibinfo {author} {\bibfnamefont {R.}~\bibnamefont
			{Movassagh}}\ and\ \bibinfo {author} {\bibfnamefont {J.}~\bibnamefont
			{Schenker}},\ }\bibfield  {title} {\bibinfo {title} {Theory of ergodic
			quantum processes},\ }\href {https://doi.org/10.1103/PhysRevX.11.041001}
	{\bibfield  {journal} {\bibinfo  {journal} {Phys. Rev. X}\ }\textbf {\bibinfo
			{volume} {11}},\ \bibinfo {pages} {041001} (\bibinfo {year}
		{2021})}\BibitemShut {NoStop}%
	\bibitem [{\citenamefont {Steele}(1989)}]{steele:1989:subadd}%
	\BibitemOpen
	\bibfield  {author} {\bibinfo {author} {\bibfnamefont {J.~M.}\ \bibnamefont
			{Steele}},\ }\bibfield  {title} {\bibinfo {title} {Kingman's subadditive
			ergodic theorem},\ }\href {http://www.numdam.org/item/AIHPB_1989__25_1_93_0}
	{\bibfield  {journal} {\bibinfo  {journal} {Ann. Inst. H. Poincar{\'{e}}
				Probab. Statist.s}\ }\textbf {\bibinfo {volume} {25}},\ \bibinfo {pages} {93}
		(\bibinfo {year} {1989})}\BibitemShut {NoStop}%
	\bibitem [{\citenamefont {Hennion}(1997)}]{hennion:1997:product}%
	\BibitemOpen
	\bibfield  {author} {\bibinfo {author} {\bibfnamefont {H.}~\bibnamefont
			{Hennion}},\ }\bibfield  {title} {\bibinfo {title} {Limit theorems for
			products of positive random matrices},\ }\href
	{https://doi.org/10.1214/aop/1023481103} {\bibfield  {journal} {\bibinfo
			{journal} {Ann. Probab.}\ }\textbf {\bibinfo {volume} {25}},\ \bibinfo
		{pages} {1545} (\bibinfo {year} {1997})}\BibitemShut {NoStop}%
	\bibitem [{\citenamefont {Macieszczak}\ \emph {et~al.}(2016)\citenamefont
		{Macieszczak}, \citenamefont {Gu\ifmmode \mbox{\c{t}}\else
			\c{t}\fi{}\ifmmode~\u{a}\else \u{a}\fi{}}, \citenamefont {Lesanovsky},\ and\
		\citenamefont {Garrahan}}]{macieszczak:2016:meta}%
	\BibitemOpen
	\bibfield  {author} {\bibinfo {author} {\bibfnamefont {K.}~\bibnamefont
			{Macieszczak}}, \bibinfo {author} {\bibfnamefont {M.}~\bibnamefont
			{Gu\ifmmode \mbox{\c{t}}\else \c{t}\fi{}\ifmmode~\u{a}\else \u{a}\fi{}}},
		\bibinfo {author} {\bibfnamefont {I.}~\bibnamefont {Lesanovsky}},\ and\
		\bibinfo {author} {\bibfnamefont {J.~P.}\ \bibnamefont {Garrahan}},\
	}\bibfield  {title} {\bibinfo {title} {Towards a theory of metastability in
			open quantum dynamics},\ }\href
	{https://doi.org/10.1103/PhysRevLett.116.240404} {\bibfield  {journal}
		{\bibinfo  {journal} {Phys. Rev. Lett.}\ }\textbf {\bibinfo {volume} {116}},\
		\bibinfo {pages} {240404} (\bibinfo {year} {2016})}\BibitemShut {NoStop}%
	\bibitem [{\citenamefont {{Lasserre}}(1995)}]{lasserre:ineq:trace}%
	\BibitemOpen
	\bibfield  {author} {\bibinfo {author} {\bibfnamefont {J.~B.}\ \bibnamefont
			{{Lasserre}}},\ }\bibfield  {title} {\bibinfo {title} {A trace inequality for
			matrix product},\ }\href {https://doi.org/10.1109/9.402252} {\bibfield
		{journal} {\bibinfo  {journal} {IEEE Trans. Autom. Control.}\ }\textbf
		{\bibinfo {volume} {40}},\ \bibinfo {pages} {1500} (\bibinfo {year}
		{1995})}\BibitemShut {NoStop}%
	\bibitem [{\citenamefont {Misra}\ and\ \citenamefont
		{Sudarshan}(1977)}]{misra:1977:zeno}%
	\BibitemOpen
	\bibfield  {author} {\bibinfo {author} {\bibfnamefont {B.}~\bibnamefont
			{Misra}}\ and\ \bibinfo {author} {\bibfnamefont {E.~C.~G.}\ \bibnamefont
			{Sudarshan}},\ }\bibfield  {title} {\bibinfo {title} {The {Zeno's} paradox in
			quantum theory},\ }\href {https://doi.org/10.1063/1.523304} {\bibfield
		{journal} {\bibinfo  {journal} {J. Math. Phys.}\ }\textbf {\bibinfo {volume}
			{18}},\ \bibinfo {pages} {756} (\bibinfo {year} {1977})}\BibitemShut
	{NoStop}%
	\bibitem [{\citenamefont {Chiribella}(2012)}]{chiribella:2012:qswitch}%
	\BibitemOpen
	\bibfield  {author} {\bibinfo {author} {\bibfnamefont {G.}~\bibnamefont
			{Chiribella}},\ }\bibfield  {title} {\bibinfo {title} {Perfect discrimination
			of no-signalling channels via quantum superposition of causal structures},\
	}\href {https://doi.org/10.1103/PhysRevA.86.040301} {\bibfield  {journal}
		{\bibinfo  {journal} {Phys. Rev. A}\ }\textbf {\bibinfo {volume} {86}},\
		\bibinfo {pages} {040301(R)} (\bibinfo {year} {2012})}\BibitemShut {NoStop}%
	\bibitem [{\citenamefont {Procopio}\ \emph {et~al.}(2015)\citenamefont
		{Procopio}, \citenamefont {Moqanaki}, \citenamefont {Ara{\'{u}}jo},
		\citenamefont {Costa}, \citenamefont {Calafell}, \citenamefont {Dowd},
		\citenamefont {Hamel}, \citenamefont {Rozema}, \citenamefont {Brukner},\ and\
		\citenamefont {Walther}}]{procopio:2015:qswitch:exp}%
	\BibitemOpen
	\bibfield  {author} {\bibinfo {author} {\bibfnamefont {L.~M.}\ \bibnamefont
			{Procopio}}, \bibinfo {author} {\bibfnamefont {A.}~\bibnamefont {Moqanaki}},
		\bibinfo {author} {\bibfnamefont {M.}~\bibnamefont {Ara{\'{u}}jo}}, \bibinfo
		{author} {\bibfnamefont {F.}~\bibnamefont {Costa}}, \bibinfo {author}
		{\bibfnamefont {I.~A.}\ \bibnamefont {Calafell}}, \bibinfo {author}
		{\bibfnamefont {E.~G.}\ \bibnamefont {Dowd}}, \bibinfo {author}
		{\bibfnamefont {D.~R.}\ \bibnamefont {Hamel}}, \bibinfo {author}
		{\bibfnamefont {L.~A.}\ \bibnamefont {Rozema}}, \bibinfo {author}
		{\bibfnamefont {{\v{C}}.}~\bibnamefont {Brukner}},\ and\ \bibinfo {author}
		{\bibfnamefont {P.}~\bibnamefont {Walther}},\ }\bibfield  {title} {\bibinfo
		{title} {Experimental superposition of orders of quantum gates},\ }\href
	{https://doi.org/10.1038/ncomms8913} {\bibfield  {journal} {\bibinfo
			{journal} {Nat. Commun.}\ }\textbf {\bibinfo {volume} {6}},\ \bibinfo {pages}
		{7913} (\bibinfo {year} {2015})}\BibitemShut {NoStop}%
	\bibitem [{\citenamefont {Rubino}\ \emph {et~al.}(2017)\citenamefont {Rubino},
		\citenamefont {Rozema}, \citenamefont {Feix}, \citenamefont {Ara{\'{u}}jo},
		\citenamefont {Zeuner}, \citenamefont {Procopio}, \citenamefont {Brukner},\
		and\ \citenamefont {Walther}}]{rubino:2017:qswitch:exp}%
	\BibitemOpen
	\bibfield  {author} {\bibinfo {author} {\bibfnamefont {G.}~\bibnamefont
			{Rubino}}, \bibinfo {author} {\bibfnamefont {L.~A.}\ \bibnamefont {Rozema}},
		\bibinfo {author} {\bibfnamefont {A.}~\bibnamefont {Feix}}, \bibinfo {author}
		{\bibfnamefont {M.}~\bibnamefont {Ara{\'{u}}jo}}, \bibinfo {author}
		{\bibfnamefont {J.~M.}\ \bibnamefont {Zeuner}}, \bibinfo {author}
		{\bibfnamefont {L.~M.}\ \bibnamefont {Procopio}}, \bibinfo {author}
		{\bibfnamefont {{\v{C}}.}~\bibnamefont {Brukner}},\ and\ \bibinfo {author}
		{\bibfnamefont {P.}~\bibnamefont {Walther}},\ }\bibfield  {title} {\bibinfo
		{title} {Experimental verification of an indefinite causal order},\ }\href
	{https://doi.org/10.1126/sciadv.1602589} {\bibfield  {journal} {\bibinfo
			{journal} {Sci. Adv.}\ }\textbf {\bibinfo {volume} {3}},\ \bibinfo {pages}
		{e1602589} (\bibinfo {year} {2017})}\BibitemShut {NoStop}%
	\bibitem [{\citenamefont {Goswami}\ \emph {et~al.}(2018)\citenamefont
		{Goswami}, \citenamefont {Giarmatzi}, \citenamefont {Kewming}, \citenamefont
		{Costa}, \citenamefont {Branciard}, \citenamefont {Romero},\ and\
		\citenamefont {White}}]{goswami:2018:qswitch:exp}%
	\BibitemOpen
	\bibfield  {author} {\bibinfo {author} {\bibfnamefont {K.}~\bibnamefont
			{Goswami}}, \bibinfo {author} {\bibfnamefont {C.}~\bibnamefont {Giarmatzi}},
		\bibinfo {author} {\bibfnamefont {M.}~\bibnamefont {Kewming}}, \bibinfo
		{author} {\bibfnamefont {F.}~\bibnamefont {Costa}}, \bibinfo {author}
		{\bibfnamefont {C.}~\bibnamefont {Branciard}}, \bibinfo {author}
		{\bibfnamefont {J.}~\bibnamefont {Romero}},\ and\ \bibinfo {author}
		{\bibfnamefont {A.~G.}\ \bibnamefont {White}},\ }\bibfield  {title} {\bibinfo
		{title} {Indefinite causal order in a quantum switch},\ }\href
	{https://doi.org/10.1103/PhysRevLett.121.090503} {\bibfield  {journal}
		{\bibinfo  {journal} {Phys. Rev. Lett.}\ }\textbf {\bibinfo {volume} {121}},\
		\bibinfo {pages} {090503} (\bibinfo {year} {2018})}\BibitemShut {NoStop}%
	\bibitem [{\citenamefont {Wei}\ \emph {et~al.}(2019)\citenamefont {Wei},
		\citenamefont {Tischler}, \citenamefont {Zhao}, \citenamefont {Li},
		\citenamefont {Arrazola}, \citenamefont {Liu}, \citenamefont {Zhang},
		\citenamefont {Li}, \citenamefont {You}, \citenamefont {Wang}, \citenamefont
		{Chen}, \citenamefont {Sanders}, \citenamefont {Zhang}, \citenamefont
		{Pryde}, \citenamefont {Xu},\ and\ \citenamefont
		{Pan}}]{wei:2019:qswtich:exp}%
	\BibitemOpen
	\bibfield  {author} {\bibinfo {author} {\bibfnamefont {K.}~\bibnamefont
			{Wei}}, \bibinfo {author} {\bibfnamefont {N.}~\bibnamefont {Tischler}},
		\bibinfo {author} {\bibfnamefont {S.-R.}\ \bibnamefont {Zhao}}, \bibinfo
		{author} {\bibfnamefont {Y.-H.}\ \bibnamefont {Li}}, \bibinfo {author}
		{\bibfnamefont {J.~M.}\ \bibnamefont {Arrazola}}, \bibinfo {author}
		{\bibfnamefont {Y.}~\bibnamefont {Liu}}, \bibinfo {author} {\bibfnamefont
			{W.}~\bibnamefont {Zhang}}, \bibinfo {author} {\bibfnamefont
			{H.}~\bibnamefont {Li}}, \bibinfo {author} {\bibfnamefont {L.}~\bibnamefont
			{You}}, \bibinfo {author} {\bibfnamefont {Z.}~\bibnamefont {Wang}}, \bibinfo
		{author} {\bibfnamefont {Y.-A.}\ \bibnamefont {Chen}}, \bibinfo {author}
		{\bibfnamefont {B.~C.}\ \bibnamefont {Sanders}}, \bibinfo {author}
		{\bibfnamefont {Q.}~\bibnamefont {Zhang}}, \bibinfo {author} {\bibfnamefont
			{G.~J.}\ \bibnamefont {Pryde}}, \bibinfo {author} {\bibfnamefont
			{F.}~\bibnamefont {Xu}},\ and\ \bibinfo {author} {\bibfnamefont {J.-W.}\
			\bibnamefont {Pan}},\ }\bibfield  {title} {\bibinfo {title} {Experimental
			quantum switching for exponentially superior quantum communication
			complexity},\ }\href {https://doi.org/10.1103/PhysRevLett.122.120504}
	{\bibfield  {journal} {\bibinfo  {journal} {Phys. Rev. Lett.}\ }\textbf
		{\bibinfo {volume} {122}},\ \bibinfo {pages} {120504} (\bibinfo {year}
		{2019})}\BibitemShut {NoStop}%
	\bibitem [{\citenamefont {Guo}\ \emph {et~al.}(2020)\citenamefont {Guo},
		\citenamefont {Hu}, \citenamefont {Hou}, \citenamefont {Cao}, \citenamefont
		{Cui}, \citenamefont {Liu}, \citenamefont {Huang}, \citenamefont {Li},
		\citenamefont {Guo},\ and\ \citenamefont
		{Chiribella}}]{guo:2020:qswitch:exp}%
	\BibitemOpen
	\bibfield  {author} {\bibinfo {author} {\bibfnamefont {Y.}~\bibnamefont
			{Guo}}, \bibinfo {author} {\bibfnamefont {X.-M.}\ \bibnamefont {Hu}},
		\bibinfo {author} {\bibfnamefont {Z.-B.}\ \bibnamefont {Hou}}, \bibinfo
		{author} {\bibfnamefont {H.}~\bibnamefont {Cao}}, \bibinfo {author}
		{\bibfnamefont {J.-M.}\ \bibnamefont {Cui}}, \bibinfo {author} {\bibfnamefont
			{B.-H.}\ \bibnamefont {Liu}}, \bibinfo {author} {\bibfnamefont {Y.-F.}\
			\bibnamefont {Huang}}, \bibinfo {author} {\bibfnamefont {C.-F.}\ \bibnamefont
			{Li}}, \bibinfo {author} {\bibfnamefont {G.-C.}\ \bibnamefont {Guo}},\ and\
		\bibinfo {author} {\bibfnamefont {G.}~\bibnamefont {Chiribella}},\ }\bibfield
	{title} {\bibinfo {title} {Experimental transmission of quantum information
			using a superposition of causal orders},\ }\href
	{https://doi.org/10.1103/PhysRevLett.124.030502} {\bibfield  {journal}
		{\bibinfo  {journal} {Phys. Rev. Lett.}\ }\textbf {\bibinfo {volume} {124}},\
		\bibinfo {pages} {030502} (\bibinfo {year} {2020})}\BibitemShut {NoStop}%
	\bibitem [{\citenamefont {Chiribella}\ \emph {et~al.}(2013)\citenamefont
		{Chiribella}, \citenamefont {D'Ariano}, \citenamefont {Perinotti},\ and\
		\citenamefont {Valiron}}]{chiribella:2013:qswitch}%
	\BibitemOpen
	\bibfield  {author} {\bibinfo {author} {\bibfnamefont {G.}~\bibnamefont
			{Chiribella}}, \bibinfo {author} {\bibfnamefont {G.~M.}\ \bibnamefont
			{D'Ariano}}, \bibinfo {author} {\bibfnamefont {P.}~\bibnamefont
			{Perinotti}},\ and\ \bibinfo {author} {\bibfnamefont {B.}~\bibnamefont
			{Valiron}},\ }\bibfield  {title} {\bibinfo {title} {Quantum computations
			without definite causal structure},\ }\href
	{https://doi.org/10.1103/PhysRevA.88.022318} {\bibfield  {journal} {\bibinfo
			{journal} {Phys. Rev. A}\ }\textbf {\bibinfo {volume} {88}},\ \bibinfo
		{pages} {022318} (\bibinfo {year} {2013})}\BibitemShut {NoStop}%
	\bibitem [{\citenamefont {Ebler}\ \emph {et~al.}(2018)\citenamefont {Ebler},
		\citenamefont {Salek},\ and\ \citenamefont
		{Chiribella}}]{ebler:2018:qswitch}%
	\BibitemOpen
	\bibfield  {author} {\bibinfo {author} {\bibfnamefont {D.}~\bibnamefont
			{Ebler}}, \bibinfo {author} {\bibfnamefont {S.}~\bibnamefont {Salek}},\ and\
		\bibinfo {author} {\bibfnamefont {G.}~\bibnamefont {Chiribella}},\ }\bibfield
	{title} {\bibinfo {title} {Enhanced communication with the assistance of
			indefinite causal order},\ }\href
	{https://doi.org/10.1103/PhysRevLett.120.120502} {\bibfield  {journal}
		{\bibinfo  {journal} {Phys. Rev. Lett.}\ }\textbf {\bibinfo {volume} {120}},\
		\bibinfo {pages} {120502} (\bibinfo {year} {2018})}\BibitemShut {NoStop}%
	\bibitem [{\citenamefont {Meschede}\ \emph {et~al.}(1985)\citenamefont
		{Meschede}, \citenamefont {Walther},\ and\ \citenamefont
		{M\"uller}}]{meschede:1985:prl:maser}%
	\BibitemOpen
	\bibfield  {author} {\bibinfo {author} {\bibfnamefont {D.}~\bibnamefont
			{Meschede}}, \bibinfo {author} {\bibfnamefont {H.}~\bibnamefont {Walther}},\
		and\ \bibinfo {author} {\bibfnamefont {G.}~\bibnamefont {M\"uller}},\
	}\bibfield  {title} {\bibinfo {title} {One-atom maser},\ }\href
	{https://doi.org/10.1103/PhysRevLett.54.551} {\bibfield  {journal} {\bibinfo
			{journal} {Phys. Rev. Lett.}\ }\textbf {\bibinfo {volume} {54}},\ \bibinfo
		{pages} {551} (\bibinfo {year} {1985})}\BibitemShut {NoStop}%
	\bibitem [{\citenamefont {Varcoe}\ \emph {et~al.}(2000)\citenamefont {Varcoe},
		\citenamefont {Brattke}, \citenamefont {Weidinger},\ and\ \citenamefont
		{Walther}}]{varcoe:2000:nat:photon}%
	\BibitemOpen
	\bibfield  {author} {\bibinfo {author} {\bibfnamefont {B.~T.~H.}\
			\bibnamefont {Varcoe}}, \bibinfo {author} {\bibfnamefont {S.}~\bibnamefont
			{Brattke}}, \bibinfo {author} {\bibfnamefont {M.}~\bibnamefont {Weidinger}},\
		and\ \bibinfo {author} {\bibfnamefont {H.}~\bibnamefont {Walther}},\
	}\bibfield  {title} {\bibinfo {title} {Preparing pure photon number states of
			the radiation field},\ }\href {https://doi.org/10.1038/35001526} {\bibfield
		{journal} {\bibinfo  {journal} {Nature}\ }\textbf {\bibinfo {volume} {403}},\
		\bibinfo {pages} {743} (\bibinfo {year} {2000})}\BibitemShut {NoStop}%
	\bibitem [{\citenamefont {Chen}\ and\ \citenamefont
		{Ye}(2011)}]{chen:2011:projection}%
	\BibitemOpen
	\bibfield  {author} {\bibinfo {author} {\bibfnamefont {Y.}~\bibnamefont
			{Chen}}\ and\ \bibinfo {author} {\bibfnamefont {X.}~\bibnamefont {Ye}},\
	}\bibfield  {title} {\bibinfo {title} {Projection onto a simplex},\ }\href
	{https://arxiv.org/abs/1101.6081} {\bibfield  {journal} {\bibinfo  {journal}
			{Preprint at arXiv:1101.6081}\ } (\bibinfo {year} {2011})}\BibitemShut
	{NoStop}%
	\bibitem [{\citenamefont {Atiya}\ and\ \citenamefont
		{Parlos}(2000)}]{atiya:2000:narma}%
	\BibitemOpen
	\bibfield  {author} {\bibinfo {author} {\bibfnamefont {A.}~\bibnamefont
			{Atiya}}\ and\ \bibinfo {author} {\bibfnamefont {A.}~\bibnamefont {Parlos}},\
	}\bibfield  {title} {\bibinfo {title} {New results on recurrent network
			training: unifying the algorithms and accelerating convergence},\ }\href
	{https://doi.org/10.1109/72.846741} {\bibfield  {journal} {\bibinfo
			{journal} {{IEEE} Trans. Neural Netw. Learn. Syst.}\ }\textbf {\bibinfo
			{volume} {11}},\ \bibinfo {pages} {697} (\bibinfo {year} {2000})}\BibitemShut
	{NoStop}%
	\bibitem [{\citenamefont {Mart\'{\i}nez-Pe\~na}\ \emph
		{et~al.}(2021)\citenamefont {Mart\'{\i}nez-Pe\~na}, \citenamefont {Giorgi},
		\citenamefont {Nokkala}, \citenamefont {Soriano},\ and\ \citenamefont
		{Zambrini}}]{pena:2021:qrc:dynamic}%
	\BibitemOpen
	\bibfield  {author} {\bibinfo {author} {\bibfnamefont {R.}~\bibnamefont
			{Mart\'{\i}nez-Pe\~na}}, \bibinfo {author} {\bibfnamefont {G.~L.}\
			\bibnamefont {Giorgi}}, \bibinfo {author} {\bibfnamefont {J.}~\bibnamefont
			{Nokkala}}, \bibinfo {author} {\bibfnamefont {M.~C.}\ \bibnamefont
			{Soriano}},\ and\ \bibinfo {author} {\bibfnamefont {R.}~\bibnamefont
			{Zambrini}},\ }\bibfield  {title} {\bibinfo {title} {Dynamical phase
			transitions in quantum reservoir computing},\ }\href
	{https://doi.org/10.1103/PhysRevLett.127.100502} {\bibfield  {journal}
		{\bibinfo  {journal} {Phys. Rev. Lett.}\ }\textbf {\bibinfo {volume} {127}},\
		\bibinfo {pages} {100502} (\bibinfo {year} {2021})}\BibitemShut {NoStop}%
	\bibitem [{\citenamefont {Bertschinger}\ and\ \citenamefont
		{Natschl\"{a}ger}(2004)}]{bertschinger:2004:edgeNN}%
	\BibitemOpen
	\bibfield  {author} {\bibinfo {author} {\bibfnamefont {N.}~\bibnamefont
			{Bertschinger}}\ and\ \bibinfo {author} {\bibfnamefont {T.}~\bibnamefont
			{Natschl\"{a}ger}},\ }\bibfield  {title} {\bibinfo {title} {Real-time
			computation at the edge of chaos in recurrent neural networks},\ }\href
	{https://doi.org/10.1162/089976604323057443} {\bibfield  {journal} {\bibinfo
			{journal} {Neural Computation}\ }\textbf {\bibinfo {volume} {16}},\ \bibinfo
		{pages} {1413} (\bibinfo {year} {2004})}\BibitemShut {NoStop}%
	\bibitem [{\citenamefont {Toyoizumi}\ and\ \citenamefont
		{Abbott}(2011)}]{toyoizumi:2011:edgepre}%
	\BibitemOpen
	\bibfield  {author} {\bibinfo {author} {\bibfnamefont {T.}~\bibnamefont
			{Toyoizumi}}\ and\ \bibinfo {author} {\bibfnamefont {L.~F.}\ \bibnamefont
			{Abbott}},\ }\bibfield  {title} {\bibinfo {title} {Beyond the edge of chaos:
			Amplification and temporal integration by recurrent networks in the chaotic
			regime},\ }\href {https://doi.org/10.1103/PhysRevE.84.051908} {\bibfield
		{journal} {\bibinfo  {journal} {Phys. Rev. E}\ }\textbf {\bibinfo {volume}
			{84}},\ \bibinfo {pages} {051908} (\bibinfo {year} {2011})}\BibitemShut
	{NoStop}%
	\bibitem [{\citenamefont {Haruna}\ and\ \citenamefont
		{Nakajima}(2019)}]{haruna:2019:shortterm}%
	\BibitemOpen
	\bibfield  {author} {\bibinfo {author} {\bibfnamefont {T.}~\bibnamefont
			{Haruna}}\ and\ \bibinfo {author} {\bibfnamefont {K.}~\bibnamefont
			{Nakajima}},\ }\bibfield  {title} {\bibinfo {title} {Optimal short-term
			memory before the edge of chaos in driven random recurrent networks},\ }\href
	{https://doi.org/10.1103/PhysRevE.100.062312} {\bibfield  {journal} {\bibinfo
			{journal} {Phys. Rev. E}\ }\textbf {\bibinfo {volume} {100}},\ \bibinfo
		{pages} {062312} (\bibinfo {year} {2019})}\BibitemShut {NoStop}%
	\bibitem [{\citenamefont {Huang}\ \emph {et~al.}(2020)\citenamefont {Huang},
		\citenamefont {Kueng},\ and\ \citenamefont
		{Preskill}}]{huang:2020:nat:measurements}%
	\BibitemOpen
	\bibfield  {author} {\bibinfo {author} {\bibfnamefont {H.-Y.}\ \bibnamefont
			{Huang}}, \bibinfo {author} {\bibfnamefont {R.}~\bibnamefont {Kueng}},\ and\
		\bibinfo {author} {\bibfnamefont {J.}~\bibnamefont {Preskill}},\ }\bibfield
	{title} {\bibinfo {title} {{Predicting many properties of a quantum system
				from very few measurements}},\ }\href
	{https://doi.org/10.1038/s41567-020-0932-7} {\bibfield  {journal} {\bibinfo
			{journal} {Nat. Phys.}\ }\textbf {\bibinfo {volume} {16}},\ \bibinfo {pages}
		{1050} (\bibinfo {year} {2020})},\ \Eprint {https://arxiv.org/abs/2002.08953}
	{2002.08953} \BibitemShut {NoStop}%
\end{thebibliography}

\providecommand{\noopsort}[1]{}\providecommand{\singleletter}[1]{#1}%

\end{document}


\title{Supplementary Material for\\
``Learning Temporal Quantum Tomography''}

\author{Quoc Hoan Tran}
\email{tran\_qh@ai.u-tokyo.ac.jp}
\affiliation{
	Graduate School of Information Science and Technology, The University of Tokyo, Tokyo 113-8656, Japan
}

\author{Kohei Nakajima}
\email{k\_nakajima@mech.t.u-tokyo.ac.jp}
\affiliation{
	Graduate School of Information Science and Technology, The University of Tokyo, Tokyo 113-8656, Japan
}
\affiliation{Next Generation Artificial Intelligence Research Center, The University of Tokyo, Tokyo 113-8656, Japan}

\date{\today}

\begin{abstract}
This supplementary material describes in detail the calculations, the experiments introduced in the main text, and the additional figures.
The equation, figure, and table numbers in this section are
prefixed with S (e.g., Eq.\textbf{~}(S1) or Fig.~S1, Table~S1),
while numbers without the prefix (e.g., Eq.~(1) or Fig.~1, Table~1) refer to items in the main text.
\end{abstract}

\maketitle
\tableofcontents
\newpage

\section{Learning the Temporal Tomography}
The proposed framework for learning the temporal tomography is a quantum extension of the classical reservoir computing (RC) to deal with quantum tasks.
In this section, we first explain the principle of RC and then introduce some quantum extensions of RC in a scheme called quantum reservoir computing (QRC).

\subsection{Reservoir Computing}
Within machine learning (ML), the reservoir computing (RC) paradigm is a particular form of classical recurrent neural networks with random connectivity that can generate high-dimensional trajectories~\cite{jaeger:2001:echo,maass:2002:reservoir,lukoeviius:2009:reservoir}.
The crucial principle of RC is based on the modeling for the representation of the input sequence by feeding the input into a dynamical system called \textit{the reservoir} to encode all relevant nonlinear dynamics.
The reservoir is connected to the output by the \textit{readout} part; only connections in this part are trained without affecting the reservoir dynamics.
The RC paradigm is based on the following assumption: no knowledge of the system model, a large collection of measurement of readout, and the possibility to predict the target from the states encoded by the reservoir.
Furthermore, the training mechanism is straightforward and computationally efficient, allowing RC to become particularly suitable for hardware implementations in various physical systems~\cite{nakajima:2020:physical}.

We briefly explain here the standard pipeline of RC.
In a general picture, the information processing in RC is described by the input-driven map $G: \bS \times \cX \to \cX \subset \mathbb{R}^K$,
where $\bS$ and $\cX$ are the input and the reservoir's state space, respectively.
If we consider an infinite discrete-time input sequence $\{\ldots, \bs_{-1}, \bs_0, \bs_1, \ldots\}$ fed into the reservoir, the reservoir state $\bx_n$ is represented by the following recurrent relation:
\begin{align}\label{eqn:driven-map}
\bx_n = G(\bs_n, \bx_{n-1}).
\end{align}
In this way, the sequence $\{\ldots, \bx_{-1}, \bx_0, \bx_1, \ldots\}$ is the nonlinear transformation of the input sequence via the input-driven map implemented by a dynamical system.
Therefore, the training in ML tasks can be separated to the reservoir so that we can reduce the computational cost by selecting a simple training procedure.

The RC paradigm can be used for both temporal and non-temporal supervised learning tasks. 
In temporal supervised learning tasks, we are given an input sequence $\{\bs_1,\ldots,\bs_L\}$ and the corresponding target sequence $\hat{\by}=\{\hat{\by}_1,\ldots,\hat{\by}_L\}$ where $\hat{\by}_k \in \bbR^d$ with $d$ is the output dimension.
An RC system with temporal information processing ability includes a readout map $h: \cX \to \bbR^d$, where the output signal $\by_n$ is simply obtained from the readout map $\by_n = h(\bx_n)$ such that $\by_n \approx \hat{\by}_n$ ($n=1,\ldots,L$).
The readout map is simply taken as a linear combination of the reservoir states as $\by_n = h(\bx_n) = \bw^\top\bx_n$,
where $\bw$ is the parameter and needs to be optimized.
The training to adjust $\bw$ is performed by minimizing the error such as the mean-square error between $\by_n$ and $\hat{\by_n}$ over $n=1,\ldots,L$:
\begin{align}
    \textup{MSE} = \dfrac{1}{L}\sum_{n=1}^L \Vert \by_n - \hat{\by}_n \Vert^2_2,
\end{align}
where $\Vert \cdot \Vert_2$ denotes the Euclidean norm between two vectors in $\bbR^d$.
For optimal training, a constant bias term $x_{n,K+1}=1$ is added to the reservoir state $\bx_n$.
A conventional approach is to optimize $\bw$ via the linear regression $\hat{\bY} = \bX\bw$, 
where $\hat{\bY}=[\hat{\by}_1\quad\ldots\quad \hat{\by}_L]^\top$ is the $L\times d$ target matrix and
$\bX$ is the $L\times (K+1)$ matrix that combines reservoir states $\bx_1, \bx_2, \ldots, \bx_L$ of training data.
The optimal value of $\bw$ is obtained via the Ridge regression in the matrix form $\boldsymbol{\hat{w}^\top} = (\bX^\top\bX + \eta\bI)^{-1}\bX^\top\hat{\bY}$.
Here, $\eta$ is a positive constant shifting the diagonals introduced to avoid the problem of the near-singular moment matrix.
The trained parameter $\hat{\bw}$ is used to generate outputs in the situation where we do not know about the target sequence and can only access the input sequence.

The information processing via the input-driven map is similar in non-temporal learning tasks, but the target in non-temporal ones is not the sequence but a specific value such as a label or a output vector corresponding with given input data.
The input data are firstly converted into a sequence to be fed into the reservoir, then the reservoir states obtained from the reservoir dynamics are considered nonlinear features to learn the mapping to the output.
The learning task can be regression or classification with respect to the property of output set as dense or discrete, respectively.
In this way, the loss function measuring the prediction error can be designed in a flexible way in which we can apply more complicated training algorithms such as support vector machine or heuristic search methods to nonlinear optimization problems.

\subsection{Quantum Reservoir Computing}
Quantum reservoir computing (QRC) is one of proposals to exploit the quantum dynamics as a reservoir. 
Here, a disordered ensemble quantum dynamics system is used as a computational resource with the possibility of processing with an exponentially large number of degrees of freedom.
The crucial idea is using a quantum system for obtaining the input-driven map defined in Eq.~\eqref{eqn:driven-map}.
This input-driven map depends on the dynamical evolution of the quantum system and the quantum measurements to extract reservoir states from the system.
For example, the reservoir can be implemented by a set of interacting qubits in the NMR system~\cite{fujii:2017:qrc,nakajima:2019:qrc}, 
a set of fermions~\cite{ghosh:2019:quantum,ghosh:2020:reconstruct,ghosh:2019:neuromorphic},
a set of interacting quantum harmonic oscillators~\cite{nokkala:2020:gaussian},
or even a single nonlinear oscillator driven by a Hamiltonian dynamics~\cite{govia:2020:quantum}.
The proof-of-principle experimental demonstrations for QRC are ongoing in the research with promising proposals for the NMR platform~\cite{negoro:2021:rcbook:NMR} and NISQ quantum computers~\cite{chen:2020:temporal,dasgupta:2020:designing}.

In our study, we model the QRC approach via the framework of repeated quantum interactions, which can be considered a general approach using the NMR system in Ref.~\cite{fujii:2017:qrc}. 
In the repeated quantum interactions, input sequence is fed via the sequence interactions between reservoir system (called a \textit{reduced quantum reservoir}) $\cS$ with a auxiliary system $\cE$. 
The dynamical evolution is modeled by the following recurrent relation (Eq.~2 in our main text)
\begin{align}
    \rho_n = \cL_{\beta_n}(\rho_{n-1}) = \tr_{\cE}[U(\rho_{n-1}\otimes \beta_n)U^{\dagger}],\label{eqn:recurr}
\end{align}
where $\rho_n$, $\beta_n$ are the states of $\cS$ and $\cE$ for the $n$th interaction, respectively.
Here, $U$ is an unitary evolution and $\cL_{\beta_n}$ is called the \textit{reduced dynamics map}, which is a completely positive and trace-preserving (CPTP) map acting on the space $\cM_{\cS}$ of density matrices in $\cS$.

One can ask where the mechanism to obtain the nonlinearity is, since even if the evolution is not unitary we still get linear dynamics.
In fact, it has been shown that introducing a coherent nonlinearity to quantum evolution means that we can obtain the ability to solve NP-hard problems~\cite{abram:1993:quantum}.
Therefore, a more likely way to introduce nonlinearities is to use projective measurements, which are clearly nonlinear operations using the projectors on the space spanned by the basis of the system used in a measurement.
In this way, if a local observable $O$ is described by a collection of projectors $\{P_j\}$ as 
$
    O = \sum_j a_jP_j,
$
where $a_j$ are the values of a physical quantity of a measurement associated with $P_j$,
the probability to obtain the measurement result $a_j$ is given by $p(a_j) = \tr[P_j\rho]$.
Then, the expectation value of the physical quantity is given by 
\begin{align}
    \sum_ja_jp(a_j) = \sum_ja_j\tr[P_j\rho] = \tr[O\rho] = \langle O\rangle_{\rho}.
\end{align}
We obtain partial information regarding the state $\rho_n$ of the quantum system $\cS$ after the $n$th interaction with the environment by measuring local observables $O_1, O_2, \ldots,O_K$ on the state of $\cS$.
Then, the $k$th element $x_{nk}$ of the reservoir states $\bx_n$ can be calculated as the expectation of the measurement results via $O_k$ on $\rho_n$:
\begin{align}\label{eqn:obs}
    x_{nk} = \tr[O_k\rho_n] = \langle O_k\rangle_{\rho_n}.
\end{align}
One can increase the dimension of $\bx_n$ by performing measurements at multiple times.
Between two inputs, $M$ cycles of the unitary evolution are processed and each of them is followed by a measurement.
$M$ is called the measurement multiplexity. Thus we obtain $MK$ elements in each reservoir state.
After obtaining the reservoir states, the training procedure in QRC is similar to the classical RC.

We note that we can also construct a universal quantum reservoir consisting of multiple and non-interacting quantum reservoirs $\cS$. For example, let us consider a quantum system $\cS$ consisting of two non-interacting reduced quantum reservoirs $\cS_a$ and $\cS_b$ with the corresponding reduced dynamics maps $\cL_a$ and $\cL_b$, respectively. Since $\cS_a$ and $\cS_b$ are not interacting, a quantum state $\rho$ in $\cS$ can be displayed as a tensor product 
\begin{align}
\rho  = \rho_a \otimes \rho_b,
\end{align}
where  $\rho_a$ and $\rho_b$ are two quantum states in $S_a$ and $S_b$, respectively. Then, the reduced dynamics map $\cL$ acting on the space $\cM_S$ can be defined as 
\begin{align}
    \rho = \rho_a \otimes \rho_b \to \cL(\rho) = \cL_a(\rho_a) \otimes \cL_b(\rho_b).
\end{align}

\subsection{Standard Quantum Process Tomography}

Before going into the details of learning the temporal tomography, we explain standard quantum process tomography (standard QPT). The following explanation is referenced from Refs.~\cite{nielsen:2011:QCQI,mohseni:2008:qpt}.

Suppose $\Omega(\cdot)$ is a time-independent quantum channel, which is a completely-positive linear and trace-preserving map from a $D_A$-dimensional Hilbert space $\cH_A$ to a $D_B$-dimensional Hilbert space $\cH_B$ (for notational simplicity, we take $\cH_A \equiv \cH_B \equiv \cH$ and $D_A=D_B=D$). 
In principle, $\Omega(\cdot)$ can be reconstructed from the measurement data using a proper inversion, since $\Omega(\cdot)$ is a linear map.
A time-independent $\Omega(\cdot)$ can be described using a set of $D^2$ time-independent operators $\{E_k\}_{k=1}^{D^2}$ such that $\sum_{k=1}^{D^2}E_k^{\dagger}E_k= I$ and 
\begin{align}\label{seqn:operator:qpt}
\Omega(\rho)=\sum_{k=1}^{D^2}E_k\rho E_k^{\dagger}
\end{align}
for an arbitrary $D\times D$-dimensional density matrix $\rho$.
To determine $E_k$ from measurement data, it is convenient to choose a description using a basis set of operators $\{\tE_k\}_{k=1}^{D^2}$ for the space of $D\times D$-dimensional linear operators.
In this case, $E_k = \sum_{m=1}^{D^2} e_{km}\tE_k$ for $D^2$ complex numbers $e_{km}$. Then, Eq.~\eqref{seqn:operator:qpt} becomes
\begin{align}\label{seqn:operator:kai}
    \Omega(\rho) = \sum_{mn} \tE_m\rho \tE_m^{\dagger} \chi_{mn},
\end{align}
where $\chi_{mn}=\sum_k e_{km}e^{*}_{kn}$.

If the basis set of operators $\{\tE_k\}$ is fixed, then the quantum channel $\Omega$ can be described by a $D^4$-dimensional complex number vector $\boldsymbol{\chi} = (\chi_{mn})$ (called the superoperator). However, due to the completeness relation $\sum_{k=1}^{D^2}E_k^{\dagger}E_k= I$, we only need $D^4 - D^2$ independent parameters to describe $\boldsymbol{\chi}$.
The idea determining $\boldsymbol{\chi}$ is to prepare $D^2$ linearly independent inputs (as the basis set of operators for the space of $D\times D$ operators) $\{\rho_k\}_{k=1}^{D^2}$ and perform state tomography for the outputs states $\Omega(\rho_k)$.
In this case, we often choose $\tE_k = \rho_k$ since $\rho_k$ are Hermitian operators and valid observables.

Considering an orthonormal basis $\{\ket{m}\}_{m=0}^{D-1}$ of $\cH$, a convenience choice for input states is $\rho_k = \ket{m}\bra{n}$. 
Given $\ket{+}=\frac{\ket{m} + \ket{n}}{\sqrt{2}}$ and $\ket{-}=\frac{\ket{m} + i \ket{n}}{\sqrt{2}}$, we can express the coherence $\ket{m}\bra{n}$ as 
\begin{align}
    \ket{m}\bra{n} = \ket{+}\bra{+} + \ket{-}\bra{-} - \dfrac{(1+i)}{2}\left(\ket{m}\bra{m} +  \ket{n}\bra{n}\right).
\end{align}
Therefore, from the experimental measurements of $\Omega(\ket{m}\bra{m})$, $ \Omega(\ket{n}\bra{n})$, and $\Omega(\ket{+}\bra{+})$ and $\Omega(\ket{-}\bra{-})$, we can determine $\Omega(\ket{m}\bra{n})$ due to the linearity of $\Omega$.

We note that every $\Omega(\rho_k)$ can be expressed as a linear combination of the basis states, as 
\begin{align}\label{sqn:map:basis}
\Omega(\rho_k)=\sum_l \lambda_{kl}\rho_l.
\end{align}
Here, parameters $\lambda_{kl}$ contain the measurement results and can be calculated from state tomography experiments as $\lambda_{kl} = \tr[\tE_k\Omega(\rho_l)]$ (since we choose $\tE_k = \rho_k$).
Generally, we can write 
\begin{align}\label{sqn:map:B}
\tE_m\rho_k\tE_n^{\dagger}=\sum_l B_{mn,lk}\rho_l,
\end{align}
where the $D^4\times D^4$-dimensional matrix $\boldsymbol{B}=(B_{mn,lk})$ is determined by the fixed choice of bases $\{\rho_k\}$ and $\{\tE_k\}$.
Combining Eq.~\eqref{seqn:operator:kai}, Eq.~\eqref{sqn:map:basis}, and Eq.~\eqref{sqn:map:B}, we have
\begin{align}
    \sum_l \sum_{mn} \chi_{mn}B_{mn,lk}\rho_l = \sum_l \lambda_{kl}\rho_l.
\end{align}
From the linear independence of the $\rho_l$, it follows that, for each $k$, we have $\sum_{mn}B_{mn,lk}\chi_{mn}=\lambda_{kl}$, or in the matrix form $\boldsymbol{B}\boldsymbol{\chi}=\boldsymbol{\lambda}$, where $\boldsymbol{\lambda}$ is the $D^4$-dimensional vector created from $\{\lambda_{kl}\}$.
Therefore, the superoperator $\boldsymbol{\chi}$ can be described by a proper linear conversion (since we know $\boldsymbol{B}$ and $\boldsymbol{\lambda}$).

In the standard QPT explained above, we must prepare $D^2$ independent inputs $\{\rho_k\}$ and perform quantum-state tomography for the corresponding outputs $\{\Omega(\rho_k)\}$. 
For each $\rho_k$, we must measure the expectation value of the $D^2$ fixed-basis operators $\{\tE_k\}$ in the state $\Omega(\rho_k)$ that requires a number copies of $\Omega(\rho_k)$ to obtain a fixed precision for $\boldsymbol{\chi}$. Therefore, the total number of required measurements for standard QPT is $O(D^4)$. 
Here, we use the term measurement with the implicit meaning of the measurement on an ensemble of identically prepared quantum systems corresponding to a given experimental setting.
Furthermore,  to solve the linear system $\boldsymbol{B}\boldsymbol{\chi}=\boldsymbol{\lambda}$, we need a huge classical resource to do the inversion of the $D^4\times D^4$-dimensional matrix $\boldsymbol{B}$.

\subsection{Learning the Temporal Tomography}
A significant different in the applications of QRC approaches compared with the conventional RC is the ability to deal with quantum tasks such as the classification of quantum states as entangled or separable~\cite{ghosh:2019:quantum},
the quantum tomography~\cite{ghosh:2020:reconstruct}, and the quantum state preparation~\cite{ghosh:2019:neuromorphic}.
While some of the QRC approaches are not proposed for quantum tasks~\cite{fujii:2017:qrc,govia:2020:quantum,nokkala:2020:gaussian}, the current proposals of QRC for quantum tasks focused on a static quantum state, which is coupled to the reservoir~\cite{ghosh:2019:quantum,ghosh:2020:reconstruct}.
Here, our approach based on repeated quantum interactions provides the ability to perform quantum tasks for a sequence of input quantum states where the outputs of these tasks depend on the past input states.

We explain the setting for the temporal quantum map $\cF$ to see the basic difference with the standard QPT. 
Given a sequence of input quantum states $\beta_1, \beta_2, \ldots$, 
a quantum device processes this input sequence via a temporal map $\cF$, where the corresponding output states are $\cF(\beta_1), \cF(\beta_2), \ldots$.
We assume that the output $\cF(\beta_n)$ only depends on a finite input history.
We can think that the action of $\cF$ to the input sequence is equivalent to a function of current and past inputs, which is often seen in classical time-series processing.
The density matrix of $\cF(\beta_n)$ can be reconstructed if we can model this action to an accessible input sequence, that is to model the dependency of the current output to the current and past inputs.
This is different with the standard QPT, which requires an accessible resource with $D^2$ independent input states and the corresponding $D^2$ output states to find the general form of the linear map.

In addition, Fig.~1(a) in our main text helps the readers capture the visual concept of the quantum map $\cF$. 
Here, we consider a stream of unknown quantum channels $\{\Omega_n\}$ applied to the input stream where there is a dependency between these channels. We note that $\cF$ is generally not a standard quantum channel, but the output of $\cF$ at each time in the output stream is considered as a coherent superposition or a convex mixture of channels at different times. For example, the latter form can be expressed as
\begin{align}\label{seqn:temporal:qpt}
    \cF(\beta_n) = \dfrac{1}{Z}\sum_{i=0}^d\eta_i\Omega_{n-i}(\beta_{n-i}),
\end{align}
where $\eta_i$ are unknown non-negative real numbers with $Z=\sum_{i=0}^d\eta_i$ to preserve the trace.
This concept of $\cF$ envisions a typical case in the future realization of quantum communication and quantum internet where  quantum data can be transmitted via time-dependent quantum channels under a time-delayed effect.
The density matrix of $\cF(\beta_n)$ can only be reconstructed if we can model the action of $\cF$ to the input sequence.

In learning the tomography of the temporal quantum map $\cF$, we are given an input sequence of states $\{\beta_1,\ldots,\beta_L\}$ and the corresponding target sequence $\hat{\by}=\{\hat{\by}_1,\ldots,\hat{\by}_L\}$, where $\hat{\by}_k$ is the real vector form to stack the real and imaginary elements of $\cF(\beta_k)$.
In the evaluation stage, we are given an input sequence $\{\beta_{L+1}, \ldots, \beta_{L+T}\}$ with the corresponding target $\{\hat{\sigma}_{L+1},\ldots,\hat{\sigma}_{L+T}\}$, where $\hat{\sigma}_i = \cF(\beta_i)$.
The reconstructed output sequence is $\{{\by}_{L+1}, \ldots, {\by}_{L+T}\}$, which is rearranged in the matrix form $\{{\sigma}_{L+1}, \ldots, {\sigma}_{L+T}\}$. Due to statistical fluctuations, there are some cases in which the reconstructed matrix $\sigma_i$ is not positive semidefinite. We project $\sigma_i$ onto the spectrahedron to obtain a positive semidefinite matrix $\sigma^{\prime}_i$ such that the trace of $\sigma^{\prime}_i$ is equal to 1 and the Frobenius norm between $\sigma_i$ and $\sigma^{\prime}_i$ is minimized~\cite{chen:2011:projection}.
This technique is considered as \textit{projected pseudo-inversion estimator}. Other proposals for a density matrix estimator can be found in a survey in Ref.~\cite{bantysh:2020:quantum}.

Regarding the learning mechanism, we assume that we have full tomography for the corresponding output states of $\cF$ in the learning and evaluation phase. 
Note that the evaluation phase is performed for evaluating the appropriate values of model parameters but is not needed after the model parameters are fixed.
The condition for accessible full tomography in the training phase is common and unavoidable in implementing supervised learning methods in quantum tomography, for example, in Refs.~\cite{ghosh:2020:reconstruct,quek:2021:adaptive}.
However, this does not defeat the potential advantage of our proposal. 
After the training procedure, we are able to reconstruct the output $\cF(\beta_n)$ in the sequence by only using a single process of measurements in the reservoir. This is performed without reconfiguring the experimental setup or doing full tomography with a number of identical copies of $\cF(\beta_n)$ for every $n$. 
Our framework is general and can be applied for different types of the temporal quantum map $\cF$ without reconfiguring the measurement bases. Those points are significant contributions of our proposal.


Regarding the concept of function $\cF$ in the existing experimental setups, we envision the situation in quantum time-series processing, where a quantum device may output a sequence of quantum states in a time-dependent manner.
For example, during the continuous processing of quantum states, the quantum device may generate temporal and input-dependent noise such as quantum noise in temporal fluctuations, which may have temporal effects on the quantum states~\cite{gehrig:2002:temporal}.
Recent experimental studies have shown that decoherence effects experienced by superconducting qubits are characterized by the amplitude damping time ($T_1$) and the dephasing time ($T_2$), which exhibit a fluctuations behavior~\cite{martinez:2021:temporal}. 
These fluctuations can be included in time-varying quantum channels, which provide a more realistic depiction of decoherence effects than static models in some cases.
Furthermore, in quantum optics, the optical quantum states defined in temporal modes, especially in non-Gaussian states, can be manipulated in a time-dependent manner using nonlinear optical processes such as three-wave mixing and quantum optical memories~\cite{brecht:2015:photon,raymer:2020:temporal}.
We can think of scattering processes where the quantum device may be optical. Then, sets of initial input mode operators evolve to output mode operators through the interaction with a nonlinear-optical medium driven by one or more external pumps.
If these external pumps are controlled in a time-dependent manner or depend on the property of the input mode operators, the quantum device will stand for a temporal quantum map.

Another example of a temporal quantum map is the so-called one-atom maser or micromaser device~\cite{meschede:1985:prl:maser,varcoe:2000:nat:photon}.
Here, excited atoms interact with the photon field inside a high-quality optical cavity.
In the situation where the photons suffer a long lifetime, the output of this device can be modeled as a function of the current input atom and the preceding atoms. This situation corresponds to the illustration in Fig.~1 of our manuscript.
The effect of a sequence of quantum input to the output of a quantum device has also been demonstrated experimentally for optical fiber channels with fluctuating birefringence~\cite{ball:2004:pra:corr,banaszek:2004:prl:corr}. In these experiments, consecutive light pulses undergo strongly correlated polarization transformation that makes the temporal quantum map.
We consider the above examples as explicit connections of our proposal with experimental models that capture the fundamental nature of quantum devices. Performing tomography for such devices differs from standard QPT, because the memory effects need to be taken into account.

\section{Convergence analysis}
Since $\cF$ can be considered as a function of past sequences of input quantum states, the framework to learn $\cF$ should have the ability to retain information about recent inputs in their activity.
This property in the classical context is known as the fading memory property, mentioned by S. Boyd and L. Chua~\cite{boyd:1985:fading}.
The reservoir needs to forget its initial state to ensure the fading memory property.
If the effect of the initial state in the reservoir is strong, the response of the reservoir to the same input sequence may be different if the reservoir starts from two different initial states.
This effect may result in the loss of reproducibility in temporal processing tasks.
Therefore, a quantum system with a reproducible temporal information processing ability must produce the trajectories that are robust to small perturbations to the system,
i.e., the computations for the same input sequence are independent of its initial state.
We refer to this as quantum echo state property (QESP), which is defined in Ref.~\cite{tran:2020:higherorder} and mentioned in the convergence property of dissipative quantum systems in Ref.~\cite{chen:2019:dissipative}.

\begin{definition}\label{def:qesp}
A quantum system used in temporal information processing tasks is said to satisfy the quantum echo sate property (QESP) with respect to the distance $\mathcal{D}$ between quantum states if for any input sequence of length $L$, it holds that $\mathcal{D}(\rho^{(1)}_L, \rho^{(2)}_L) \to 0$ as $L\to \infty$, where $\rho^{(1)}_L, \rho^{(2)}_L$ are the states after $L$ steps corresponding with different initial states $\rho^{(1)}_0, \rho^{(2)}_0$.
\end{definition}
In the following parts, we consider $\mathcal{D}$ as the trace distance, or the Schattern $1$-norm, defined by $\mathcal{D}(\rho, \rho^{\prime})=\Vert \rho - \rho^{\prime} \Vert = \tr[|\rho - \rho^{\prime}|]$ between quantum states $\rho$ and $\rho^{\prime}$, where we define $|A|=\sqrt{A^\dagger A}$ to be the positive square root of $A^\dagger A$ for any matrix $A$.
Since a CPTP map is a contracting map w.r.t. the trace distance, the density matrices in quantum repeated interactions satisfy decreasing system distinguishability~\cite{nielsen:2011:QCQI} as
\begin{align}
\Vert \rho^{(1)}_n, \rho^{(2)}_n \Vert \leq \Vert\rho^{(1)}_{n-1}, \rho^{(2)}_{n-1}\Vert, 
\end{align}
where $\rho^{(1)}_n, \rho^{(2)}_n$ are the states after $n$ interactions corresponding with different initial states $\rho^{(1)}_0, \rho^{(2)}_0$.
We denote $\Phi_n = \cL_{\beta_n}\cL_{\beta_{n-1}} \ldots \cL_1$, then $\Phi_n$ is a CPTP map
and $\rho_n = \Phi_n(\rho_0)$.
We are interested in the action of $\Phi_n$ to study the dynamics of $\cS$, particularly in asymptotic behavior in the limit $n\to \infty$.

Since the reduced dynamics maps $\cL_{\beta_n}$ do not have in general a common invariant state, we firstly investigate the behavior of the reservoir at ergodic limits. We use here the results developed in Ref.~\cite{bruneau:2010:infinite} (Theorem 1.3) and Ref.~\cite{nechita:2010:random} (Theorem 5.2, Proposition 6.3) to get the following result.

\begin{result}[Ergodicity of the reduced quantum reservoir]
Let $\{\beta_n\}$ be a sequence of i.i.d. random quantum states under the main hypothesis that the induced CPTP maps $\cL_{\beta_n}$ have a unique invariant state. 
Then, for all initial states $\rho_0$, the reduced quantum reservoir $S$ satisfies the ergodic limit
\begin{align}
    \lim_{N\to\infty} \dfrac{1}{N}\sum_{n=1}^N\rho_n = \lim_{N\to\infty} \dfrac{1}{N} \Phi_n(\rho_0)= \rhes,
\end{align}
where $\rhes \in \cM_{\cS}$ is the unique invariant state of the deterministic map $\cL_{\mathbb{E}[\beta]}$.
\end{result}

Next, motivated by the theory of ergodic quantum processes in Refs.~\cite{movassagh:2019:ergodic,movassagh:2020:theory}, we derive the following theorem as a sufficient condition for the QESP of a QR.

\begin{result}\label{theorem:qesp}
The quantum reservoir $\cS$ satisfies the quantum echo state property if the input sequence $\{\beta_n\}$ is a random ergodic sequence under the main hypothesis that for some $n$, $\Phi_n$ is a strictly positive CPTP map with a positive probability. Here, a CPTP map $\cL: \cM_{\cS} \to \cM_{\cS}$ is called strictly positive if it sends every positive semidefinite matrix in $\cM_{\cS}$ to a positive definite matrix.
\end{result}

\textit{Proof.---}
Since $\{\beta_n\}$ is an ergodic sequence the sequence of $\cL_{\beta_n}$ is also ergodic.
We drop the notion $\beta$ for an easy description of the reduced dynamics maps $\cL_{\beta_n}$ as $\cL_n$.
We assume that the sequence of $\cL_n$ is drawn from an ensemble $\Omega$.
Consider the shift map $\cT: \Omega \to \Omega$ such that $\cT(\cL_1, \cL_2, \ldots) = (\cL_2, \cL_3,\ldots)$. 
Due to the ergodic property, $\cT$ is ergodic map, that is, the dynamics generated by $\cT$ starting from a typical sequence will cover $\Omega$ with full measure.
We can write the composition form as 
\begin{align}
    \cL_{m+n}\ldots\cL_{m+1} = \cT^{m}\circ \Phi_n,
\end{align}
for any $n, m \geq 1$.

For a CPTP map $\cL:\cM_{\cS}\to \cM_{\cS}$, let $c(\cL)$ be the smallest number such that $\Vert \cL(\rho) - \cL(\sigma) \Vert \leq c(\cL) \Vert \rho - \sigma \Vert$ for any $\rho, \sigma \in \cM_{\cS}$.
Because $\cL$ is contracting w.r.t. $\Vert \cdot \Vert$, such $c(\cL)$ (called the \textit{contraction coefficient} of $\cL$) always exists and $0 < c(\cL) \leq 1$.
We denote $c_n = c(\Phi_n)$ and have the following inequalities:
\begin{align}
    c(\Phi_{m+n}) &\leq c(\Phi_{m}) c(\cL_{m+n}\ldots\cL_{m+1})\\
    \ln(c(\Phi_{m+n})) &\leq \ln(c(\Phi_{m})) + \ln(c(\cL_{m+n}\ldots\cL_{m+1})) \\
    \ln(c(\Phi_{m+n})) &\leq \ln(c(\Phi_{m})) + \ln(c(\cT^{m}\circ \Phi_n))
\end{align}
As $\ln(c(\Phi_{n})) \leq 0$, the subadditive ergodic theorem guarantees~\cite{steele:1989:subadd} that the following limit exists:
\begin{align}\label{eqn:lim}
    \lim_{n\to \infty} \dfrac{1}{n}\ln(c(\Phi_{n})) = \kappa,
\end{align}
where $\kappa \leq 0$ is given by the limit of expectations over $\Omega$ (Ref.~\cite{hennion:1997:product}, Lemma 3.2)
\begin{align}
    \kappa = \lim_{n\to \infty} \dfrac{1}{n}\mathbb{E}[\ln(c(\Phi_{n}))] = \inf_{n}\dfrac{1}{n}\mathbb{E}[\ln(c(\Phi_{n}))].
\end{align}

By the hypothesis of the strictly positive map, for some $n$ we have $c(\Phi_{n}) < 1$ with a positive probability. Therefore, there exists $n\geq 1$ such that $-\infty \leq \dfrac{1}{n}\mathbb{E}[\ln(c(\Phi_{n}))] < 0$, thus $\kappa < 0$.
From Eq.~\eqref{eqn:lim}, there exists a positive $\mu < 1$, a constant $A > 0$, and $n_ 0 \geq 1$ such that $c(\Phi_{n}) \leq A\mu^n$ for all $n\geq n_0$.
Then, 
\begin{align}\label{eqn:ineq:qesp}
    \Vert \rho^{(1)}_n &- \rho^{(2)}_n \Vert  = \Vert \Phi_n(\rho^{(1)}_0) - \Phi_n(\rho^{(2)}_0)  \leq c(\Phi_{n}) \Vert \rho^{(1)}_0, \rho^{(2)}_0 \Vert \leq A\mu^n\Vert \rho^{(1)}_0, \rho^{(2)}_0 \Vert
\end{align}
for all $n\geq n_0$, thus $\Vert \rho^{(1)}_n, \rho^{(2)}_n \Vert \to 0$ as $n\to \infty$.
Therefore, we obtain the QESP. Furthermore, if the system starts from two different states, then the trace distance between the states will decay exponentially to zero. 

In some practical situations, result~\ref{theorem:qesp} may be too restrictive to be used.
We can relax the definition~\ref{def:qesp} as the following definition.
\begin{definition}
Given a positive $\varepsilon$, a quantum reservoir is said to satisfy the $\varepsilon$-QESP if for any input sequence there exists a smallest number $T(\varepsilon)$ such that $\Vert \Phi_n(\rho^{(1)}_0) - \Phi_n(\rho^{(2)}_0) \Vert < \varepsilon$ for all $n \geq T(\varepsilon)$ and all different initial states $\rho^{(1)}_0$, $\rho^{(2)}_0$.
The maximum $T(\varepsilon)$ for all input sequences is defined as the $\varepsilon$-QESP time scale of the quantum reservoir.
\end{definition}

The $\varepsilon$-QESP time scale indicates the number of time steps that the quantum reservoir must wait to forget its initial state.
If we can evaluate this quantity, we can know when to start the learning process after a number of transient time steps.
We demonstrate that the $\varepsilon$-QESP and the corresponding time scale can be evaluated via the spectrum of the reduced dynamics maps $\cL_n$.
First, we consider the case of the constant input $\beta_n=\beta$ for all $n$.
We have $\cL_n\equiv \mathcal{L}$ for all $n$ and $\rho_n=\mathcal{L}^{n}(\rho_0)$.
Any density matrix can be converted into the vector form, which allows us to define a linear space of matrices (Fock-Liouville space) associated with an inner product.
Consider an arbitrary density matrix 
$
\rho=\sum_{i,j}\rho_{i,j}\ket{i}\bra{j}.
$
We can write its vectorized form using the Choi-Jamiolkowski isomorphism as
$
    \vc(\rho)=\sum_{i,j}\rho_{i,j}\ket{j}\otimes\ket{i},
$
which is the same as stacking columns of $\rho$.
The reduced dynamics map $\cL$ can now be expressed as a matrix $\tilde{\cL}$, such that $\vc(\rho_n)=\tilde{\cL}\vc(\rho_{n-1})$.
The spectrum of $\cL$ can be written as $1=|\lambda_1|\geq |\lambda_2| \geq \ldots \geq |\lambda_s|$, where $\lambda_j$ is the $j$th eigenvalue of $\tilde{\cL}$. Because $\cL$ can be a non-Hermitian map, it can have both left and right eigenmatrices. We denote $L_j$ and $R_j$ as the left and right eigenmatrices corresponding to eigenvalue $\lambda_j$, respectively.
We choose the normalization such that $\tr(L_kR_l) = \delta_{kl}$ for all $1\leq k,l \leq s$,
and $\tr[R_1]=1$ (i.e., $L_1$ being the identity $L_1=I$).
Since $\vc(\rho_n) = \tilde{\cL}^n(\vc(\rho_0))$, we have
\begin{align}\label{eqn:representation}
    \vc(\rho_n) = \vc(R_1) + \sum_{j=2}^sb_j\lambda_j^n\vc(R_j),
\end{align}
where $b_j = \vc(L_j)^{\dagger}\vc(\rho_0)=\tr[L_j\rho_0]$.
Therefore, if $|\lambda_2|^{-1} > 1$, the steady state $\rhss$ of the system is unique ($\rhss=R_1$) with the convergence rate depending on the magnitude of $|\lambda_2|^{-1}$, i.e., $\Vert \rho_n - \rhss \Vert \sim e^{-n/N}$ with $N \sim \frac{1}{\ln(|\lambda_2|^{-1})}$.
The QESP is satisfied that we can evaluate the $\varepsilon$-QESP time scale as $O(\frac{\ln(\varepsilon^{-1})}{\ln(|\lambda_2|^{-1})})$.

Next, we consider the case when the input is a sequence of i.i.d. density matrices.
The result~\ref{theorem:qesp} provides us the following proposition, which is stated in our main text.
\begin{proposition}\label{prop:random}
If there exist $n_0$ such that $\mu = \max(c_n) < 1$ for all $n \geq n_0$, where $0 < c_n \leq 1$ is the  contraction coefficient of $\cL_{\beta_n}$, then the $\varepsilon$-QESP time scale can be evaluated as $O(\frac{\ln(\varepsilon^{-1})}{\ln(\mu^{-1})})$.
\end{proposition}
\textit{Proof.---} 
In the proof of result~\ref{theorem:qesp}, for a CPTP map $\cL:\cM_{\cS}\to \cM_{\cS}$, the contraction coefficient $c(\cL)$ denotes the smallest number such that $\Vert \cL(\rho) - \cL(\sigma) \Vert \leq c(\cL) \Vert \rho - \sigma \Vert$ for any $\rho, \sigma \in \cM_{\cS}$.
Such $c(\cL)$ always exists and $0 < c(\cL) \leq 1$.
Since  $\mu = \max(c_n) < 1$ for all $n\geq n_0$, we have $c(\Phi_n) \leq A\mu^n < 1$ for some constant $A$ for all $n \geq n_1 \geq n_0$ for some $n_1$. 
Equation~\eqref{eqn:ineq:qesp} shows us the result in the proposition.

Proposition~\ref{prop:random} gives us a consequence that to obtain a sufficient check for the QESP and to know the $\varepsilon$-QESP time scale, we only need to investigate the contracting coefficient via the spectrum of the reduced dynamics map for random inputs.

\begin{figure}
		\includegraphics[width=14cm]{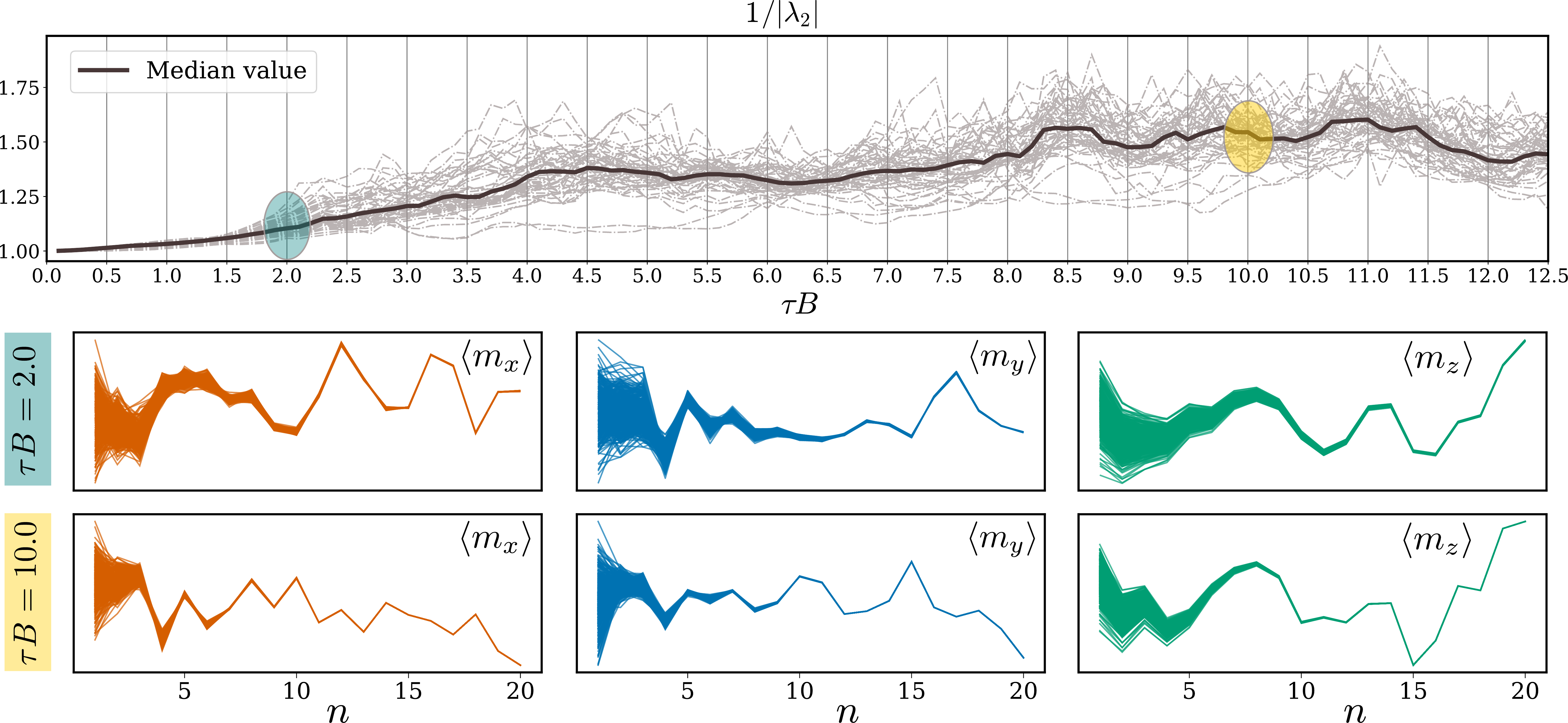}
		\protect\caption{(Top) The values of $1/|\lambda_2|$ for 100 random reduced dynamics maps $\cL_{\beta}$ with $N_m=4, N_e=2, \alpha=1, J=1,$ and $J/B=1$. The solid line depicts the median value of the ensemble of dynamics maps. (Middle and bottom) The trajectories of the average magnetization generated with 500 random initial states of $\cS$ at $\tau B = 2.0$ and $\tau B = 10.0$.
		\label{fig:dynamics:spec}}
\end{figure}

To numerically investigate the transient dynamics, we consider a specific system modeled by the transverse field Ising model, where the Hamiltonian is given by 
\begin{align}\label{eqn:ising}
    H = -\sum_{i> j=1}^NJ_{i,j}\hsig^x_i\hsig^x_j - B\sum_j^N\hsig^z_j.
\end{align}
Here, $B$ is the natural frequency of $1/2$-spins represented for qubits. 
$\hsig_j^\gamma$ $(\gamma \in \{x, y, z\})$ are the Pauli operators measuring the qubit $j$ along the $\gamma$ direction,
which can be described as an $N$-tensor product of $2\times 2$-matrices as
\begin{align}
\hsig_j^\gamma = \bI\otimes\ldots\otimes\underbrace{\hsig^{\gamma}}_\text{$j$-index}\otimes\ldots\otimes\bI,
\end{align}
where
$
\bI = \left[\begin{array}{cc}1 & 0\\0 & 1\end{array}\right]$,
$
\hsig^x = \left[\begin{array}{cc}0 & 1\\1 & 0\end{array}\right]$,
$
\hsig^y = \left[\begin{array}{cc}0 & -i\\i & 0\end{array}\right]$,
and
$
\hsig^z = \left[\begin{array}{cc}1 & 0\\0 & -1\end{array}\right].
$
The coupling coefficients $J_{ij}$ between spins can be randomly selected or fixed depending on the distance of interaction.
To describe our QR, we assume that the auxiliary system $\cE$ includes the first $N_e$ spins
where the remaining $N_m=N-N_e$ spins form the reservoir $\cS$. 
We present the setting of \textit{power-law decaying} for $J_{ij}$ and frequency $B$ corresponding with the physical implementation in a trapped-ion quantum simulation~\cite{porras:2004:trapped,kim:2009:trapped,jurcevic:2014:trapped}.
Here, $J_{ij}=J|i-j|^{-\alpha}/N(\alpha)$ with an interaction strength $J$, power coefficient $0 < \alpha < 3$, and normalization constant 
$
N(\alpha) = \dfrac{1}{N-1} \sum_{i>j}|i-j|^{-\alpha}.
$
The coupled system $(\cS, \cE)$ is a closed system with the total Hamiltonian $H$ unchanged during each interaction.
We consider the model parameters $\alpha=1.0, J=B=1.0$, and the unitary $U=\exp(-i\tau H)$, where $\tau$ is the interaction time.

The top panel of Fig.~\ref{fig:dynamics:spec} shows the values of $1/|\lambda_2|$ according to the normalized interaction time $\tau B$ for 100 random reduced dynamics maps $\cL_{\beta}(\rho)=\tr_{\cE}[U(\rho\otimes\beta)U^{\dagger}]$ with $N_m=4, N_e=2$.
Here, $\beta$ is drawn from an ensemble of Haar random pure states of dimension $2^{N_e}$. 
The QESP is satisfied in values of $\tau B$ such as all values of $1/|\lambda_2|$ are higher than 1. 
Furthermore, the higher $1/|\lambda_2|$  leads to the smaller $\varepsilon$-QESP time scale.
We then fix a random input sequence and investigate the trajectories of the average magnetization $\langle m_{\gamma}\rangle = \tr[\hat{m}_{\gamma}\rho_n]$, where $\hat{m}_{\gamma}=\frac{1}{N_m}\sum_j\hsig^{\gamma}_j$ are the average spin operators at $\gamma \in \{x, y, z\}$ directions in $\cS$.
The trajectories begin with 500 random initial states of $\cS$. The fluctuations reduce as the number $n$ of interactions increases. This reduction is faster as the $\varepsilon$-QESP scale is smaller, as illustrated in Fig.~\ref{fig:dynamics:spec} with $\tau B = 2.0$ and $\tau B = 10.0$.

\section{Metastability analysis}
In this section, we derive the metastability of the reservoir dynamics.
Metastability is a characteristic feature of the dynamics of a slow relaxing system with the partial relaxation into long-lived states before eventual decay to the true stationary state.
Metastability appears due to the separation of time scales in the dynamics; therefore, different initial states of the system will relax to different metastable states in the transient dynamics before eventual relaxation to the true stationary state.
The metastability in open quantum systems has been studied in Ref.~\cite{macieszczak:2016:meta}. Here, we formulate the metastability in our quantum reservoir framework.

First, we consider the case of constant input $\beta_n=\beta$ for all $n$.
The reduced dynamics map is represented by a fixed CPTP map $\cL$ with the spectrum $1=|\lambda_1|\geq |\lambda_2| \geq \ldots |\lambda_s|$, where $\lambda_j$ is the $j$th eigenvalue of $\cL$. 
We observe that each $\lambda_j$ such that $|\lambda_j| < 1$ represents a time scale of the system.
Let us assume that $|\lambda_2| < 1$ and there is a large separation between $|\lambda_m|$ and $|\lambda_{m+1}|$ in the spectrum of $\cL$ as
$
1=|\lambda_1|\geq \ldots |\lambda_m| \gg |\lambda_{m+1}| \geq \ldots |\lambda_s|.
$
This separation corresponds to two scales defined by 
$N_1 = \dfrac{1}{\ln(|\lambda_{m+1}|^{-1})}$ and $\quad N_2 = \dfrac{1}{\ln(|\lambda_{m}|^{-1})}$.
The range $N_1 \ll n \ll N_2$ defines the range where the metastability occurs and the system relaxes into a state in metastable manifold (MM), which we can consider as an effective dimensional reduction.
In the metastable regime, the dynamics will be described via the motion in the MM towards the true stationary state, which is reached at $n \gg N_2$.

The MM is a convex subset of the system states on which the long transient dynamics takes place.
From Eq.~\eqref{eqn:representation}, the projection $\cP$ of $\rho_n$ onto the MM is given by 
\begin{align}
\cP\rho_n = \rhss + \sum_{j=2}^m\tr[\rho_nL_j]R_j.
\end{align}
The evolution of dynamics in this MM is described by the effective reduced dynamics map $\cL_{\textup{eff}}=\cP\cL\cP$, which is the projection of $\cL$ on the MM. 
In this sense, MM can be regarded as being $(m-1)$ dimensional simplex, but each point in this manifold represents a density matrix in $\cM_{\cS}$.
The general structure for MM is discussed in Ref.~\cite{macieszczak:2016:meta}, where MM is determined by a subset in $\bbR^{m-1}$ given by the coefficients bounded by the maximum and minimum eigenvalues of the relevant left eigenmatrices $L_j$ for $1\leq j \leq m$.

From the perspective in temporal processing of the quantum reservoir, the higher-dimensional the MM the longer effects of the reservoir's initial states remained in the processing stream to hamper the QESP.
In contrast, if the dynamics is characterized in a low-dimensional MM, the effects of the initial states are reduced faster and the local observables remain in a relaxation behavior that only depends on a finite number of history input states.
This is favorable for the fading memory property of the system and enhances the performance of temporal learning tasks.

To understand the transient dynamics, we consider a specific case when the large separation occurs at $m=2$.
Here, the MM is a one-dimensional simplex, and the metastable states for this case can be considered as a linear combination of the states  at end points of the MM, which are called \textit{extreme metastable states} (EMSs)~\cite{macieszczak:2016:meta}. These EMSs are defined as follows:
\begin{align}
    \rhtd_1 = \rhss + \cmax R_2, \quad \rhtd_2 = \rhss + \cmin R_2,
\end{align}
where $\cmax$ and $\cmin$ are the maximal and minimal eigenvalues of $L_2$,
which are real numbers since $L_2$ is a Hermitian matrix.
Because $L_2$ and $\rhss$ are different eigenmodes of $\cL$, we obtain the orthogonality as $\tr[L_2\rhss]=0$.
From Lemma II.1 in Ref.~\cite{lasserre:ineq:trace} and since $L_2$ is a Hermitian matrix and $\rhss$ is a Hermitian and positive semi-definite matrix, we obtain the following inequality: 
\begin{align}
    \cmin \tr[\rhss] \leq \tr[L_2\rhss] \leq \cmax \tr[\rhss].
\end{align}
Because $\tr[\rhss]=1$ and $L_2$ is not a zero matrix, we have $\cmin \leq 0$, $\cmax \geq 0$, and $\cmax - \cmin > 0$.

The projection of a quantum state $\rho$ onto the MM is a linear combination of the extreme metastable states as
\begin{align}
\cP\rho = p^{(1)}\rhtd_1 + p^{(2)}\rhtd_2,
\end{align}
where $p^{(1,2)}=\tr[P_{1,2}\rho]$, and $P_{1,2}$ are the matrices defined by
\begin{align}
    P_1 = \dfrac{L_2 - \cmin I}{\Delta v_2} \textup{ and } \quad P_2 &= \dfrac{-L_2 + \cmax I}{\Delta v_2},
\end{align}
where $\Delta v_2 = \cmax - \cmin$.
These matrices satisfy $P_{1,2} \geq 0$ and $P_1 + P_2 = I$. 
Therefore  $P_{1,2}$ constitute a POVM. Furthermore, the EMSs $\rhtd_1$ and $\rhtd_2$ are approximately distjoint (see Supplemental Material in Ref.~\cite{macieszczak:2016:meta}).

As the number of interactions $n \gg N_1$, the quantum reservoir  dynamics can be captured by the effective long transient dynamics in the MM. 
The reservoir state corresponding with the measurement of an observable $O$ after the $n$th interaction can be calculated as
\begin{align}
    \langle O\rangle_{\rho_n} = \sum_i p^{(i)}_n\tr[O\rhtd_i].
\end{align}

In our quantum reservoir, the projection of $\rho_n$ after the $n$th interaction is
\begin{align}\label{eqn:mm:evol}
    \cP\rho_n = p^{(1)}_n\rhtd_1 + p^{(2)}_n\rhtd_2.
\end{align}
We can project both the left and right side of the formula $\rho_{n+1}=\cL(\rho_n)$ to the MM and obtain
\begin{align}
p^{(1)}_{n+1}\rhtd_1 + p^{(2)}_{n+1}\rhtd_2 = p^{(1)}_n\cL(\rhtd_1) + p^{(2)}_n\cL(\rhtd_2).
\end{align}
Since $\lambda_2$ is the eigenvalue of $\Tilde{\cL}$ with the corresponding right eigenmatrix $R_2$, we have
\begin{align}\label{eqn:mm:evol2}
    \cL(\rhtd_1) &= \rhss + \cmax\lambda_2R_2  = \lambda_2\rhtd_1 + (1-\lambda_2)\rhss\\
    &= \frac{1}{\Delta v_2}\left\{ [\lambda_2\cmax - \cmin]\rhtd_1 + (1-\lambda_2)\cmax\rhtd_2\right\}\nonumber,
\end{align}
\begin{align}\label{eqn:mm:evol3} 
    \cL(\rhtd_2) &= \rhss + \cmin\lambda_2R_2 = \lambda_2\rhtd_2 + (1-\lambda_2)\rhss\\
    &= \frac{1}{\Delta v_2}\left\{ (1-\lambda_2)\cmin\rhtd_1 + [\cmax - \lambda_2\cmin]\rhtd_2 \right\}\nonumber.
\end{align}
From Eqs.~\eqref{eqn:mm:evol}--\eqref{eqn:mm:evol3}, the evolution dynamics on the MM can be described by the dynamics of $\bp_n =\left[\begin{array}{c}p^{(1)}_n\\p^{(2)}_n\end{array}\right]$ as
\begin{align}
    \bp_{n+1} = \dfrac{1}{\Delta v_2}\left[\begin{array}{cc}\lambda_2\cmax - \cmin & -(1-\lambda_2)\cmin\\(1-\lambda_2)\cmax & \cmax - \lambda_2\cmin\end{array}\right]\bp_{n},
\end{align}
or
\begin{align}\label{eqn:evol:eff}
    \Delta\bp_{n}  = \bp_{n+1} - \bp_n = A_{\textup{eff}}\bp_n,
\end{align}
where
\begin{align}
    A_{\textup{eff}} = \dfrac{1-\lambda_2}{\Delta v_2}\left[\begin{array}{cc}-\cmax & -\cmin\\\cmax & \cmin\end{array}\right].
\end{align}
Because $\cmax \geq 0$ and $\cmin \leq 0$ are real numbers, from Eq.~\eqref{eqn:evol:eff} we note that $\lambda_2$ must be real. 
Furthermore, since we assume that $|\lambda_2| < 1$, the matrix $A_{\textup{eff}}$ has non-positive diagonal terms and non-negative off-diagonal terms. The sum of entries in each column in $A_{\textup{eff}}$ is equal to 0.
Therefore, $A_{\textup{eff}}$ can be considered as a generator of a discrete-time Markov chain induced from states in the MM.
This generator describes the classical stochastic dynamics between two metastable states $\rhtd_1$ and $\rhtd_2$, and for $n\to \infty$ we obtain the probabilities corresponding to the stationary state $\rhss$. 

\begin{figure}
		\includegraphics[width=12.0cm]{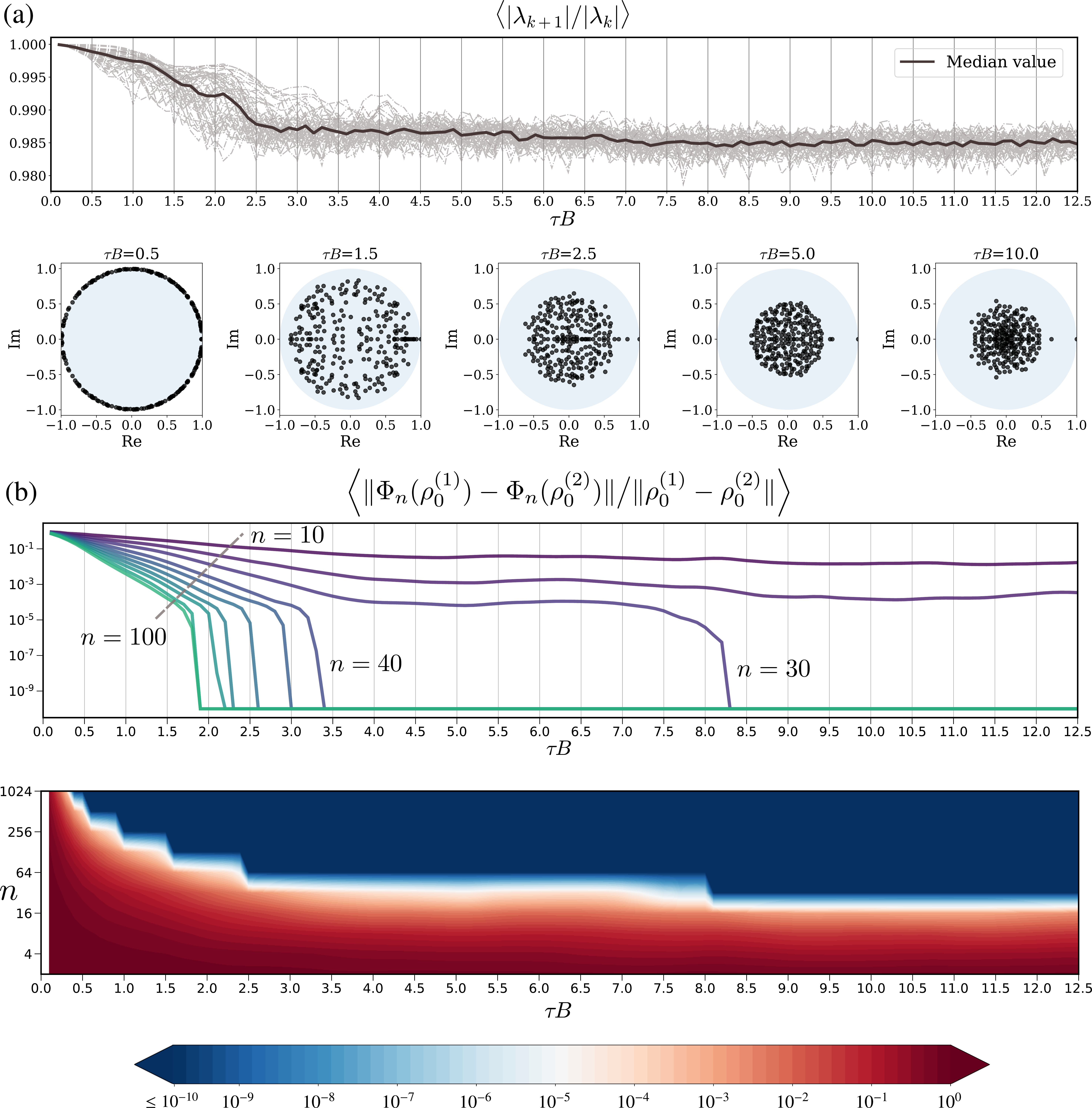}
		\protect\caption{(a) (Top) The values of $\left \langle |\lambda_{k+1}|/|\lambda_k| \right \rangle$ for 100 random reduced dynamics maps $\cL_{\beta}$ with $N_m=4, N_e=2, \alpha=1.0$, and $J=B=1.0$. The solid line depicts the median value of the ensemble of dynamics maps. (Bottom) The distribution of eigenvalues in the unit disk at $\tau B = 0.5, 1.5, 2.5, 5.0$, and $10.0$. 
		(b) (Top) The distance ratio between the trace distance of final states $\Phi_n(\rho^{(1)}_0), \Phi_n(\rho^{(2)}_0)$ and the trace distance of initial states $\rho^{(1)}_0, \rho^{(2)}_0$ for the number of interaction steps going from $n=10$ (purple line), $n=20,30,\ldots$ to $n=100$ (green line). This ratio is averaged over 100 different pairs of initial states keeping $\| \rho^{(1)}_0 - \rho^{(2)}_0\| > 0.5$. All values below the threshold of $10^{-10}$ are kept to this minimum value for clear visualization.
		(Bottom) The distance ratio as the function of $\tau B$ and $n$. The levels of the ratio are indicated by the color bar.
		\label{fig:dynamics:ratio}}
\end{figure}

\begin{figure}
		\includegraphics[width=12.0cm]{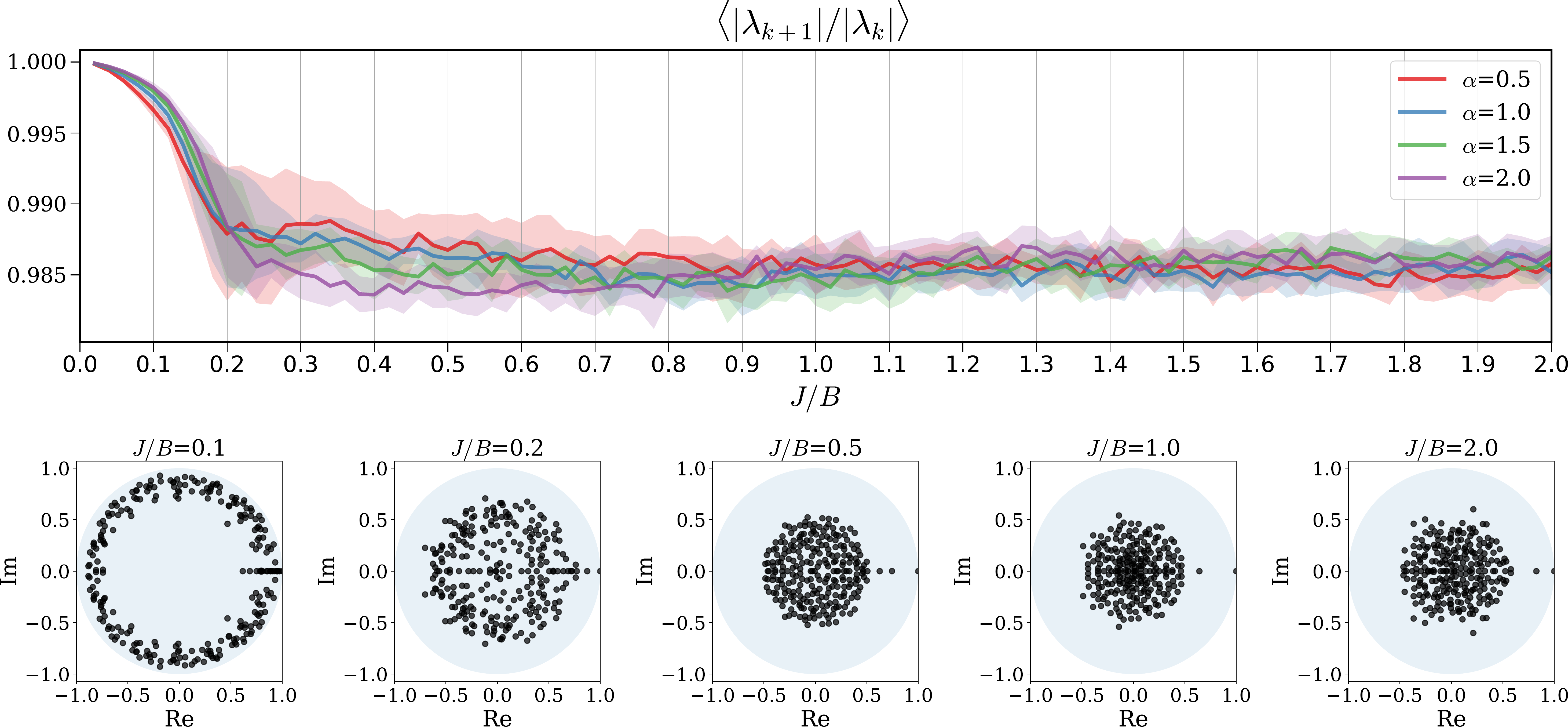}
		\protect\caption{(Top) Values of $\left \langle |\lambda_{k+1}|/|\lambda_k| \right \rangle$ as the function of $J/B$ and $\alpha$ with $\tau B = 10.0$. The solid lines and the shaded areas indicate the median values and the confidence intervals (one standard deviation) calculated in the ensemble of 100 random $\cL_{\beta}$.
		(Bottom) Examples of the distribution in the unit disk of eigenvalues of the reduced dynamics maps $\cL_{\beta}$ at $\alpha=1.0, J/B = 0.1, 0.2, 0.5, 1.0$, and $2.0$. 
		\label{fig:dynamics:JB}}
\end{figure}

In the numerical analysis with the model in Eq.~\eqref{eqn:ising}, we further investigate the eigenvalues' distribution of the reduced dynamics map $\cL_{\beta}$.
We first measure the ratios in absolute values between two consecutive eigenvalues as $r_k=|\lambda_{k+1}|/|\lambda_k|$,
which indicate the separation of time scales in the dynamics.
In Fig.~\ref{fig:dynamics:ratio}(a), the top panel shows the expectation value of $r_k$ over $k$  according to the normalized time $\tau B$ for 100 random $\cL_{\beta}$ with $N_m=4, N_e=2$.
The bottom panel of Fig.~\ref{fig:dynamics:ratio}(a) displays exemplary distributions of eigenvalues for $\tau B = 0.5, 1.5, 2.5, 5.0$, and $10.0$.
To examine the convergence property, in the top panel of Fig.~\ref{fig:dynamics:ratio}(b), we plot the distance ratio between the trace distance of final states $\Phi_n(\rho^{(1)}_0), \Phi_n(\rho^{(2)}_0)$ and the trace distance of initial states $\rho^{(1)}_0, \rho^{(2)}_0$ for the number of interaction steps going from $n=10$ (purple line), $n=20,30,\ldots$ to $n=100$ (green line). This ratio is averaged over 100 different pairs of initial states maintaining $\| \rho^{(1)}_0 - \rho^{(2)}_0\| > 0.5$.
In the bottom panel of Fig.~\ref{fig:dynamics:ratio}(b), the calculated ratios are displayed as the function of $\tau B$ and the number $n$ of interactions, where the color bar indicates the levels of ratios.
Similarly, Fig.~\ref{fig:dynamics:ratio}(c) indicates values of $\left \langle |\lambda_{k+1}|/|\lambda_k| \right \rangle$ as the function of $J/B$ and $\alpha$ with $\tau B = 10.0$, and examples of the eigenvalues' distribution of the reduced dynamics maps $\cL_{\beta}$ at $\alpha=1.0, J/B = 0.1, 0.2, 0.5, 1.0$, and $2.0$. 
 
If the eigenvalues are close to the unit border or $r_k$ are close to 1, 
$\cL_{\beta}$ reduces toward a unitary map and the QR relaxes into a state on a high-dimensional MM.
Such long transient dynamics makes the QR retain the information of its initial states on the MM.
In terms of fading memory, if $\cL_{\beta}$ reduces toward a unitary map, then only a little information of the input state $\beta$ is remained in the reservoir state after applying the reduced dynamics map $\cL_\beta$.
Therefore, the fading memory property is not enhanced in this ``more unitary" behavior.
We observe the dynamical transition ($1.5< \tau B < 2.5$) from high values to a stable range of $\left \langle r_k \right \rangle$.
In the stable range, more eigenvalues are moved near the center of the unit disk, the effects of initial states are reduced, and the dynamics will be characterized by lower-dimensional MMs in a ``more ergodic" phase.
This phase guarantees enough information in the input states to be entangled with the reservoir. 
The local observables then become functions of a finite number of past input states with the fading memory property.
As demonstrated in the bottom panel of Fig.~\ref{fig:dynamics:ratio}(b), the convergence property is also enhanced in this phase.
We observe the similar dynamical transition in Fig.~\ref{fig:dynamics:JB},
where the dynamics goes from more unitary ($J/B < 0.1$) with slower relaxation to more ergodic behavior ($J/B > 0.2$) with faster relaxation.

\clearpage
\section{Quantum  memory capacity}

The capacity to reconstruct the previous $d$ steps of the input states is evaluated via the square of the distance correlation~\cite{gabor:2007:corr} between the output $\{\sigma_n\}$ and the target $\{\hat{\sigma}_n\}=\{\beta_{n-d}\}$:
\begin{align}
    \cR^2(d) = \dfrac{\cV^2(\{\sigma_n\}, \{\hat{\sigma}_n\})}{\sqrt{\cV^2(\{\sigma_n\}, \{\sigma_n\})\cV^2(\{\hat{\sigma}_n\}, \{\hat{\sigma}_n\})}}.
\end{align}
Here, $\cV^2(\{\rho_n\}, \{\sigma_n\})$ represents the squared distance covariance of random sequences of density matrices $\{\rho_n\}, \{\sigma_n\}$.

The squared distance covariance $\cV^2(\{\rho_n\}, \{\sigma_n\})$ is calculated from all pairwise distances $A(\rho_j, \rho_k)$ and $A(\sigma_j, \sigma_k)$ for $j, k = 1,2,\ldots,n$.
Here, the distance $A(\rho, \sigma) = \arccos{F(\rho, \sigma)}$ for given density matrices $\rho$ and $\sigma$ is defined as the angle induced from the fidelity $F(\rho, \sigma)=\tr[\sqrt{\sqrt{\sigma}\rho\sqrt{\sigma}}]$.
We construct the distance matrices for $\{\rho_n\}$ and $\{\sigma_n\}$ as $(R_{jk})$ and $(S_{jk})$ with the elements $R_{jk}=A(\rho_j, \rho_k)$ and $S_{jk}=A(\sigma_j, \sigma_k)$.
We take all double centered distances
\begin{align}
    r_{j,k} &= R_{j,k} - \bar{R}_{j.} - \bar{R}_{.k} + \bar{R}_{..},\\
    s_{j,k} &= S_{j,k} - \bar{S}_{j.} - \bar{S}_{.k} + \bar{S}_{..},
\end{align}
where $\bar{R}_{j.}$ and $\bar{R}_{.k}$ are the $j$th row mean and the $k$th column mean, respectively,
and $\bar{R}_{..}$ is the grand mean of the distance matrix $(R_{jk})$ (the same notations for $S$).
The squared distance covariance is the arithmetic average of the products $r_{j,k}s_{j,k}$
\begin{align}
    \cV^2(\{\rho_n\}, \{\sigma_n\}) = \dfrac{1}{n^2}\sum_{j=1}^n\sum_{k=1}^nr_{j,k}s_{j,k}.
\end{align}

The distance covariance and correlation are developed for measuring the dependence and testing independence between two random vectors~\cite{gabor:2007:corr}.
In the definition of $\cR^2(d)$, we are interested in the empirical distance covariance and correlation to study the serial dependence between $\{\sigma_n\}$ and $\{\hat{\sigma}_n\}=\{\beta_{n-d}\}$.
Here, $0\leq \cR^2(d) \leq 1$, and $\cR^2(d) = 1$ if we can find some linear transformation from the output sequence $\{\sigma_n\}$ to the target sequence $\{\hat{\sigma}_n\}$.
In contrast, $\cR^2(d)=0$ implies that the system cannot reconstruct the previous $d$ steps of the inputs because the output and the target sequences are completely independent.
We define $\cR^2(d)$ as the \textit{quantum memory function} of the quantum reservoir to represent the fraction of distance variance explainable in a sequence of states by other.
We then define the \textit{quantum memory capacity} as 
$
    \textup{QMC} = \sum_{d=0}^{\infty}\cR^2(d)
$
to measure how much distance variance of the delay input states can be recovered from reconstructed output states, summed over all delays.

QMC is motivated from the  memory capacity in classical reservoir computing~\cite{jaeger:2001:short} and is generally proposed here as a standard quantity to compare the temporal processing capacity of quantum devices.
We note that without the condition of the quantum states, $\sigma_n$ and $\hat{\sigma}_n$ can be written in the vector form, and QMC can be defined in a same way with the classical  memory capacity in Ref.~\cite{jaeger:2001:echo} and Definition 1 in Ref.~\cite{dambre:2012:nonlinear}.
In this situation, QMC is bounded by the number of elements ($MK$) in each reservoir state if the input states are i.i.d.
We hypothesize that this bound is still true for the definition of $\cR^2(d)$ and QMC in our framework.

\begin{figure}
		\includegraphics[width=14cm]{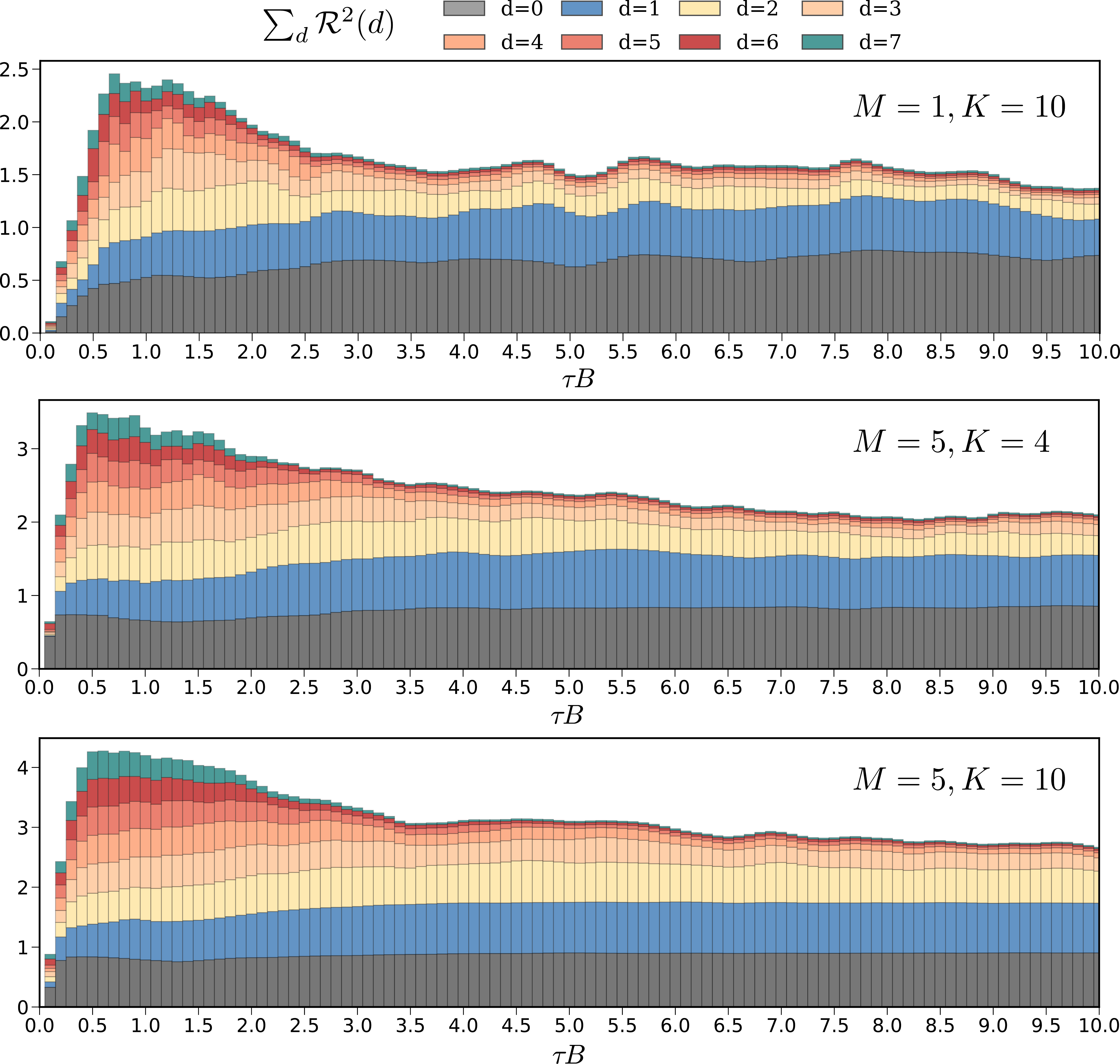}
		\protect\caption{The quantum memory capacity $\sum_d\cR^2(d)$ broken down in different $d$ $(0\leq d \leq 7)$ according to $\tau B$ with $\alpha=1.0, J=1.0, J/B=1.0, N_e=2, N_m=4$, and different values of measurement multiplexity $M$ and number $K$ of observables.
		\label{fig:qmem}}
\end{figure}

Figure~\ref{fig:qmem} shows QMC as a function of $\tau B$ broken down in values of $d$ ($0\leq d \leq 7$) with the model parameters $\alpha=1.0, J=1.0, J/B=1.0,$ and $N_e=2, N_m=4$.
Here, the number of observables is set to $K=N_m$ if we only select the observables as spin projections $O_j=\hsig^z_j$ for all $j$, and to $K=N_m(N_m+1)/2$ if we further select the observables as two-spins correlations $\hsig^z_i\hsig^z_j$ for all $i < j$.
The first 1000 time steps are excluded for initial transients to satisfy the QESP.
The training and evaluating are performed with 3000 and 1000 time steps, respectively.
The value of $\cR^2(d)$ is averaged over different runs with ten random trials of the initial state and input sequence. 
This value is faded out as increasing the delay $d \geq 7$.
This value is increased as increasing the measurement multiplexity ($M$) and the number of observables ($K$) to obtain high fidelities in tomography tasks.
We note that increasing $M$ may be more realistic since it appears difficult to find many physical quantities that can be measured in the physical implementation.
However, related to the quantum Zeno effect~\cite{misra:1977:zeno}, we should not set the value $\tau B / M$ as too small, since in this case, the time evolution slows down and the frequent measurements force the quantum state to remain in a projected subspace.

\begin{figure}
		\includegraphics[width=14cm]{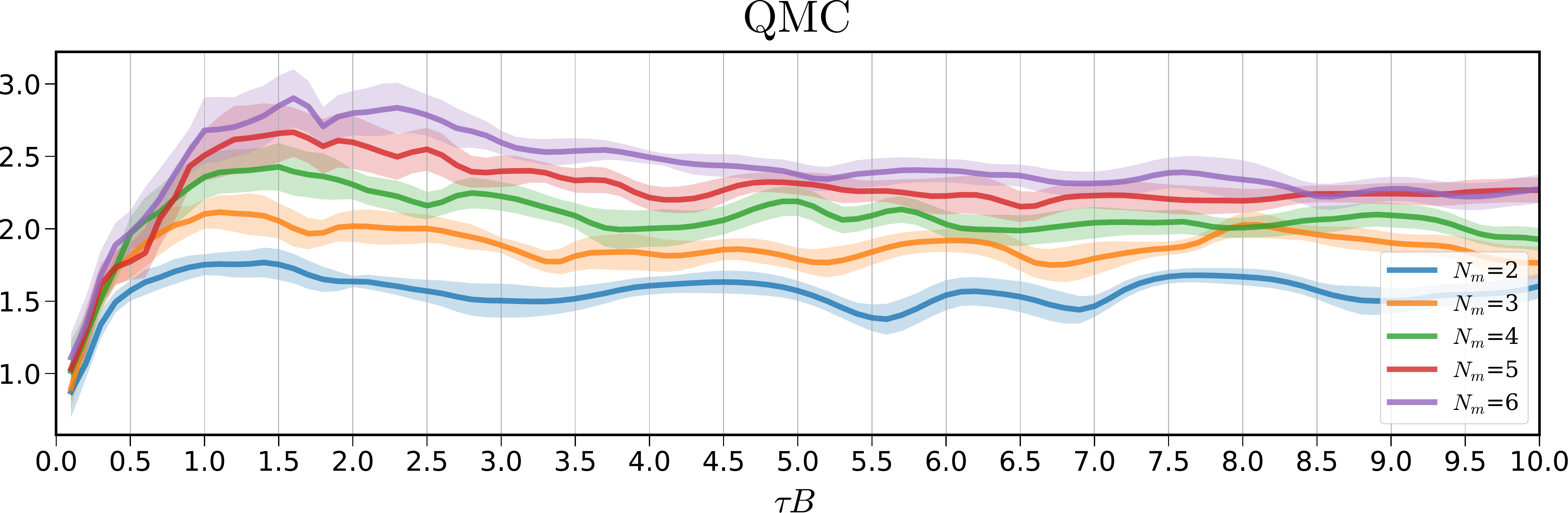}
		\protect\caption{Quantum  memory capacity (QMC) for sequences of i.i.d. two-qubit states of the quantum reservoir modeled in Eq.~\eqref{eqn:ising} according to the normalized interaction time $\tau B$ and the number $N_m$ of qubits in the reservoir. Other model parameters are $\alpha=1.0, J=1.0,$ and $J/B=1.0$.
		QMC is calculated until $d_{\textup{max}}=20$.
		The lines depict the average QMC across 10 sequences of i.i.d. input states. The shaded areas indicate the confidence intervals (one standard deviation) calculated in the same ensemble of runs.
		\label{fig:qmc:nspins}}
\end{figure}

\begin{figure}
		\includegraphics[width=11.5cm]{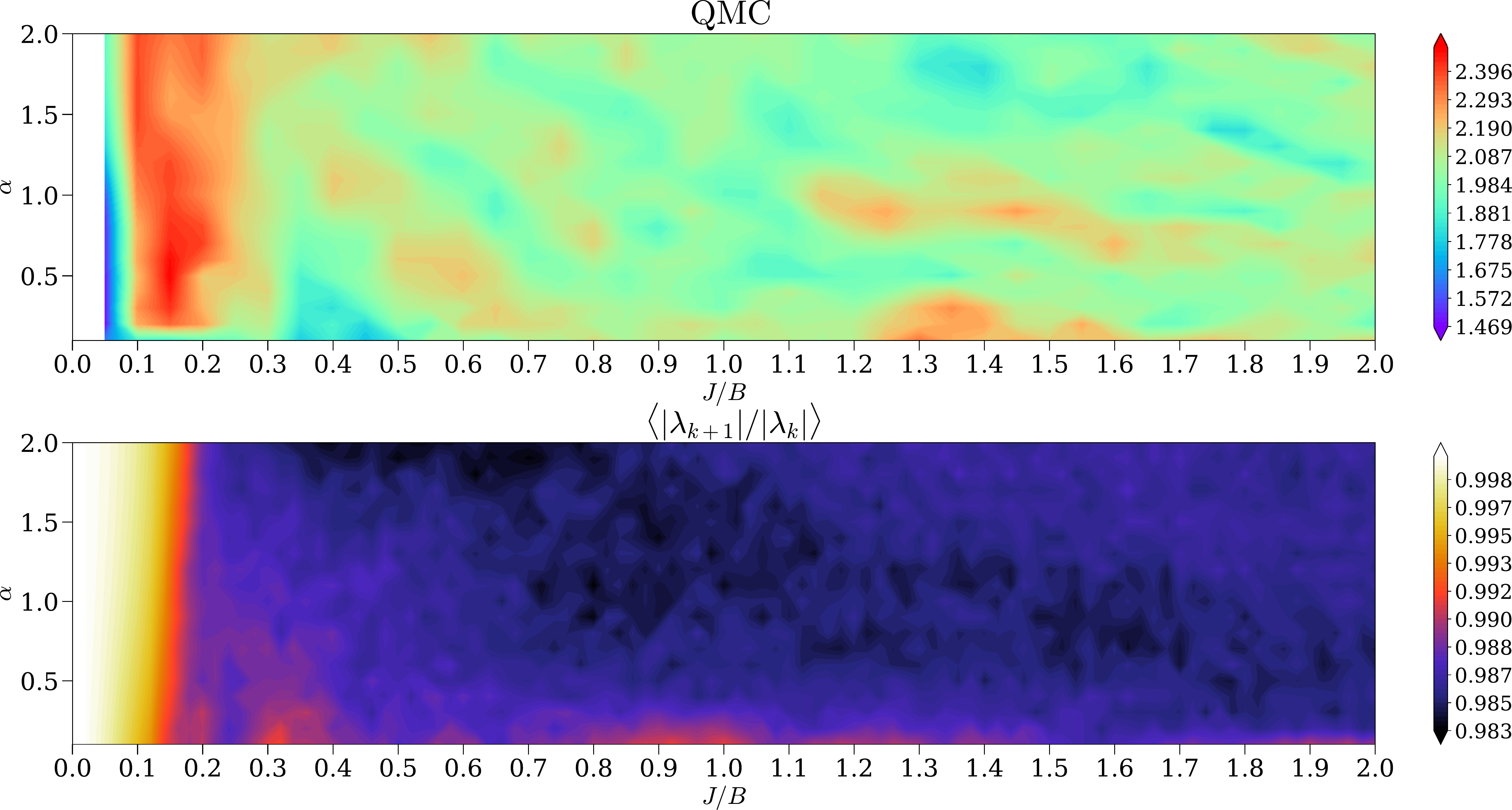}
		\protect\caption{(Top) Quantum  memory capacity (calculated until $d_{\textup{max}}=10$) as the function of model parameters $\alpha$ and $J/B$ with $N_e=2, N_m=4,$ and $\tau B = 10.0$.
		(Bottom) The value $\left \langle |\lambda_{k+1}|/|\lambda_k| \right \rangle$ over $k$ as the function of model parameters.
		\label{fig:qmem:tau10}}
\end{figure}

\begin{figure}
		\includegraphics[width=11.5cm]{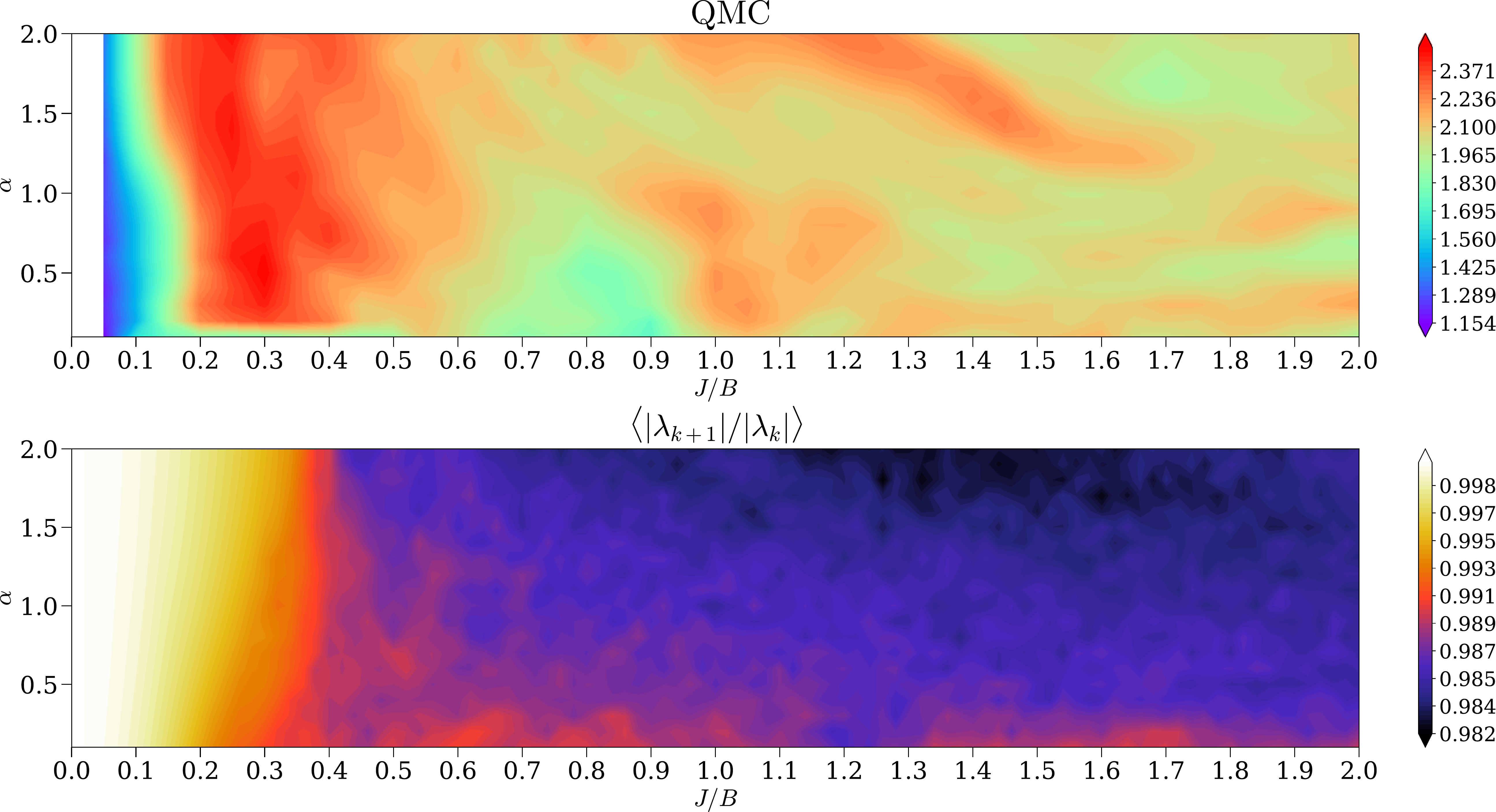}
		\protect\caption{(Top) Quantum  memory capacity (calculated until $d_{\textup{max}}=10$) as the function of model parameters $\alpha$ and $J/B$ with $N_e=2, N_m=4,$ and $\tau B = 5.0$.
		(Bottom) The value $\left \langle |\lambda_{k+1}|/|\lambda_k| \right \rangle$ over $k$ as the function of model parameters.
		\label{fig:qmem:tau5}}
\end{figure}

\begin{figure}
		\includegraphics[width=11.5cm]{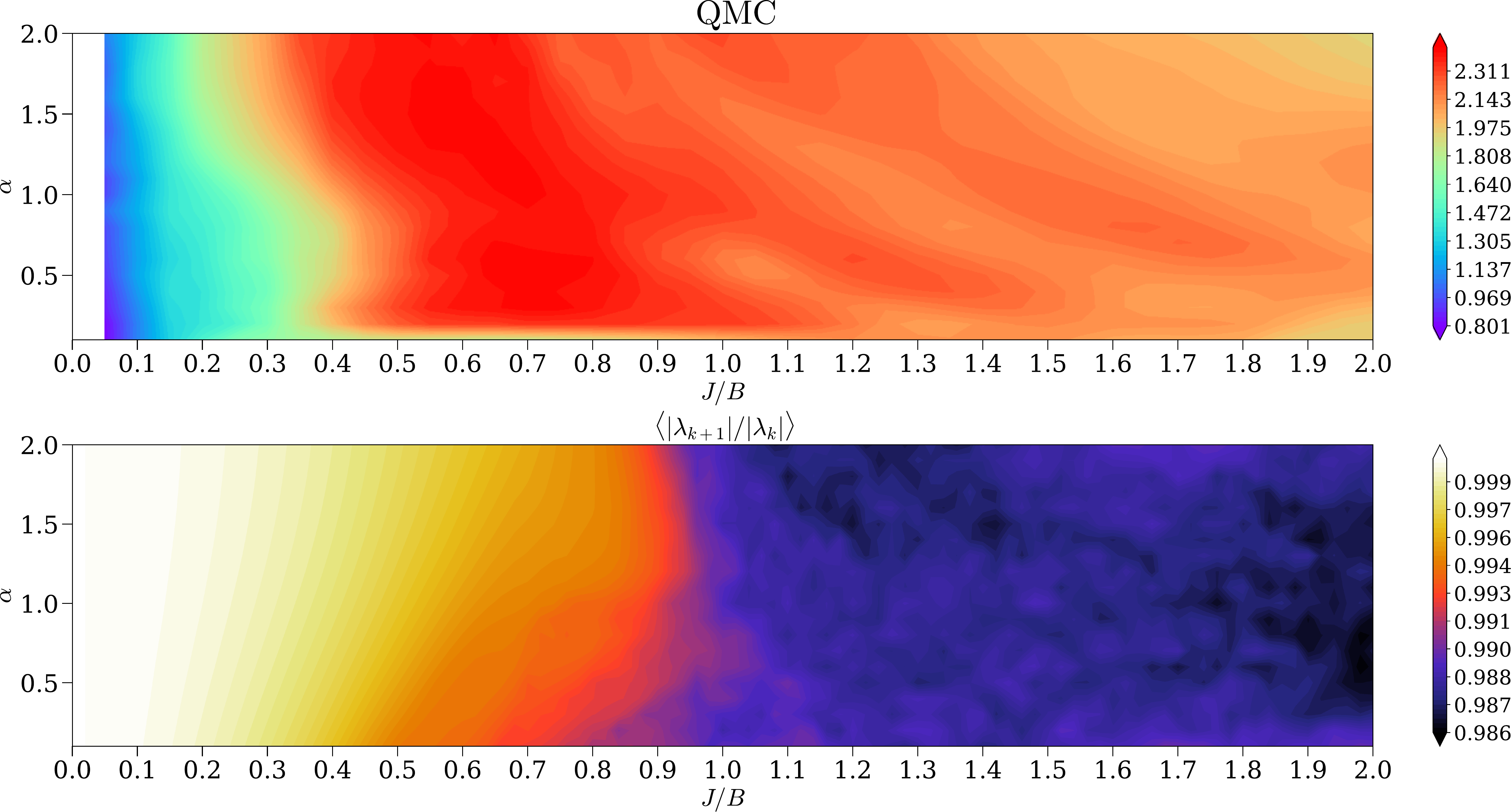}
        \protect\caption{(Top) Quantum  memory capacity (calculated until $d_{\textup{max}}=10$) as the function of model parameters $\alpha$ and $J/B$ with $N_e=2, N_m=4,$ and $\tau B = 2.0$.
		(Bottom) The value $\left \langle |\lambda_{k+1}|/|\lambda_k| \right \rangle$ over $k$ as the function of model parameters.
		\label{fig:qmem:tau2}}
\end{figure}

We plot other numerical results for QMC in Figs.~\ref{fig:qmc:nspins}--\ref{fig:qmem:tau2}.
In these numerical simulations, the first 1000 time steps are excluded for initial transients to satisfy the QESP.
The training and evaluating are performed with 3000 and 100 time steps, respectively.
The capacity is averaged over ten random trials of input sequences and initial states.

\clearpage
\section{Results on the temporal tomography tasks}
\subsection{Convex mixture of quantum channels in times}\label{sec:convex:mixture}
Given a sequence of input states $\beta_1, \beta_2, \ldots$ in a Hilbert space $\cH_A$ with the dimension $D_A$, we consider the temporal map $\cF$ as
\begin{align}\label{eqn:temporal:qpt}
    \cF(\beta_n) = \dfrac{1}{Z}\sum_{i=0}^d\eta_i\Omega_{n-i}(\beta_{n-i}),
\end{align}
where $\Omega_1, \Omega_2,\ldots$ is a sequence of unknown quantum channels from $\cH_A$ to another Hilbert space,
and $\eta_0, \eta_1, \ldots, \eta_d$ are unknown non-negative real numbers with $Z=\sum_{i=0}^d\eta_i$ to preserve the trace.
Our objective is to characterize $\cF$ from the measurement data.
If $\eta_i=1$ for all $i$, we can consider $\cF$ as a quantum version of the simple moving average filter for a sequence of quantum channels $\Omega_1, \Omega_2,\ldots$ acting on the input states.
If $\eta_i=(d+1-i)$ for all $i$, we have a weighted moving average filter.
If $\eta_d=1$ and $\eta_i=0$ for $i\neq d$, we have a memory-based reconstruction of the channel applied on $d$-delay quantum states.
In our demonstration, we set $\Omega_n$ as a time-dependent depolarizing quantum channel $\Omega_n(\beta) =p_n\frac{I}{D} + (1-p_n)\beta$.
The depolarizing probability $p_n$ is formulated as the $r$th-order nonlinear dynamical output 
$
p_n = \kappa p_{n-1} + \eta p_{n-1}\left( \sum_{j=0}^{r-1} p_{n-j-1}\right) + \gamma u_{n-r+1}u_n + \delta,
$
where $r=10, \kappa=0.3, \eta=0.04, \gamma=1.5$, and $\delta=0.1$.
Here, $\{u_n\}$ is a random sequence of scalar values in $[0, 0.2]$ to set $p_n$ into the stable range in $[0, 1]$.
The sequence $\{p_n\}$ resembles the NARMA benchmark~\cite{atiya:2000:narma}, which is commonly used for evaluating the computational capability of temporal processing with long time dependence.
Furthermore, the sequence $\{u_n\}$ is randomly generated with the same random seed used for generating the input sequence $\{\beta_n\}$; therefore, $\{p_n\}$ depends on $\{\beta_n\}$.
This setting makes the quantum channel $\Omega_n$ dependent on to the current input and the output of previous channels, then the tomography task requires memory to characterize these channels.

\begin{figure}
		\includegraphics[width=14cm]{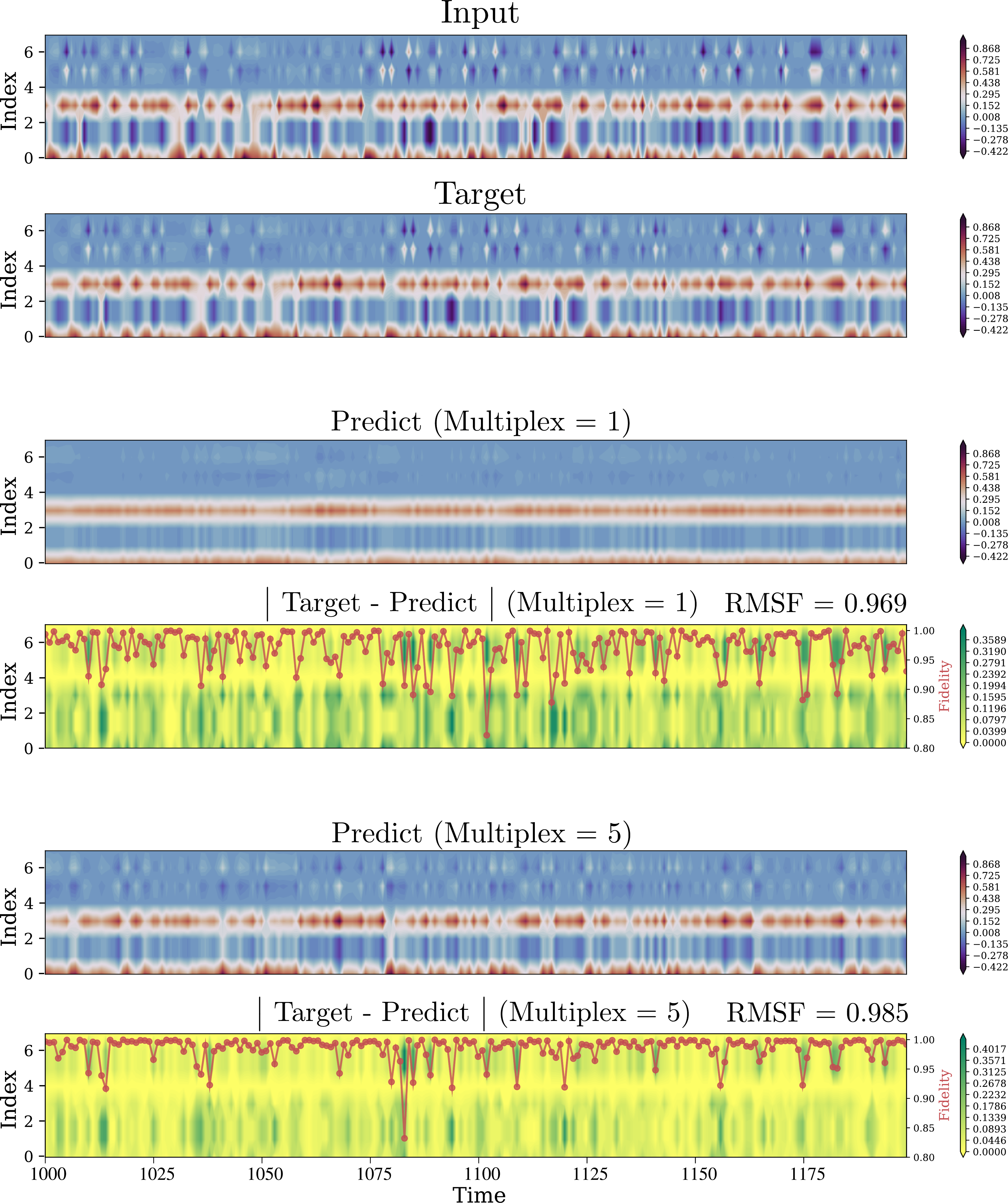}
		\protect\caption{Contour plots of a temporal forecast with the absolute difference between the target and the prediction for the reconstruction of delayed depolarizing quantum  map with delay time $d=5$, $N_m=5$, and $N_e=1$.
		Other model parameters are $\alpha=1.0, J=1.0, J/B=1.0$, and $\tau B = 2.0$.
		At each time point, the $2\times 2$ density matrix is represented as a vector with $8$ elements by stacking the real and imaginary parts.
		The red points in the plots with green-yellow color maps (with right y-axis) represent the fidelities at each time point between the target and the predicted state.
		\label{fig:qpt:multiplex-delay}}
\end{figure}

Figure~\ref{fig:qpt:multiplex-delay} illustrates an example of a memory-based reconstruction of the delayed depolarizing quantum map ($\eta_d=1$ and $\eta_i=0$ for $i\neq d$) with delay time $d=5, N_m=5, N_e=1$, and the measurement multiplexity $M=1, 5$. 
Other model parameters are $\alpha=1.0, J/B=1.0$.
Here, the first $t_{\textup{buffer}}=500$ time steps are excluded for initial transients to satisfy the QESP.
The training and evaluating are performed in the range $(t_{\textup{buffer}}, t_{\textup{train}}]$ and $(t_{\textup{train}}, t_{\textup{val}}]$, respectively, where $t_{\textup{train}}=1000$ and $t_{\textup{val}}=1200$.
At each time point, the $D\times D$ density matrix is represented as a vector with $2D^2$ elements by stacking the real and imaginary parts.
The plots with a green-yellow color map in Fig.~\ref{fig:qpt:multiplex-delay} show the absolute error between the target and the predicted vector at $M=1$ and $M=5$.
The red points in these plots indicate the fidelities between the target and the predicted quantum states.
We confirm that we can nearly reconstruct the previous depolarizing quantum channel almost perfectly as the fidelities are above 95\% (with $M=5$).

\begin{figure}
		\includegraphics[width=14cm]{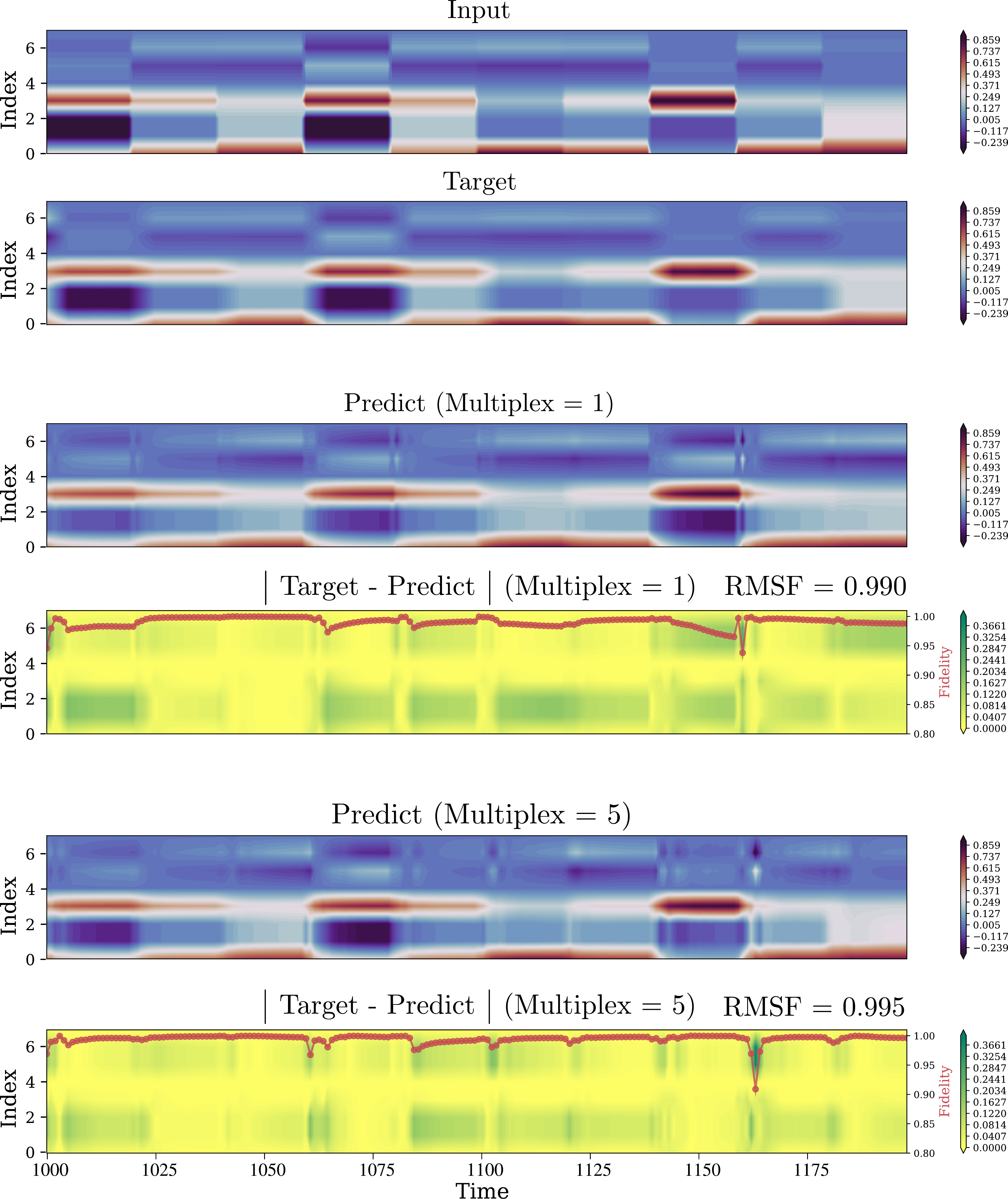}
		\protect\caption{Contour plots of a temporal forecast with the absolute difference between the target and the prediction for the reconstruction of a simple moving average filter for a sequence of depolarizing quantum channels with delay time $d=5$, $N_m=5$, and $N_e=1$.
		Other model parameters are $\alpha=1.0, J=1.0, J/B=1.0$, and $\tau B = 2.6$.
		At each time point, the $2\times 2$ density matrix is represented as a vector with $8$ elements by stacking the real and imaginary parts.
		The red points in the plots with green-yellow color maps (with right y-axis) represent the fidelities at each time point between the target and the predicted state.
		Here, the input states jump to a new random quantum state every 20 time steps, thus introducing temporal dependencies between input states.
		\label{fig:qpt:multiplex-filter}}
\end{figure}

Figure~\ref{fig:qpt:multiplex-filter} illustrates an example of a simple moving average filter ($\eta_i=1$ for all $i$) for a sequence of quantum channels $\Omega_1, \Omega_2,\ldots$ acting on the input states with the delay time $d=5, N_m=5, N_e=1$, and the measurement multiplexity $M=1, 5$. 
Other parameters are $\alpha=1.0, J/B=1.0, t_{\textup{buffer}}=500, t_{\textup{train}}=1000, $ and $t_{\textup{val}}=1200$.
Here, the input states jump to a new random quantum state every 20 time steps, thus introducing temporal dependencies between input states.
Our framework can reconstruct this intriguing tomography of a simple moving average filter.

\begin{figure}
		\includegraphics[width=18cm]{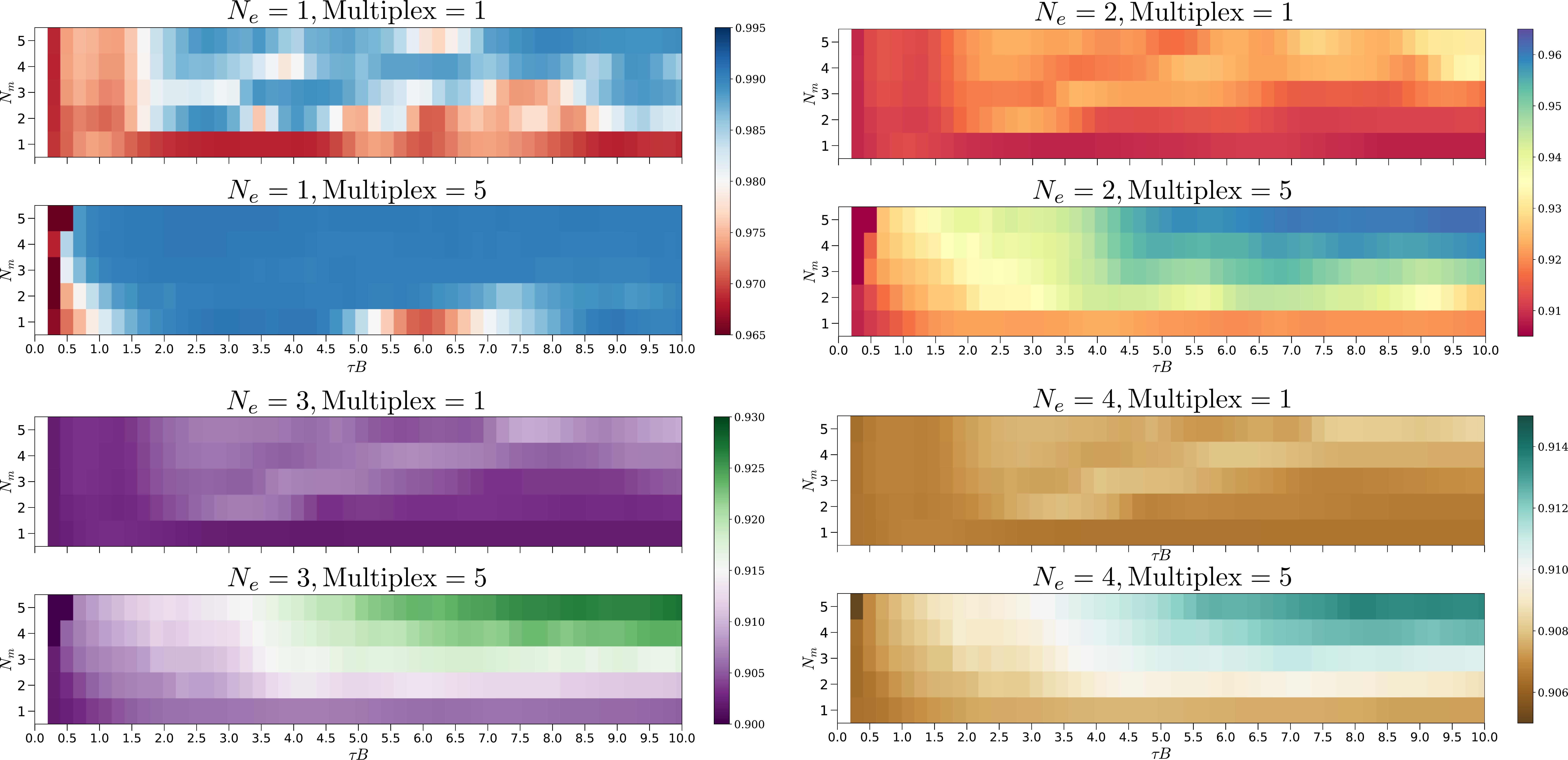}
		\protect\caption{Fidelities as functions of $N_m, N_e$, and the normalized interaction time $\tau B$ for the memory-based reconstruction of delayed depolarizing quantum map $\cF(\beta_n)=\Omega_{n-d}(\beta_{n-d})$ with delay $d=1$. The levels of fidelity values are indicated by the color bars.
		\label{fig:qpt:delay-qubits}}
\end{figure}

\begin{figure}
\begin{center}
    		\includegraphics[width=18cm]{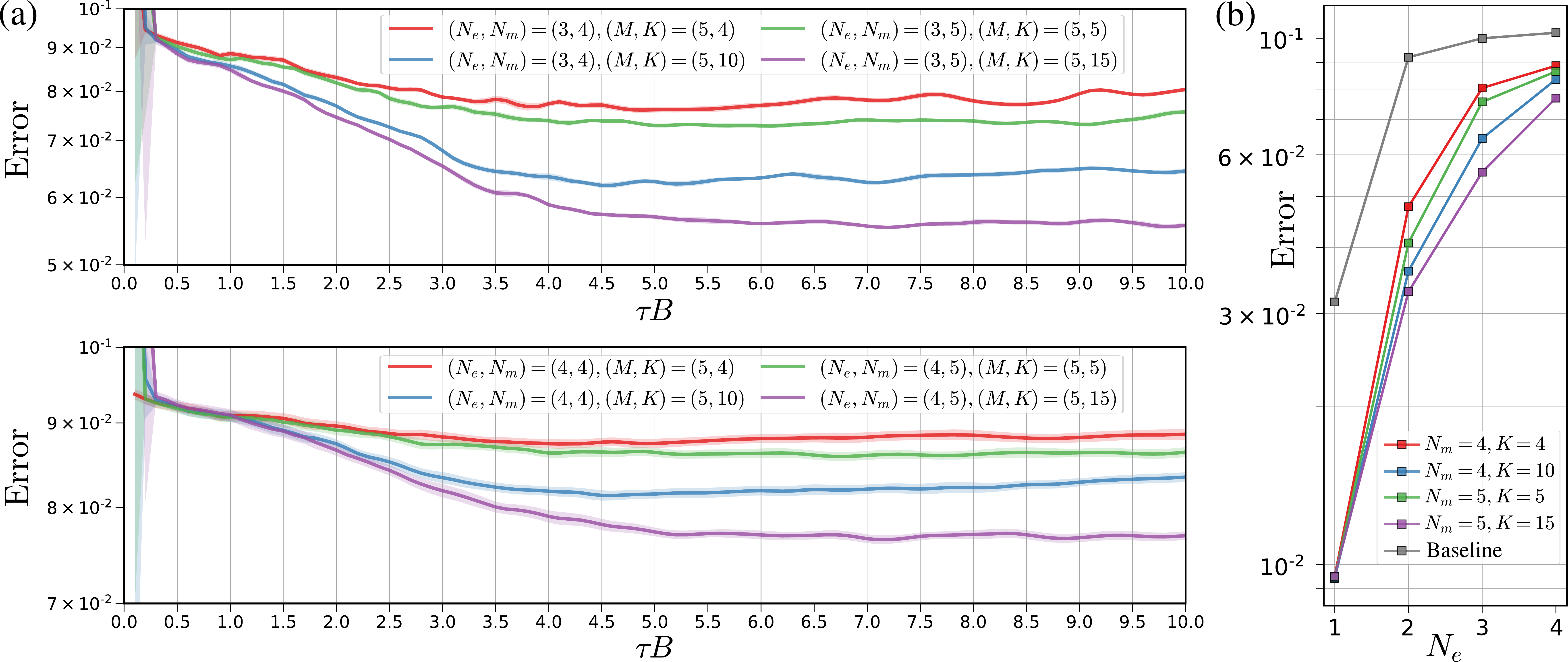}
		\protect\caption{(a) Tomography errors ($1.0 - \textup{RMSF}$) for input states of $N_e=3, 4$ qubits and the quantum reservoir of $N_m=4, 5$ qubits as functions of $\tau B$ for the reconstruction of delayed depolarizing quantum  map $\cF(\beta_n)=\Omega_{n-d}(\beta_{n-d})$ with $d=1$.
		Other model parameters are $\alpha=1.0, J/B=1.0$.
		(b) Tomography errors of the baseline and our method for different $N_e, N_m$, and $K$ at $\tau B = 10.0$. 
		\label{fig:qpt:nspins}}
\end{center}
\end{figure}

Figure~\ref{fig:qpt:delay-qubits} and Fig.~\ref{fig:qpt:nspins} illustrate the other results for the tomography task $F(\beta_n) = \Omega_{n-1}(\beta_{n-1})$ demonstrated in the main text.
The first $t_{\textup{buffer}}=1000$ time steps are excluded for initial transients to satisfy the QESP,
the training and evaluating are performed in the range $(t_{\textup{buffer}}, t_{\textup{train}}]$ and $(t_{\textup{train}}, t_{\textup{val}}]$, respectively, where $t_{\textup{train}}=4000$ and $t_{\textup{val}}=5000$.
Other model parameters are $\alpha=1.0, J/B=1.0$.
Figure~\ref{fig:qpt:delay-qubits} illustrates the calculated RMSF as functions of $N_m, N_e$, and $\tau B$ with the measurement multiplexity $M=1, 5$, and the number of observables is $K=N_m$. 
The RMSF is averaged over ten different runs with random trials of the input sequence and initial state.
We confirm that increasing $N_m$ and $M$ indeed scales the fidelity. 


Figure~\ref{fig:qpt:nspins}(a) illustrates the calculated error (1.0 - RMSF) as functions of $\tau B$ with the measurement multiplexity $M=5$ for different values of $N_e, N_m$ and $K$.
Here, the number of observables is set to $K=N_m$ if we only select the observables as spin projections $O_j=\hsig^z_j$ for all $j$, and to $K=N_m(N_m+1)/2$ if we further select the observables as two-spins correlations $\hsig^z_i\hsig^z_j$ for all $i < j$.
The error is averaged over ten different runs with random trials of the input sequence and initial state.
The fidelity is larger than 94\% even with $N_e=3$ qubits and $N_m = 4,5$.

We further compare our method with the classical baseline method, which uses linear regression for the sequence of input density matrices to obtain the sequence of output density matrices.
Therefore, this baseline method does not have memory effect to emulate the recurrent relation in the temporal quantum map.
In the classical baseline method, we assume that we can obtain full tomography of input states, which is not required in our method. Instead of using the measurement results in the quantum reservoir, reservoir state $\bx_n$ is constructed directly from the vector form of $\beta_n$ by stacking the real and imaginary parts in the corresponding density matrix. The readout map and training process are the same as our method. In Fig.~\ref{fig:qpt:nspins}(b), we compare our method with the baseline method in terms of tomography error ($1.0 - \textup{RMSF}$) for the reconstruction task of $\cF(\beta_n)=\Omega_{n-1}(\beta_{n-1})$ for different values of $N_e, N_m$, and $K$.
We confirm that our method with short-term memory property outperforms the baseline method for all settings of $N_e, N_m$, and $K$.
We note that our method does not need to perform full tomography for input states as the baseline method but rather perform measurements with a single and controllable setup for a quantum reservoir to obtain the reservoir states to train the recurrent relation in the temporal quantum map.

\clearpage
\subsection{Quantum switch: a superposition of quantum channels in times}

We perform the tomography of a novel quantum primitive called the quantum switch, which is a a superposition of two alternative orders of quantum channels.
It is proposed in Ref.~\cite{chiribella:2012:qswitch} and then followed by experimental demonstrations~\cite{procopio:2015:qswitch:exp,rubino:2017:qswitch:exp,goswami:2018:qswitch:exp,wei:2019:qswtich:exp,guo:2020:qswitch:exp}.
In the classical counterpart in the communication network, a switch is an operation of control that can route a target system through two classical channels $C_1$ and $C_2$ in series following one causal order ($C_1$ then $C_2$) or the other ($C_2$ then $C_1$).
The big difference in the quantum switch is that it induces entirely new quantum trajectories, where the order of the two operators is indefinite.

\begin{figure}
		\includegraphics[width=16cm]{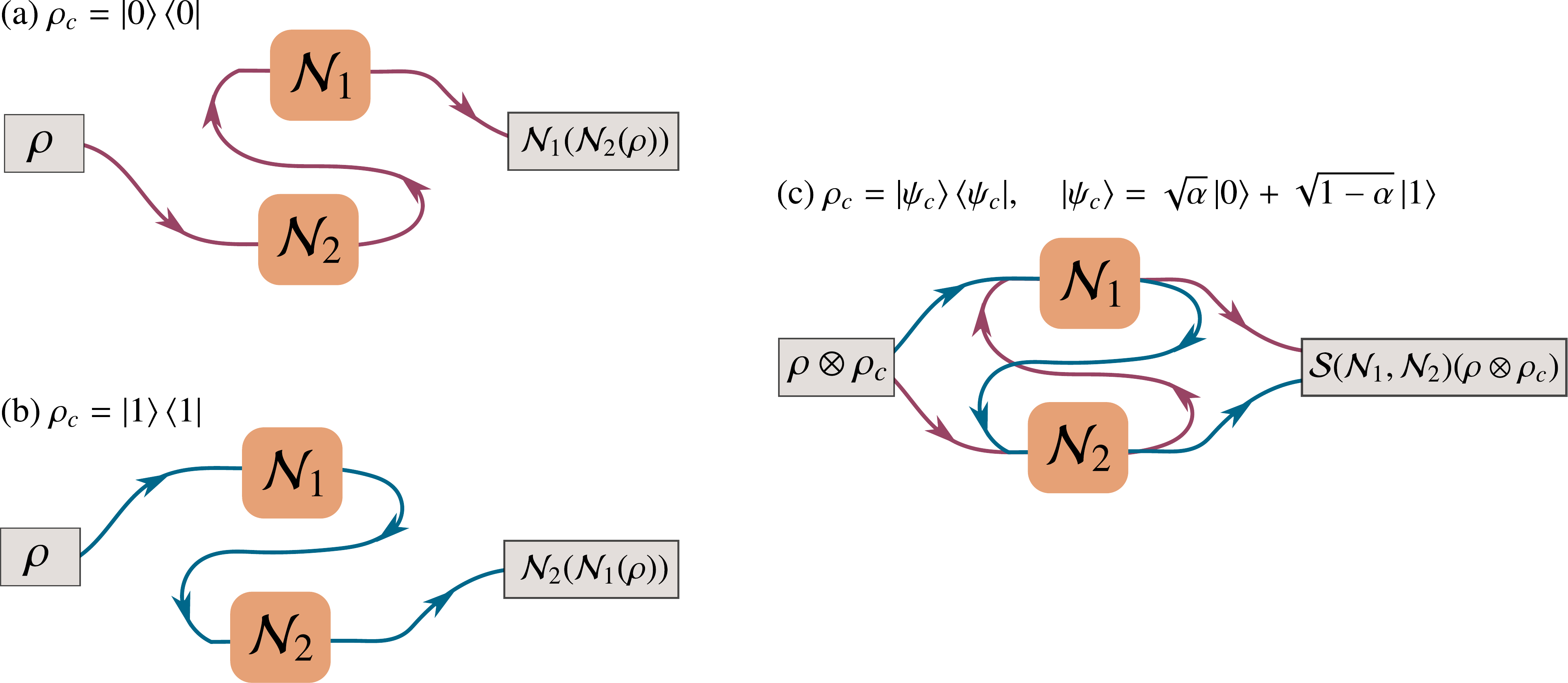}
		\protect\caption{
The concept of a quantum switch $\cS(\cN_1, \cN_2)$ with two quantum channels $\cN_1$ and $\cN_2$. 
Depending on an independent control state $\rho_c$, the quantum switch acts on a state $\rho$ to return the output.
(a) If $\rho_c = \ket{0}\bra{0}$, the quantum switch returns the state $\cN_1\cN_2(\rho)\otimes \ket{0}\bra{0}$. (b) If $\rho_c = \ket{1}\bra{1}$, the quantum switch returns the state $\cN_2\cN_1(\rho)\otimes \ket{1}\bra{1}$.
(c) When the control qubit is in a superposition of $\ket{0}$ and $\ket{1}$, the quantum switch returns a correlated state as the result of $\cN_1$ and $\cN_2$ acting on $\rho$ in a quantum superposition of two alternative orders.
We said that quantum channels $\cN_1$ and $\cN_2$ are in a superposition of causal orders whose input and output are $\rho \otimes \rho_c$ and $\cS(\cN_1, \cN_2)(\rho \otimes \rho_c)$, respectively.
		\label{fig:qswitch}}
\end{figure}

Technically, a quantum switch takes two quantum channels $\cN_1$ and $\cN_2$ as inputs and creates a new channel $\cS(\cN_1, \cN_2)$, which uses the channels $\cN_1$ and $\cN_2$ in an order that is entangled with the state of a control qubit.
Under an independent control state $\rho_c$, the quantum switch acts on a state $\rho$ as $\cS(\cN_1, \cN_2)(\rho \otimes \rho_c)$.
If $\rho_c = \ket{0}\bra{0}$, the channel $\cS(\cN_1, \cN_2)$ returns the state $\cN_1\cN_2(\rho)\otimes \ket{0}\bra{0}$ [Fig.~\ref{fig:qswitch}(a)]; if $\rho_c = \ket{1}\bra{1}$, the channel $\cS(\cN_1, \cN_2)$ returns the state $\cN_2\cN_1(\rho)\otimes \ket{1}\bra{1}$ [Fig.~\ref{fig:qswitch}(b)].
When the control qubit is in a superposition of $\ket{0}$ and $\ket{1}$, the channel returns a correlated state as the result of $\cN_1$ and $\cN_2$ acting on $\rho$ in a quantum superposition of two alternative orders~[Fig.~\ref{fig:qswitch}(c)].
It was shown that the quantum switch cannot be realized where the order of the two channels $\cN_1$ and $\cN_2$  is fixed~\cite{chiribella:2013:qswitch}.
This intriguing setup is used to perform communication tasks that are impossible in conventional quantum Shannon theory.
For example, it was shown that the quantum switch created from two identical copies of a completely depolarizing channel could transmit information that is impossible when the order is fixed or even correlated with a classical variable~\cite{ebler:2018:qswitch}.

We denote the Kraus operators of channels $\cN_1$ and $\cN_2$ as $\{K^{(1)}_i\}$ and $\{K^{(2)}_j\}$, respectively. 
We can express $\cN_1 = \sum_{i}K^{(1)}_i\rho K^{(1)\dagger}_i$ and $\cN_2 = \sum_{j}K^{(2)}_j\rho K^{(2)\dagger}_j$.
The Kraus operators of $\cS(\cN_1, \cN_2)$ acting on a target quantum state $\rho$ and a control state $\rho_c$ are defined as
\begin{align}
    W_{ij} = K^{(1)}_iK^{(2)}_j \otimes \ket{0}\bra{0} + K^{(2)}_jK^{(1)}_i \otimes \ket{1}\bra{1}.
\end{align}
The action of the quantum switch via the channel $\cS(\cN_1, \cN_2)$ is given by
\begin{align}
    \cS(\cN_1, \cN_2)(\rho \otimes \rho_c) = \sum_{i,j}W_{ij}(\rho \otimes \rho_c)W^{\dagger}_{ij}.
\end{align}
Here, we consider $\cN_1$ and $\cN_2$ as two depolarizing channels, which are given by
\begin{align}
    \cN_1(\rho) &= q_1 \tr[\rho]\frac{I}{D} + (1-q_1)\rho = (1-q_1)\rho + \dfrac{q_1}{D^2}\sum_{i=1}^{D^2}U_i\rho U^{\dagger}_i = \dfrac{q_1}{D^2}\sum_{i=0}^{D^2}U_i\rho U^{\dagger}_i,\\
    \cN_2(\rho) &= q_2 \tr[\rho]\frac{I}{D} + (1-q_2)\rho = (1-q_2)\rho + \dfrac{q_2}{D^2}\sum_{j=1}^{D^2}V_j\rho V^{\dagger}_j = \dfrac{q_2}{D^2}\sum_{j=0}^{D^2}V_j\rho V^{\dagger}_j,
\end{align}
where $\{U_i\}_{i=1}^{D^2}$ are unitary operators which form an orthonormal basis of the space of $D\times D$ matrices. We also consider the orthonormal basis $\{V_j\}_{j=1}^{D^2}$ of the space of $D\times D$ matrices for a convenient notation.
Here, we introduce the notation $U_0 = \dfrac{D\sqrt{1-q_1}}{\sqrt{q_1}} I$ and $V_0 = \dfrac{D\sqrt{1-q_2}}{\sqrt{q_2}} I$.
Then we can define the extension Kraus operators for $\cN_1$ and $\cN_2$ as $K^{(1)}_i = \dfrac{\sqrt{q_1}}{D}U_i$ for $i=0,1,\ldots,D^2$, and $K^{(2)}_j = \dfrac{\sqrt{q_2}}{D}V_j$ for $j=0,1,\ldots,D^2$.
We can express the Kraus operators of $\cS(\cN_1, \cN_2)$ as
\begin{align}
    W_{ij} = \dfrac{\sqrt{q_1q_2}}{D^2}\left( U_iV_j \otimes \ket{0}\bra{0} + V_jU_i \otimes \ket{1}\bra{1}\right).
\end{align}

We consider the input $\rho \otimes \rho_c$ with the control state $\rho_c = \ket{\psi_c}\bra{\psi_c}$, where $\psi_c = \sqrt{\alpha}\ket{0} + \sqrt{1-\alpha}\ket{1}$ ($0\leq \alpha \leq 1$). The output of the quantum switch is given by
\begin{align}\label{eqn:qswitch}
    \cS(\cN_1, \cN_2)(\rho \otimes \rho_c) = {A}^{00}\otimes\ket{0}\bra{0} + {A}^{01}\otimes\ket{0}\bra{1} + {A}^{10}\otimes\ket{1}\bra{0} + {A}^{11}\otimes\ket{1}\bra{1}, 
\end{align}
where
\begin{align}
    {A}^{00} &= \alpha\dfrac{q_1q_2}{D^4} \sum_{i=0}^{D^2}\sum_{j=0}^{D^2}U_iV_j\rho V_j^{\dagger}U_i^{\dagger} = \alpha\cN_1\cN_2(\rho) \\
    {A}^{01} &= A^{10} = \sqrt{\alpha (1-\alpha)}\dfrac{q_1q_2}{D^4}\sum_{i=0}^{D^2}\sum_{j=0}^{D^2}U_iV_j\rho U_i^{\dagger}V_j^{\dagger},\\
        &= \sqrt{\alpha (1-\alpha)}\dfrac{q_1q_2}{D^4}\left(\sum_{i=1}^{D^2}\sum_{j=1}^{D^2}U_iV_j\rho U_i^{\dagger}V_j^{\dagger} + \sum_{j=1}^{D^2}U_0V_j\rho U_0^{\dagger}V_j^{\dagger} + \sum_{i=1}^{D^2}U_iV_0\rho U_i^{\dagger}V_0^{\dagger} + U_0V_0\rho U_0^{\dagger}V_0^{\dagger} \right)\\
        &= \sqrt{\alpha (1-\alpha)} \left( \dfrac{q_1q_2}{D^2}\sum_{i=1}^{D^2}\tr[\rho U_i^{\dagger}]\dfrac{U_i}{D} + \dfrac{q_1(1-q_2)}{D^2}\sum_{j=1}^{D^2}V_j\rho V_j^{\dagger} + \dfrac{q_2(1-q_1)}{D^2}\sum_{i=1}^{D^2}U_i\rho U_i^{\dagger} +  (1-q_1)(1-q_2)\rho \right)\\
        &= \sqrt{\alpha (1-\alpha)} \left( \dfrac{q_1q_2}{D^2}\rho + q_1(1-q_2)\dfrac{I}{D} + q_2(1-q_1)\dfrac{I}{D} + (1-q_1)(1-q_2)\rho\right)\\
    {A}^{11} &= (1-\alpha)\dfrac{q_1q_2}{D^4} \sum_{i=0}^{D^2}\sum_{j=0}^{D^2}V_jU_i\rho U_i^{\dagger}V_j^{\dagger} = (1-\alpha)\cN_2\cN_1(\rho).
\end{align}

In our demonstration, we consider the pure input states $\{\beta_n=\ket{\psi_n}\bra{\psi_n}\}$, where $\psi_n$ is a superposition $\sqrt{u_n}\ket{0}+\sqrt{1-u_n}\ket{1}$ with $\{u_n\}$ randomly generated in $[0, 1]$.
Given two delay values $d_t$ and $d_c$, we consider $\cF$ as a temporal quantum switch that acts on the state $\rho = \beta_{n-d_t}$ under the control state $\rho_c = \beta_{n-d_c}$ with $\cN_1 = \Omega_{n-d_c}$ and $\cN_2=\Omega_{n-d_t}$.
Here, we set $\Omega_n$ as the time-dependent depolarizing quantum channel $\Omega_n(\beta) =p_n\frac{I}{D} + (1-p_n)\beta$,
where the depolarizing probability $p_n$ is formulated as the nonlinear dynamical output of NARMA model~\cite{atiya:2000:narma}: 
$
p_n = \kappa p_{n-1} + \eta p_{n-1}\left( \sum_{j=0}^{r-1} p_{n-j-1}\right) + \gamma v_{n-r+1}v_n + \delta,
$
where $r=10, \kappa=0.3, \eta=0.04, \gamma=1.5$, $\delta=0.1$, and $v_n=0.2u_n$ to keep $0\leq p_n \leq 1$.

By applying Eq.~\eqref{eqn:qswitch}, we have the explicit form of $\cF(\beta_n)$ as
\begin{align}
    \cF(\beta_n) &= \cS(\Omega_{n-d_c}, \Omega_{n-d_t})(\beta_{n-d_t}\otimes \beta_{n-d_c}) \\
    &= {A}^{00}_n\otimes\ket{0}\bra{0} + {A}^{01}_n\otimes\ket{0}\bra{1} + {A}^{10}_n\otimes\ket{1}\bra{0} + {A}^{11}_n\otimes\ket{1}\bra{1},
\end{align}
where
\begin{align}
    {A}^{00}_n &= u_{n-d_c}\Omega_{n-d_c}\Omega_{n-d_t}(\beta_{n-d_t}),\label{eqn:temporal:switch1}\\
    {A}^{01}_n &= {A}^{10}_n = \sqrt{u_{n-d_c}(1-u_{n-d_c})}\left(  \dfrac{p_{n-d_c}p_{n-d_t}}{D^2} + (1-p_{n-d_c})(1-p_{n-d_t}) \right) \beta_{n-d_t} + \nonumber\\ &\sqrt{u_{n-d_c}(1-u_{n-d_c})}\left(p_{n-d_c} + p_{n-d_t} - 2p_{n-d_c}p_{n-d_t}\right)\dfrac{I}{D},\label{eqn:temporal:switch2}\\
    {A}^{11}_n &= (1-u_{n-d_c})\Omega_{n-d_t}\Omega_{n-d_c}(\beta_{n-d_t})\label{eqn:temporal:switch3}.
\end{align}
To reconstruct this kind of temporal quantum map, at time $n$ we need to memorize the information of $\beta_{n-d_t}$ and $\beta_{n-d_c}$, also as $p_{n-d_c}$ and $p_{n-d_t}$ (for channels $\Omega_{n-d_c}$ and $\Omega_{n-d_t}$).

\begin{figure}
		\includegraphics[width=17cm]{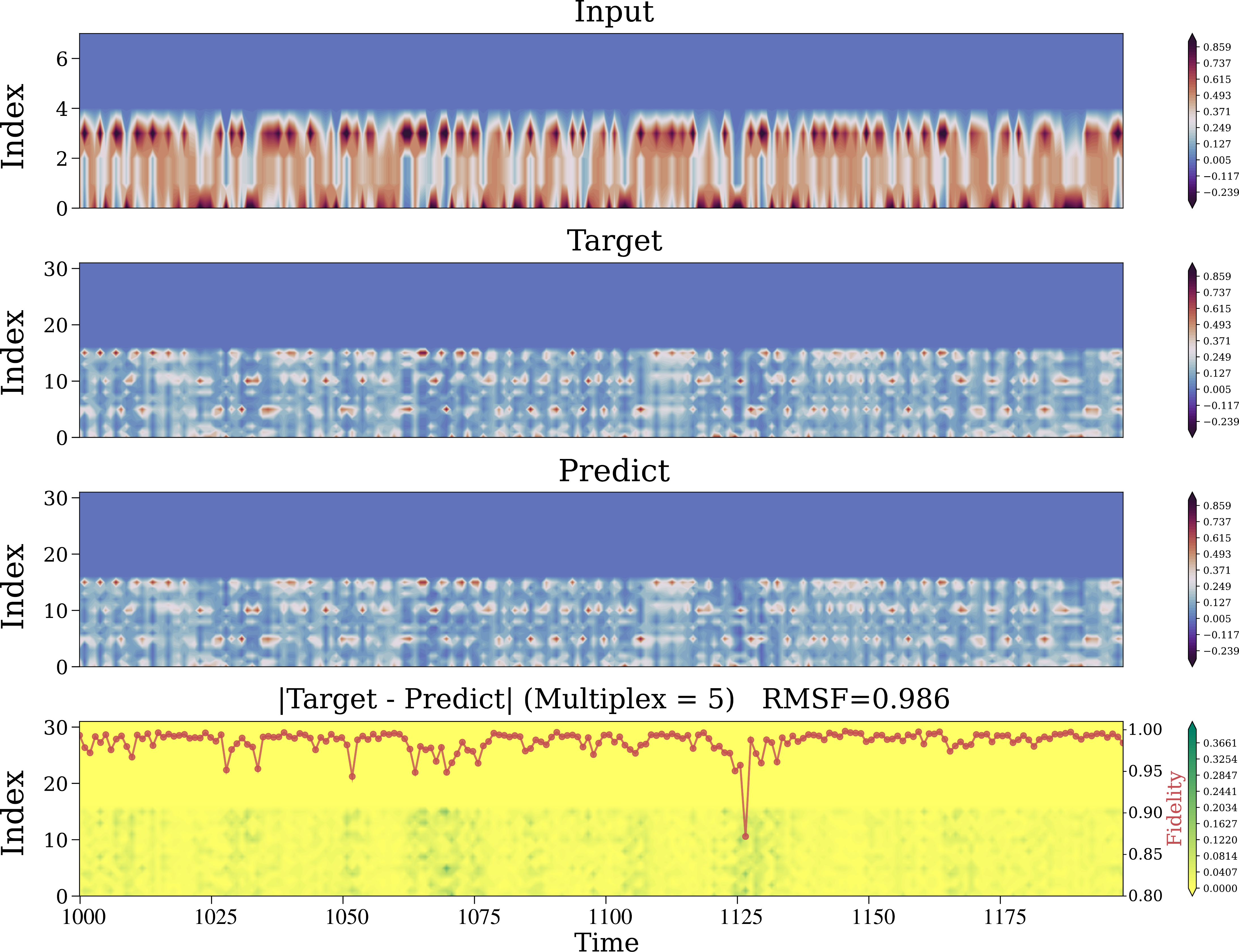}
		\protect\caption{Contour plots of a temporal forecast with the absolute difference between the target and the prediction for the reconstruction of the quantum switch with $d_c=2$ and $d_t=3$.
		Other model parameters are $N_e=1, N_m=5,$ $M=5, K=N_m(N_m+1)/2$, $\alpha=1.0, J=1.0, J/B=1.0$, and $\tau B = 10.0$.
		We consider the pure input states $\{\beta_n=\ket{\psi_n}\bra{\psi_n}\}$, where $\psi_n$ is a superposition $\sqrt{u_n}\ket{0}+\sqrt{1-u_n}\ket{1}$ with $\{u_n\}$ randomly generated in $[0, 1]$.
		The output of the temporal map $\cF$ at input $\beta_n$ is $\cS(\Omega_{n-d_c}, \Omega_{n-d_t})(\beta_{n-d_t}\otimes \beta_{n-d_c})$.
		At each time point, the  density matrix is represented as a vector by stacking the real and imaginary parts.
		The red points in the plot with a green-yellow color maps (with right y-axis) represent the fidelities at each time point between the target and the predicted state.
		\label{fig:qpt:channel:super}}
\end{figure}

Figure~\ref{fig:qpt:channel:super} illustrates an example of a reconstruction of a temporal quantum switch with the control delay time $d_c=2$, the target delay time $d_t=3$, the measurement multiplexity $M=5$, and the number of observables $K=N_m(N_m+1)/2$. 
Other model parameters are $\alpha=1.0, J=B=1.0$, and $\tau B = 10.0$.
Here, the first $t_{\textup{buffer}}=500$ time steps are excluded for initial transients to satisfy the QESP.
The training and evaluating are performed in the range $(t_{\textup{buffer}}, t_{\textup{train}}]$ and $(t_{\textup{train}}, t_{\textup{val}}]$, respectively, where $t_{\textup{train}}=1000$ and $t_{\textup{val}}=1200$.
At each time point, the density matrix is represented as a vector by stacking the real and imaginary parts.
The plot with a green-yellow color map in Fig.~\ref{fig:qpt:channel:super} shows the absolute error between the target and the predicted vector.
The red points in this plot indicate the fidelities between the target and the predicted quantum states.
We confirm that we can nearly reconstruct this temporal map almost perfectly as the fidelities are above 95\%.

\begin{figure}
		\includegraphics[width=9cm]{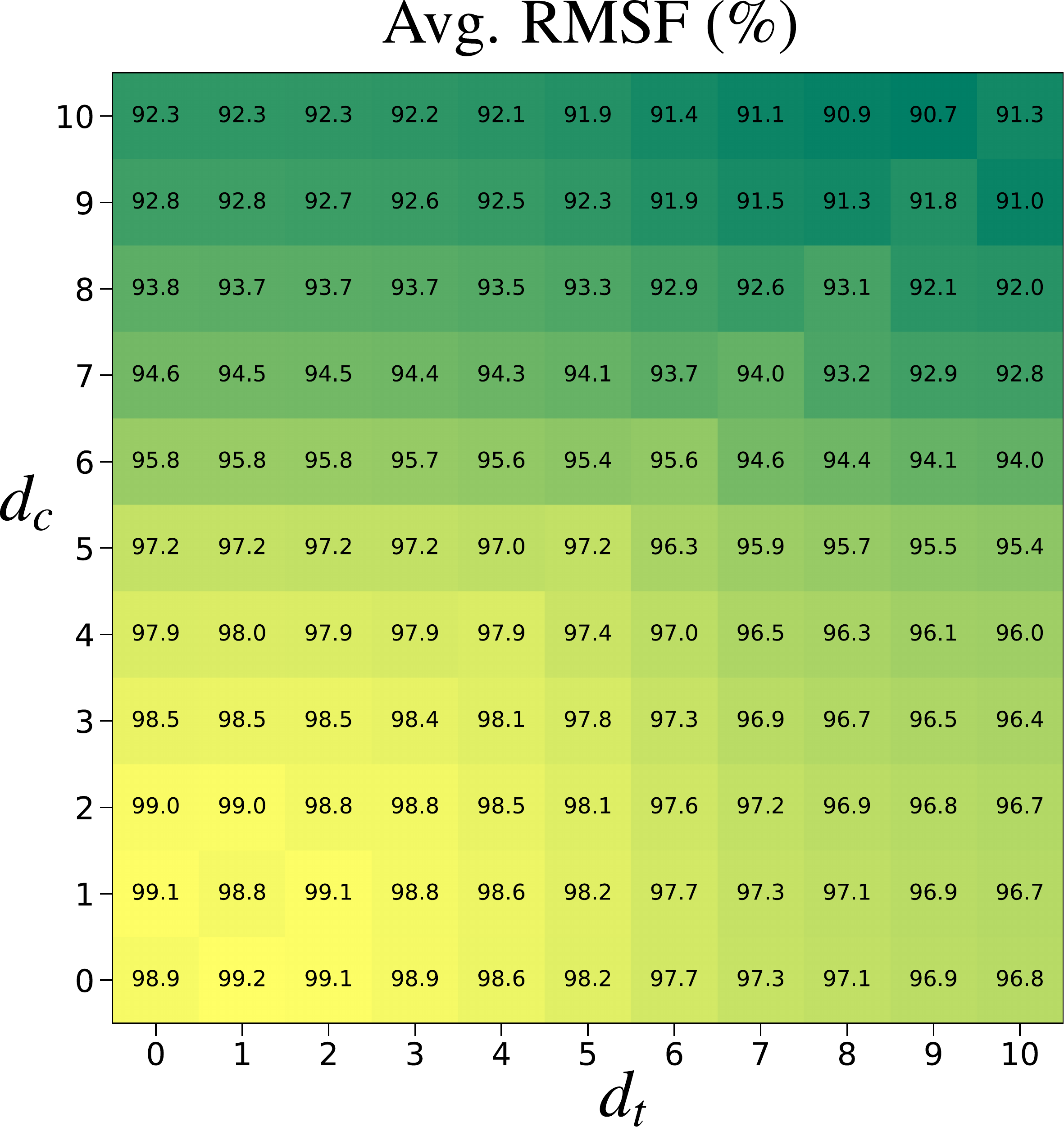}
		\protect\caption{The average RMSF in the temporal quantum switch task as a function of $d_c$ and $d_t$ over ten different runs with random trials of the input sequence and initial state. We keep other parameters the same as the parameters in Fig.~\ref{fig:qpt:channel:super}.
		\label{fig:qpt:switch-heat}}
\end{figure}

Figure~\ref{fig:qpt:switch-heat} illustrates the average RMSF as a function of $d_c$ and $d_t$ over ten different runs with random trials of the input sequence and initial state.
Due to the effect of the short-term memory, the performance decreases as we increase $d_c$ or $d_t$.
Figure~\ref{fig:qpt:switch-heat} also shows that we need more information on $\beta_{n-d_c}$ than $\beta_{n-d_t}$ to increase the performance as more complicated information of $\beta_{n-d_c}$ appears in Eqs.~\eqref{eqn:temporal:switch1}-\eqref{eqn:temporal:switch3}.

\subsection{Temporal Entangler}

Finally, we demonstrate the tomography of a temporal entangler of past channel’s outputs or a temporal entangler as a Bell state creator from past inputs.
First, we perform the tomography of a temporal entanglement creator for two outputs $\Omega_{n-d_1}(\beta_{n-d_1})$ and $\Omega_{n-d_2}(\beta_{n-d_2})$, given two delays $d_1$ and $d_2$, where $\Omega_n$ is generated in the same way with the tasks in Section~\ref{sec:convex:mixture}.
Here, the temporal map $\cF$ is given by
\begin{align}
    \cF(\beta_n) = U(\Omega_{n-d_1}(\beta_{n-d_1}) \otimes \Omega_{n-d_2}(\beta_{n-d_2}))U^{\dagger},
\end{align}
where $U$ is an arbitrary unitary applying to the product state $\Omega_{n-d_1}(\beta_{n-d_1}) \otimes \Omega_{n-d_2}(\beta_{n-d_2})$.
In our demonstration, we consider $\{\beta_n\}$ to be a randomly generated sequence of density matrices of one qubit states. We consider $U=\exp(-iHt)$ with $t=10.0$ and $H=h_{12}\hsig^x_1\hsig^x_2 + (2.0+g_1)\hsig^z_1 + (2.0+g_2)\hsig^z_2$. Here, $h_{12}, g_1,$ and $g_2$ are randomly generated in $[-0.5, 0.5]$, and $\hsig_j^\gamma$ $(\gamma \in \{x, y, z\})$ are the Pauli operators measuring the qubit $j$ along the $\gamma$ direction in a two-qubits system.
To reconstruct this kind of temporal quantum map, at time $n$ we need to memorize the information of $\beta_{n-d_1}$ and $\beta_{n-d_2}$, also as channels $\Omega_{n-d_1}$ and $\Omega_{n-d_2}$.

Second, we perform the tomography of a temporal Bell state creator for two input states $\beta_{n-d_1}$ and $\beta_{n-d_2}$.
Here, we consider $\beta_n=\ket{b_n}\bra{b_n}$, where $\{b_n\}$ is a random binary sequence ($b_n\in \{0, 1\}$). The temporal map $\cF$ is given by
\begin{align}
    \cF(\beta_n) = \ket{\psi(b_{n-d_1}, b_{n-d_2})}\bra{\psi(b_{n-d_1}, b_{n-d_2})},
\end{align}
where $\psi(b_{n-d_1}, b_{n-d_2}) = \left( \dfrac{\ket{0, b_{n-d_2}} + (-1)^{b_{n-d_1}}\ket{1, \bar{b}_{n-d_2}} }{\sqrt{2}} \right)$ is a Bell state.
To reconstruct this kind of temporal quantum map, at time $n$ we need to memorize the information of $\beta_{n-d_1}$ and $\beta_{n-d_2}$ and then emulate the quantum circuit to create Bell state from two-qubit input $\ket{b_{n-d_1}b_{n-d_2}}$.

\begin{figure}
		\includegraphics[width=17cm]{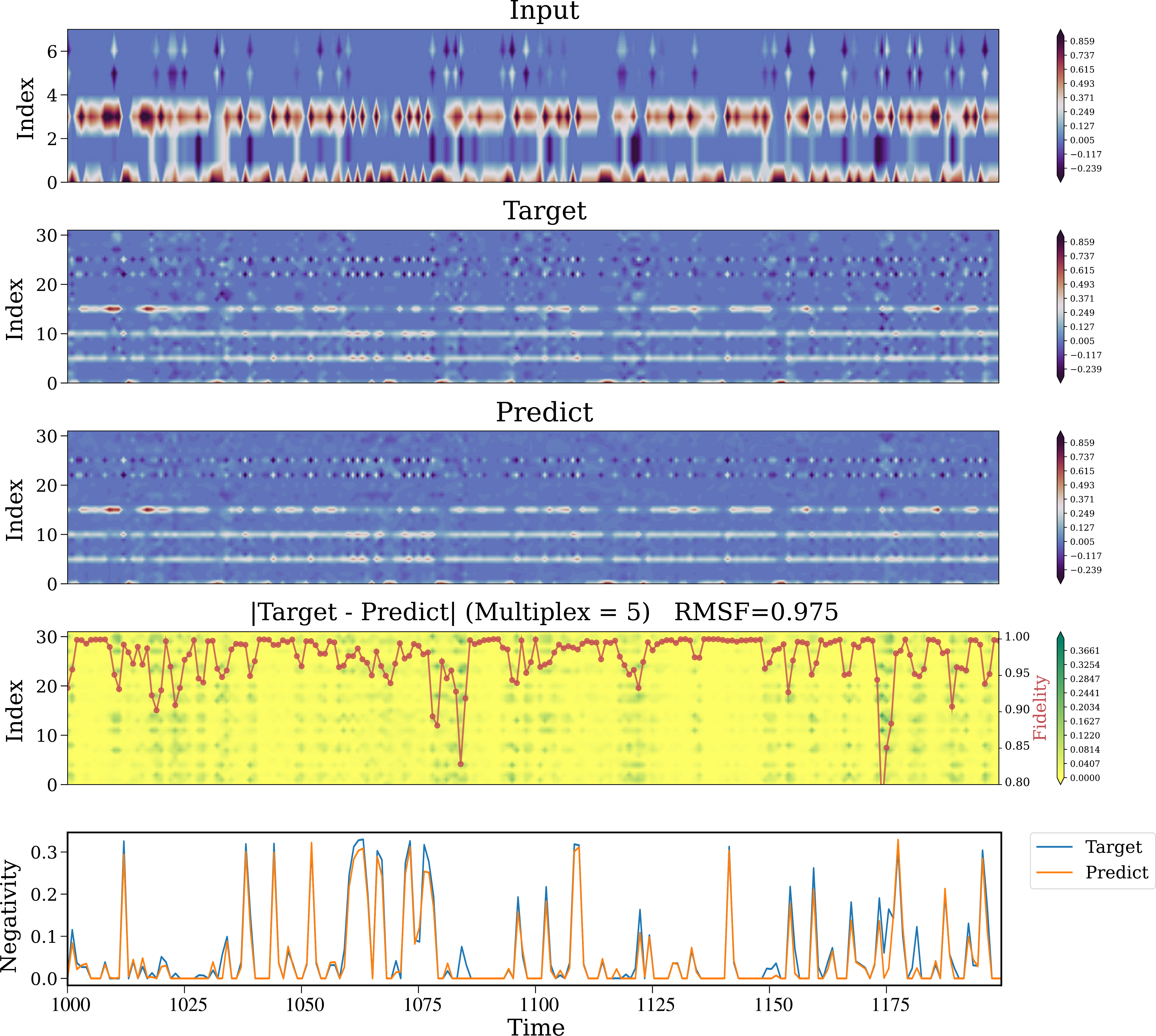}
		\protect\caption{Contour plots of a temporal forecast with the absolute difference between the target and the prediction for the reconstruction of the temporal entangler of channels' outputs.
		Model parameters are $N_e=1, N_m=5,$ $M=5, K=N_m(N_m+1)/2$, $\alpha=1.0, J=B=1.0$, and $\tau B = 2.0$.
		The output of the temporal map $\cF$ at input $\beta_n$ is $U\left(\Omega_{n-d_1}(\beta_{n-d_1}) \otimes \Omega_{n-d_2}(\beta_{n-d_2})\right)U^{\dagger}$.
		At each time point, the  density matrix is represented as a vector by stacking the real and imaginary parts.
		The red points in the plot with a green-yellow color map (with right y-axis) represent the fidelities at each time point between the target and the predicted state.
		The last plot shows the negativities of the target and predict states at each time point.
		\label{fig:qpt:channel:ent}}
\end{figure}

\begin{figure}
		\includegraphics[width=17cm]{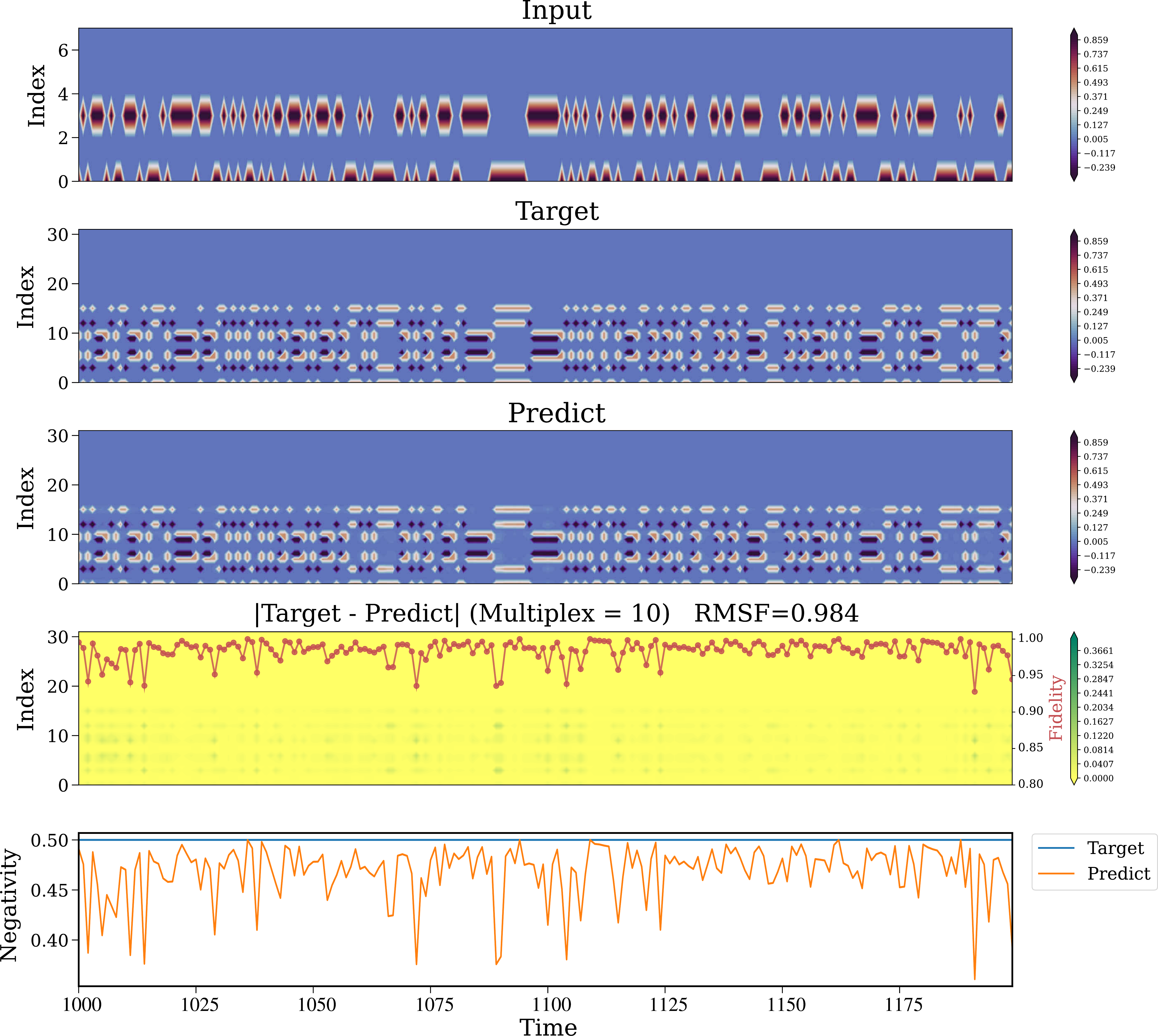}
		\protect\caption{Contour plots of a temporal forecast with the absolute difference between the target and the prediction for the reconstruction of the temporal Bell state creator of past inputs.
		Model parameters are $N_e=1, N_m=5,$ $M=10, K=N_m(N_m+1)/2$, $\alpha=1.0, J=B=1.0$, and $\tau B = 2.0$.
		At each time point, the  density matrix is represented as a vector by stacking the real and imaginary parts.
		The red points in the plot with a green-yellow color map (with right y-axis) represent the fidelities at each time point between the target and the predicted state.
		The last plot shows the negativities of the target and predict states at each time point.
		\label{fig:qpt:channel:Bell}}
\end{figure}

Figures~\ref{fig:qpt:channel:ent} and \ref{fig:qpt:channel:Bell} illustrate tomography examples of entanglers at delay times $d_1=0$ and $d_2=1$.
Here, model parameters are $N_e=1, N_m=5,$ $\alpha=1.0, J=B=1.0$, and $\tau B = 10.0$.
The first $t_{\textup{buffer}}=500$ time steps are excluded for initial transients to satisfy the QESP.
The training and evaluating are performed in the range $(t_{\textup{buffer}}, t_{\textup{train}}]$ and $(t_{\textup{train}}, t_{\textup{val}}]$, respectively, where $t_{\textup{train}}=1000$ and $t_{\textup{val}}=1200$.
At each time point, the density matrix is represented as a vector by stacking the real and imaginary parts.
The plots with green-yellow color maps in Fig.~\ref{fig:qpt:channel:ent} and Fig.~\ref{fig:qpt:channel:Bell} show the absolute error between the target and the predicted vector.
The red points in these plots indicate the fidelities between the target and the predicted quantum states.
Furthermore, we plot the negativities of the target and predict states at each time point in the last plots of  Fig.~\ref{fig:qpt:channel:ent} and Fig.~\ref{fig:qpt:channel:Bell}. 
Here, negativity is a measure of quantum entanglement, which can be defined in terms of a density matrix $\rho$ as $\cN(\rho)=\dfrac{\Vert \rho^{\Gamma_A} \Vert_1 - 1}{2}$. Here, $\rho^{\Gamma_A}$ is the partial transpose of $\rho$ with respect to subsystem $A$, and $\Vert \cdot \Vert_1$ denotes the trace norm.
In our demonstrations, we compute the negativities of the two-qubit target states with respect to the subsystem of the first qubit.
We confirm that the negativities of the target states and predicted states fit very well in Fig.~\ref{fig:qpt:channel:ent}. In the task of temporal Bell state creator, we can reconstruct the state with large negativities, which are relatively near the target negativities of the Bell states.

\begin{figure}
		\includegraphics[width=18cm]{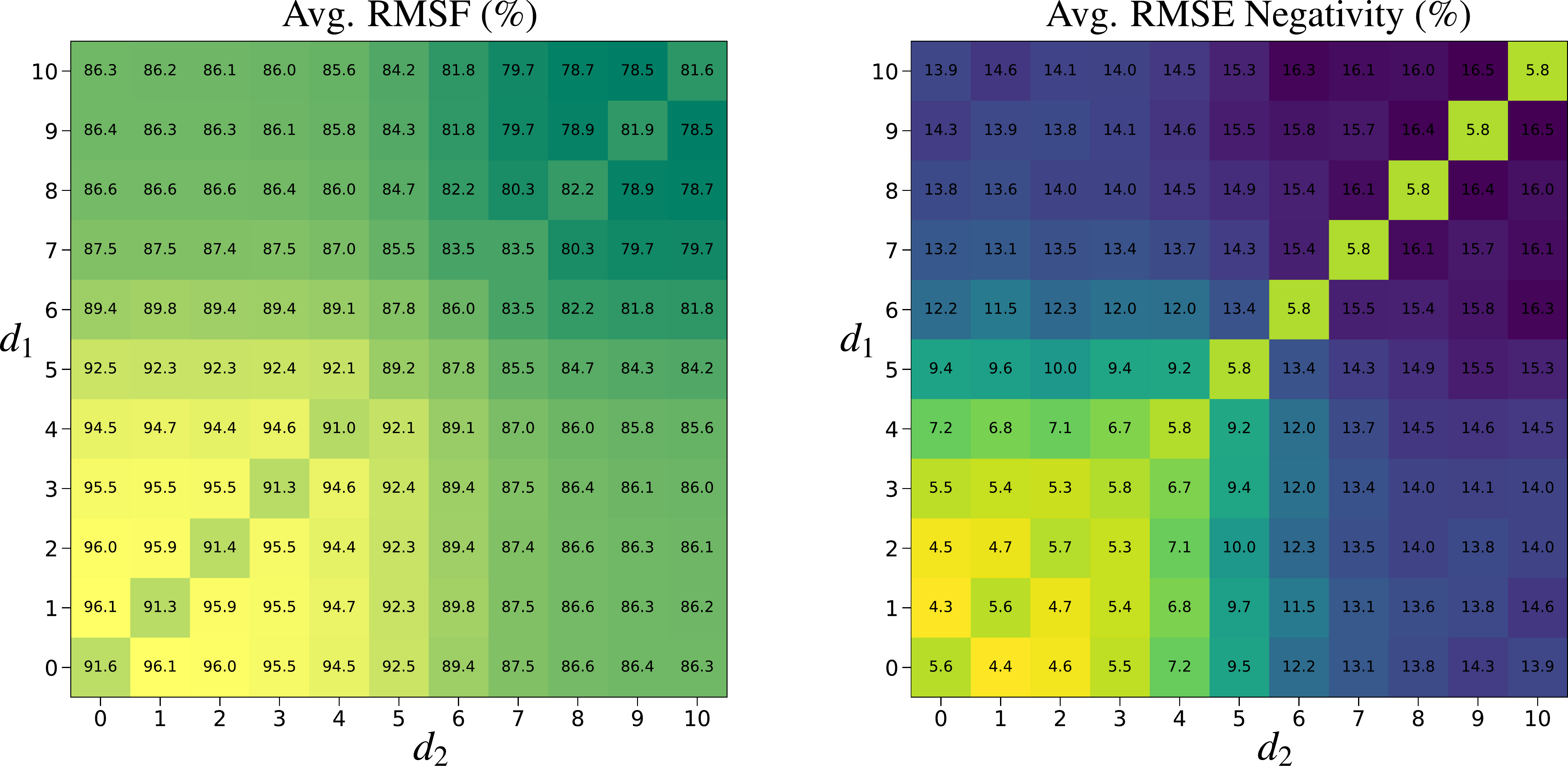}
		\protect\caption{The average RMSF and the average RMSE Negativity at combinations of $d_1$ and $d_2$ in the tomography task of the temporal entangler of channels' outputs over ten different runs with random trials of the input sequence and initial state. We set $\tau B = 10.0$ and keep other parameters the same as the parameters in Fig.~\ref{fig:qpt:channel:ent}.
		\label{fig:qpt:channel-delay}}
\end{figure}

\begin{figure}
		\includegraphics[width=18cm]{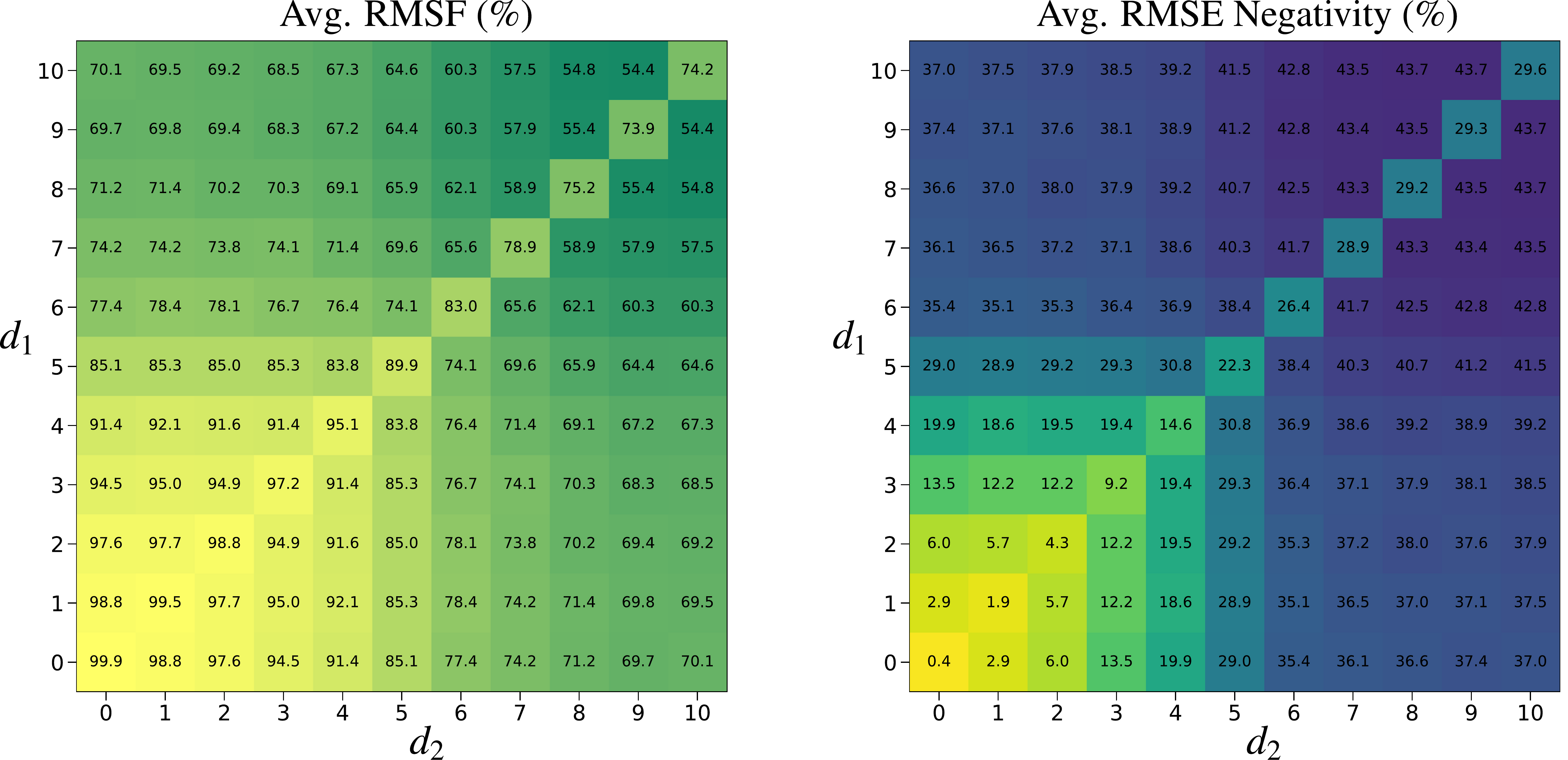}
		\protect\caption{The average RMSF and the average RMSE Negativity at combinations of $d_1$ and $d_2$ in the tomography task of the temporal Bell state creator over ten different runs with random trials of the input sequence and initial state. We set $\tau B = 10.0$ and keep other parameters the same as the parameters in Fig.~\ref{fig:qpt:channel:Bell}.
		\label{fig:qpt:Bell-delay}}
\end{figure}

Figures~\ref{fig:qpt:channel-delay} and~\ref{fig:qpt:Bell-delay} illustrates the average RMSF and the average Root Mean Square Error (RMSE) of negativities as a function of $d_1$ and $d_2$ over ten different runs with random trials of the input sequence and initial state.
Due to the effect of the short-term memory, the performance decreases as we increase $d_1$ or $d_2$.

\clearpage

\providecommand{\noopsort}[1]{}\providecommand{\singleletter}[1]{#1}%